\documentclass[aps,pre,twocolumn,superscriptaddress,longbibliography,nofootinbib,showkeys]{revtex4-1}
\pdfoutput=1

\usepackage[utf8]{inputenc}
\usepackage[T1]{fontenc}
\usepackage[american]{babel}
\usepackage{amsmath, amssymb}
\usepackage{csquotes}
\usepackage{enumerate}
\usepackage{graphicx}
\usepackage{mathtools}
\usepackage{color}
\usepackage{xcolor}
\usepackage[separate-uncertainty]{siunitx}
\usepackage{braket}
\usepackage{booktabs}
\usepackage{calc}
\usepackage{array}
\usepackage{enumitem}
\usepackage[percent]{overpic}

\usepackage{pdfpages}
\usepackage{etoolbox} 
\makeatletter
\patchcmd{\@outputpage@head}{\@ifx{\LS@rot\@undefined}{}{\LS@rot}}{}{}{}
\makeatother

\usepackage{pdflscape}

\usepackage[activate={true,nocompatibility}]{microtype}
\usepackage[colorlinks=true, linkcolor=blue, citecolor=blue, urlcolor=blue]{hyperref}
\hypersetup{
	pdftitle={Numerical-experimental observation of shape bistability of red blood cells flowing in a microchannel},
	pdfauthor={Achim Guckenberger, Alexander Kihm, Thomas John, Christian Wagner, Stephan Gekle},
	pdfkeywords={Hemodynamics, Red blood cells, Bistability, Phase diagram, Shapes, Croissant, Slipper, Experiments, Simulations, Boundary Integral Method}
}

\begin{document}
	\title{Numerical-experimental observation of shape bistability of red blood cells flowing in a microchannel}

	\author{Achim Guckenberger}
	\affiliation{Biofluid Simulation and Modeling, Fachbereich Physik, Universit\"at Bayreuth, Bayreuth}
	
	\author{Alexander Kihm}
	\affiliation{Experimental Physics, Saarland University, 66123, Saarbr\"ucken, Germany}
	
	\author{Thomas John}
	\affiliation{Experimental Physics, Saarland University, 66123, Saarbr\"ucken, Germany}

	\author{Christian Wagner}
	\affiliation{Experimental Physics, Saarland University, 66123, Saarbr\"ucken, Germany}
	\affiliation{Physics and Materials Science Research Unit, University of Luxembourg, Luxembourg, Luxembourg}

	\author{Stephan Gekle}
	\affiliation{Biofluid Simulation and Modeling, Fachbereich Physik, Universit\"at Bayreuth, Bayreuth}
	
	\date{November 19, 2017}

	\begin{abstract}
Red blood cells flowing through capillaries assume a wide variety of different shapes owing to their high deformability.
Predicting the realized shapes is a complex field as they are determined by the intricate interplay between the flow conditions and the membrane mechanics.
In this work we construct the shape phase diagram of a single red blood cell with a physiological viscosity ratio flowing in a microchannel.
We use both experimental in-vitro measurements as well as 3D numerical simulations to complement the respective other one.
Numerically, we have easy control over the initial starting configuration and natural access to the full 3D shape.
With this information we obtain the phase diagram as a function of initial position, starting shape and cell velocity.
Experimentally, we measure the occurrence frequency of the different shapes as a function of the cell velocity to construct the experimental diagram which is in good agreement with the numerical observations.
Two different major shapes are found, namely croissants and slippers.
Notably, both shapes show coexistence at low ($< \SI{1}{mm/s}$) and high velocities ($> \SI{3}{mm/s}$) while in-between only croissants are stable.
This pronounced bistability indicates that RBC shapes are not only determined by system parameters such as flow velocity or channel size, but also strongly depend on the initial conditions.

	\end{abstract}

	\maketitle

\section{Introduction}

Red blood cells (RBCs) are the major constituent of mammalian blood and therefore determine the majority of its flow properties.
One of the most amazing features of RBCs is their deformability, allowing them to squeeze through channels with diameters much smaller than their own equilibrium size \cite{Freund2013, Picot2015, Salehyar2016}.
Another consequence of their deformability is the wide range of stationary and non-stationary shapes assumed by the RBCs in microchannel flows with dimensions similar to or slightly larger than the RBC equilibrium radius \cite{Fedosov2014, Aouane2014, Tahiri2013}.
Understanding and being able to predict these shapes is of high importance for a variety of reasons.
From a fundamental point of view, it serves as the foundation in a bottom-up approach to understand the properties of red blood cell suspensions which are chiefly determined by single particle behavior \cite{Vitkova2008, Fedosov2011, Kruger2013, Thiebaud2014, Katanov2015, Lanotte2016}.
From an applied perspective, a series of recent investigations have devised promising approaches for sorting cells based on their mechanical properties either in lateral displacement devices \cite{Henry2016} or using high-speed video microscopy \cite{Otto2015}.
Finally, knowledge of the precise cell shape is also essential for accurately measuring geometric properties of cells \cite{Merola2017}.

\begin{figure*}[t]
	\centering
	\begin{overpic}[height=2.1cm]{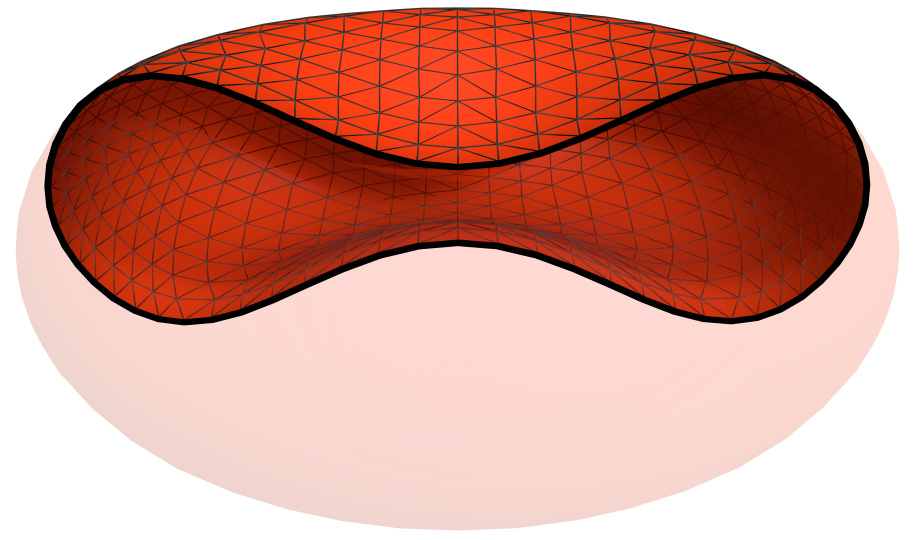}
		\put(0,55) {(a)}
	\end{overpic}
	\hfill
	\begin{overpic}[height=2.65cm]{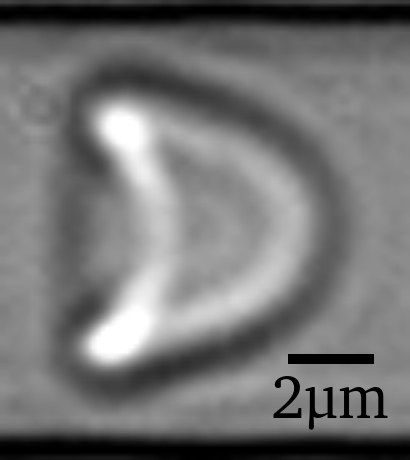}
		\put(70,80) {\color{white}(b)}
	\end{overpic}
	\begin{overpic}[height=2.65cm]{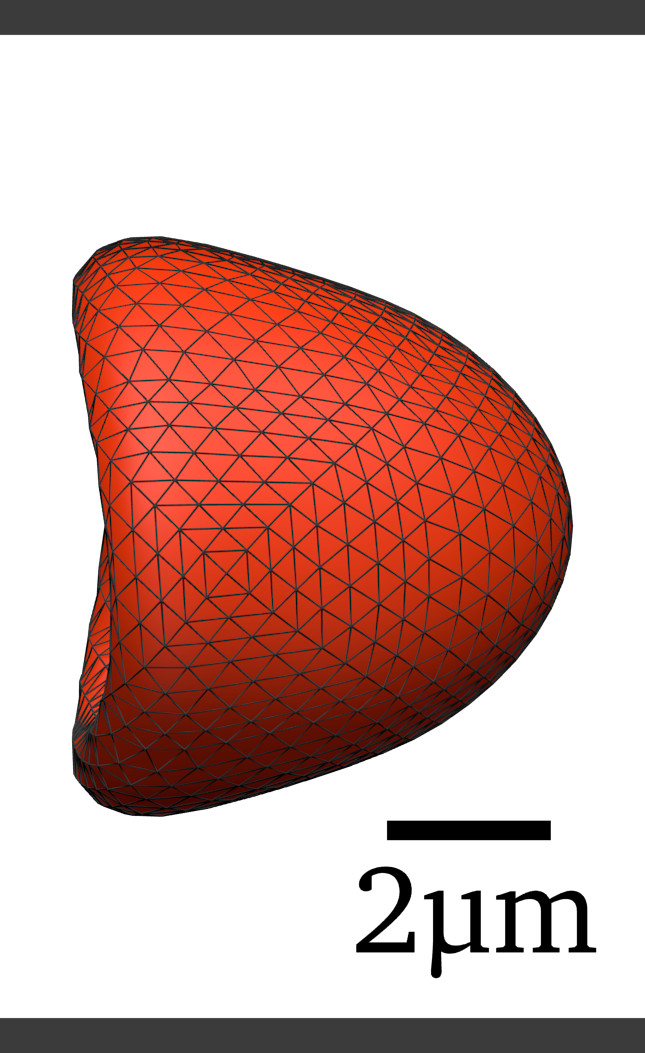}
		\put(45,80) {(c)}
	\end{overpic}
	\begin{overpic}[height=2.65cm]{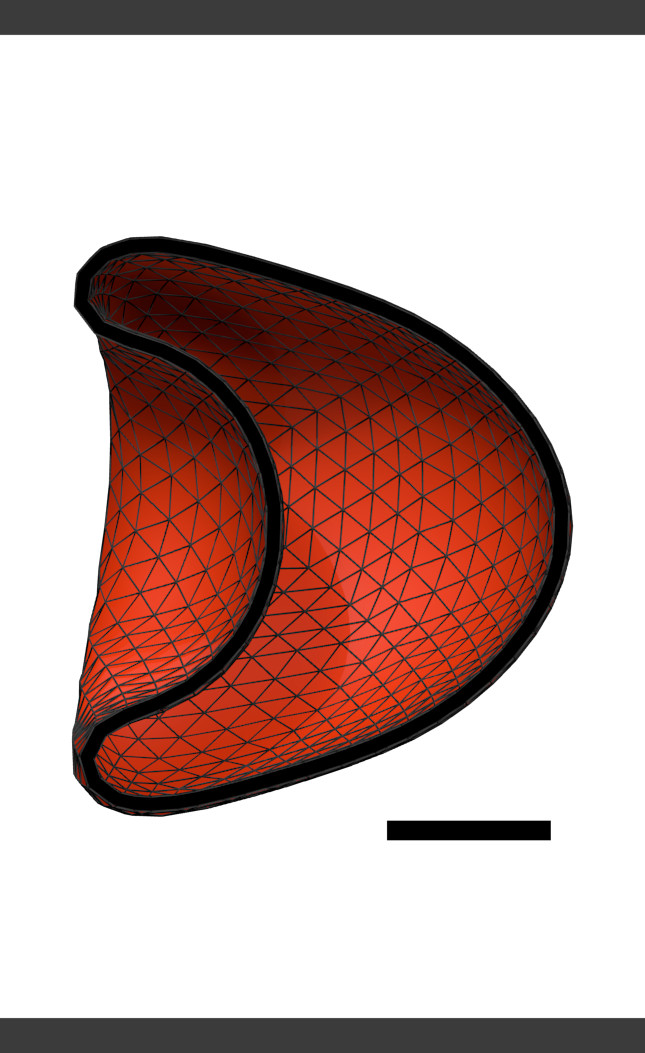}
		\put(45,80) {(d)}
	\end{overpic}
	\hfill
	\begin{overpic}[height=2.65cm]{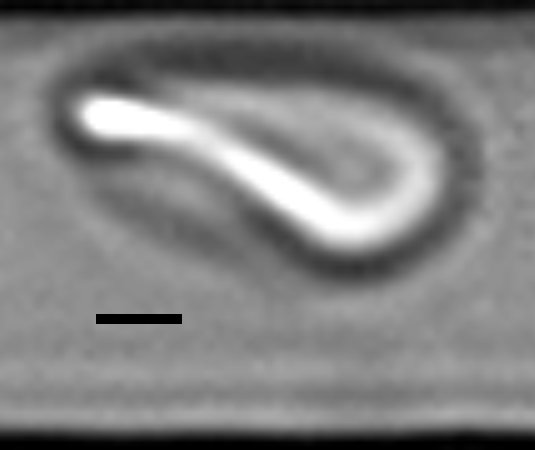}
		\put(85,65) {\color{white}(e)}
	\end{overpic}
	\begin{overpic}[height=2.65cm]{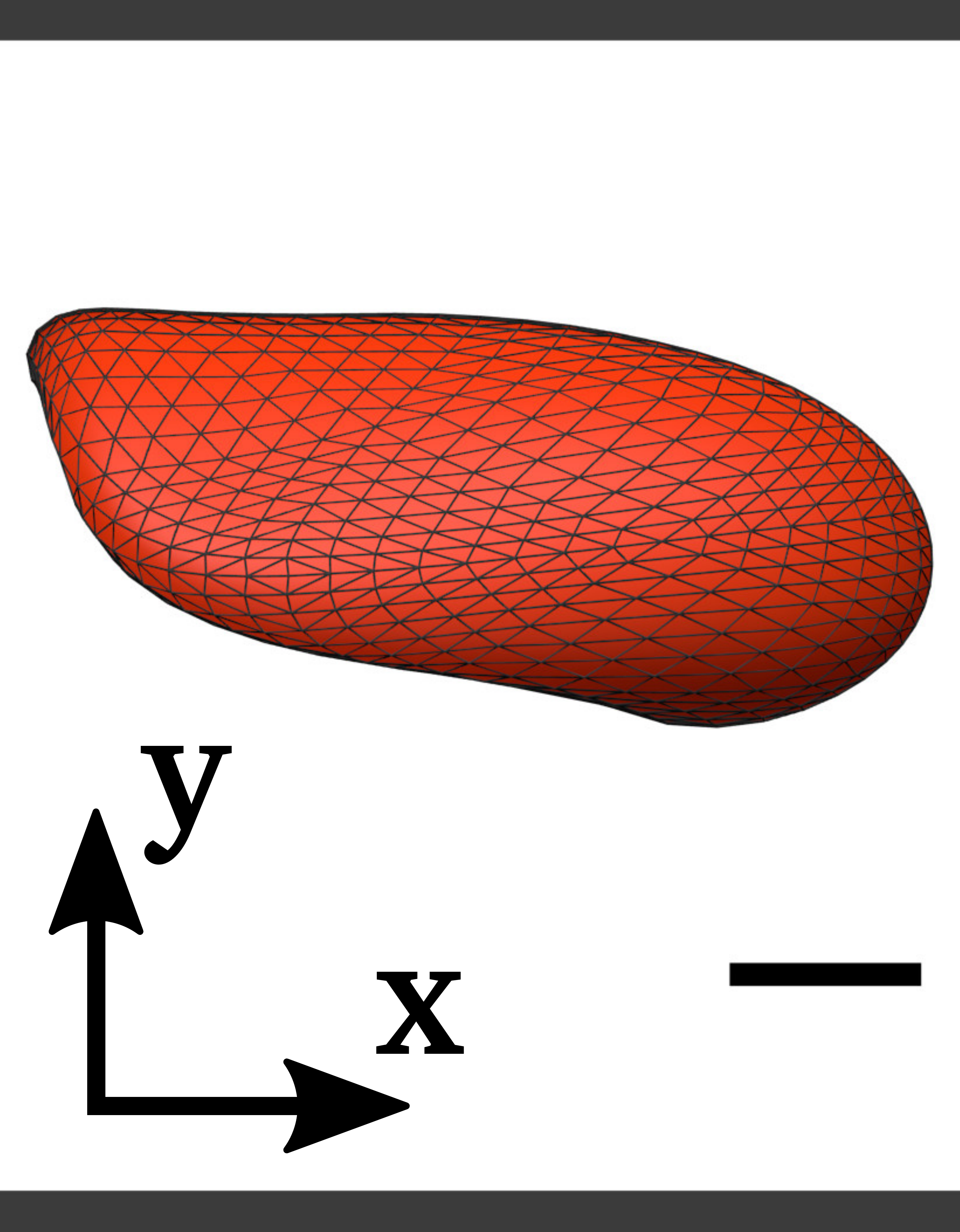}
		\put(55,80) {(f)}
	\end{overpic}
	\begin{overpic}[height=2.65cm]{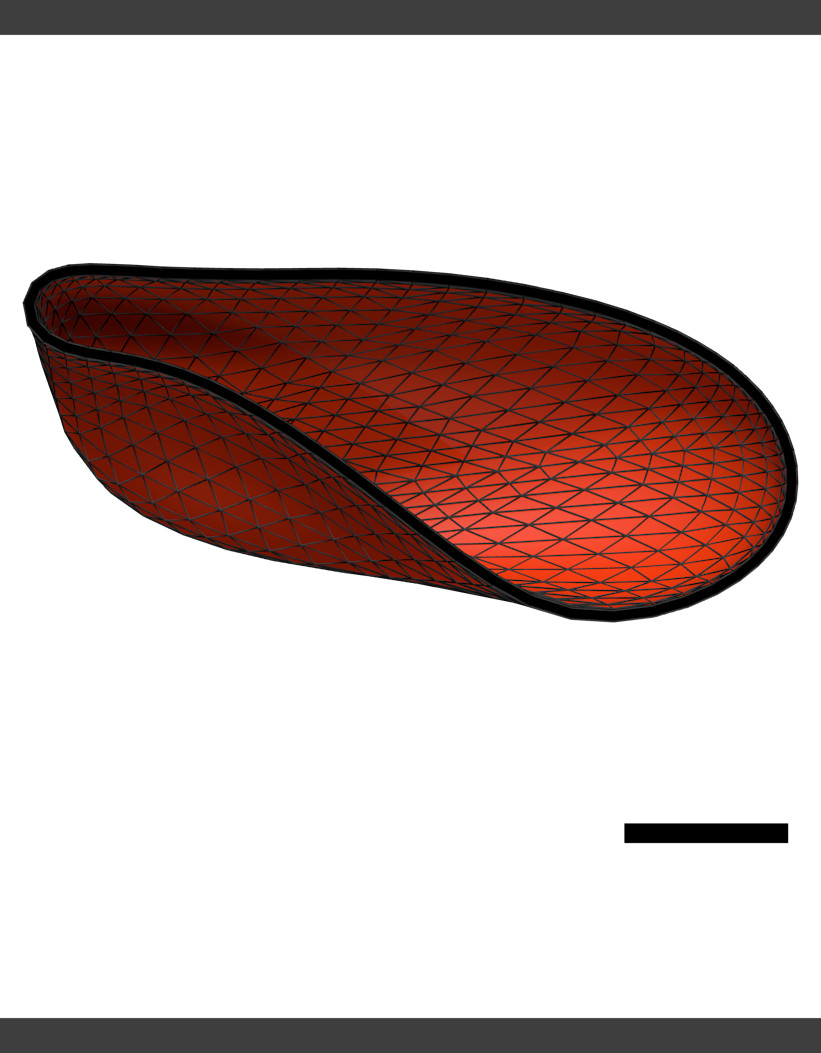}
		\put(55,80) {(g)}
	\end{overpic}
	\caption{%
		Typical RBC shapes from simulations and experiments.
		(a)~The typical discocyte shape employed in some of the simulations as the starting shape. Half of it was made transparent for illustration purposes. Its horizontal diameter is \SI{8}{\micro\meter}.
		(b)~A typical croissant observed in the experiments when applying a pressure drop of \SI{100}{mbar} (cell velocity \SI{0.98 +- 0.07}{mm/s}).
		(c)~A croissant with a velocity of $\approx \SI{1.1}{mm/s}$ obtained from the numerical simulations.
		(d)~The cross-section of the croissant from~(c).
		(e)~A slipper from the experiments at \SI{500}{mbar} (cell velocity \SI{5.16 +- 0.11}{mm/s}).
		(f)~A typical slipper from the simulations with a cell velocity of $\approx \SI{5.2}{mm/s}$.
		(g)~The cross-section of the slipper from~(f).
		The black lines on the shapes from the simulations depict the mesh.
		The bottom and top black lines in all figures are the walls ($L_y \approx \SI{12}{\micro\meter}$ apart), while the small black lines are scale bars of length $\SI{2}{\micro\meter}$.
		The flow is in the positive $x$-direction (except in figure~(a) where no flow exists).
	}
	\label{fig:TypicalShapes}
\end{figure*}
The most frequently observed shapes of RBCs in microchannel flows are the so-called \enquote{croissant} and \enquote{slipper} shapes. 
Examples are depicted in figure~\ref{fig:TypicalShapes}.
Some researchers refer to croissants also as parachutes, although here we prefer the term croissant since our shapes are not perfectly rotationally symmetric (similar to the ones found by \citet{Farutin2014a}).
Probably one of the earliest experimental study on isolated red blood cells in flow was performed by Gaehtgens \textit{et al.}~\cite{Gaehtgens1980}, where slippers as well as parachutes have been found depending on the diameter of the cylindrical channel.
Suzuki \textit{et al.}~\cite{Suzuki1996} presented a phase diagram of parachutes and slippers as a function of velocity and confinement in a cylindrical tube. Slippers dominated at smaller diameters and higher velocities.
Secomb \textit{et al.}~\cite{Secomb2007} compared experiments with 2D simulations in cylindrical channels of $\SI{8}{\micro\meter}$ diameter for a cell velocity of approximately \SI{1.25}{mm/s}.
Furthermore, two other publications~\cite{Faivre2006,Abkarian2008} considered the flow of RBCs at very low viscosity ratios of $\lambda \lesssim 0.27$. 
They presented a phase diagram showing parachutes and slippers, where the velocity was varied in the very high regime of $10$ to $\SI{170}{mm/s}$.
Tomaiuolo \textit{et al.}~\cite{Tomaiuolo2009} found parachutes at smaller and slippers at higher velocities in cylindrical channels of $\SI{10}{\micro\meter}$ diameter.
A subsequent study \cite{Tomaiuolo2011} as well as \citet{Prado2015} considered the transient during start-up of the flow.
\citet{Cluitmans2014} detected croissants at lower ($\lesssim \SI{5}{mm/s}$) and slippers at higher velocities ($\gtrsim \SI{10}{mm/s}$) in rectangular channels with widths $\leqslant \SI{10}{\micro\meter}$.
Moreover, \citet{Quint2017} found a stable slipper and a metastable croissant at the same set of parameters in a wider channel of $\SI{25}{\micro\meter} \times \SI{10}{\micro\meter}$.
Other publications presenting experiments in channel flow also touch the subject of RBC shapes but focus on other aspects such as the methodology \cite{Hochmuth1970, Seshadri1970, Zharov2006, Tomaiuolo2007, Guido2009, Gorthi2012, Lanotte2014, Tomaiuolo2016}, dense suspensions and cell interactions \cite{Claveria2016, Guest1963, Skalak1969, Gaehtgens1980, Kubota1996, Abkarian2008, Tomaiuolo2012,Wagner2013,Brust2014,Tomaiuolo2016} or use vastly larger channel diameters \cite{Goldsmith1972,Lanotte2016}.

Numerical simulations and semi-analytical calculations of isolated particles in microchannels mostly studied axisymmetric RBCs \cite{Secomb1986,Secomb1987,Secomb2001} or 2D vesicles \cite{Secomb1982, Kaoui2009,Kaoui2011,Kaoui2012,Shi2012a,Tahiri2013,Lazaro2014,Aouane2014}.
The numerical work by \citet{Aouane2014}, for example, identified a large amount of dynamics including deterministic chaos.
The first full 3D simulation of single cells with a realistic RBC model (but with a ratio of inner to outer viscosity of $\lambda = 1$) was conducted by Noguchi and Gompper \cite{Noguchi2005} who used a cylindrical tube with a diameter of $\SI{9.2}{\micro\meter}$.
They found the typical discocyte shape below and parachutes above a critical velocity which depends on the elastic parameters.
A subsequent study by the same group additionally explored this threshold as a function of confinement \cite{McWhirter2011}.
Moreover, Fedosov \textit{et al.}~\cite{Fedosov2014} presented very detailed phase-diagrams where the velocity and confinement was varied for three different sets of elastic moduli and a viscosity ratio of $\lambda = 1$.
They observed four distinct regions where snaking, tumbling, slippers and parachutes occurred.
Recently, \citet{Ye2017} considered the shapes of an RBC with $\lambda = 1$ in rectangular microchannels (with width \SI{10}{\micro\meter} and aspect ratios $1$ to $2$) for the three cell velocities $4$, $20$ and $\SI{100}{mm/s}$ and observation times up to $\approx \SI{0.03}{s}$.
Snapshots after this short initial transient showed parachutes or slightly slipper-like shapes.

Bistability, i.e.\ the observation of two different stable shapes depending on the initial condition but at otherwise identical system parameters, was barely considered so far.
It was observed only numerically for simpler situations such as close-to-spherical vesicles in unbounded Poiseuille flow \cite{Farutin2014a} or near a single wall \cite{Kaoui2009a}, for a 2D RBC model in bounded Poiseuille flow \cite{Secomb2007}, for the initial transient of a red blood cell in a rectangular channel \cite{Ye2017} or for simple shear flows \cite{Cordasco2014, Peng2014, Sinha2015, Lanotte2016}.
No systematic experimental investigations exist for cells flowing in microchannels.
Moreover, the 3D simulations and experimental investigations that were mentioned above and that consider the RBC shapes in microchannels in more detail all used a viscosity ratio of $\lambda \leqslant 1$, although 2D simulations showed that choosing a physiologically more realistic value of $\lambda \approx 5$ \cite{Cokelet1968} can significantly affect RBC dynamics \cite{Kaoui2012,Tahiri2013}.

Here we present a detailed systematic experimental-numerical study on the steady-state shape of isolated red blood cells in a rectangular microchannel.
We use the physiological viscosity ratio of $\lambda = 5$ appropriate for healthy human red blood cells in the microcirculation \cite{Cokelet1968}.
The initial position is varied in the simulations directly, while experimentally we determine it via measurements at the channel entrance.
Our central finding is that the initial starting position of the RBC has a decisive influence on the final steady-state shape of the red blood cell.

We begin by outlining our experimental and numerical methods in section~\ref{sec:methods}.
Afterwards, the results from our experiments (section~\ref{sec:exp}) and simulations (section~\ref{sec:sims}) are presented, while section~\ref{sec:comparison} is dedicated to their detailed comparison.
Finally, we conclude our work in section~\ref{sec:conclusion}.

\section{Methods}\label{sec:methods}

\subsection{Experimental setup}\label{sec:method:exp}

The sample preparation and experimental setup is mostly identical to the one used recently by \citet{Claveria2016}.
In short, human red blood cells were obtained from healthy donors by needle-prick and used within three hours.
After appropriate preparation \cite{Claveria2016}, they are suspended in a phosphate buffered saline (PBS) and bovine serum albumin solution which has a viscosity of approximately \SI{1}{mPa.s}.
The viscosity ratio of the cells is therefore $\lambda \approx 5$ \cite{Quint2017}.
This value corresponds to the typical physiological value of healthy red blood cells in blood plasma \cite{Cokelet1968}.
The RBCs are pumped through rectangular, PDMS-based channels by a high-precision pressure device (Elveflow OB 1, MK II) with pressure drops ranging from $20$ to $\SI{1000}{mbar}$ at room temperature.
The channels have a cross-section width of $L_y = \SI{11.9 +- 0.3}{\micro\meter}$ and a height of $L_z = \SI{9.7 +- 0.3}{\micro\meter}$ without any applied pressure drop
and are thus similar to the vessel diameters found in the microvascular system \cite{Popel2005}.
We use rectangular rather than cylindrical channels since they are easier to manufacture, are therefore prevalent in lab-on-a-chip devices and have the merit that cells are not rotated randomly around their axis due to the missing rotational symmetry. 
The latter property greatly simplifies the microscopic observation and analysis of the RBCs.

The hematocrit (volume percentage of RBCs) in the reservoir before the inlet is always \mbox{$\lesssim \SI{1.0}{\percent}$}, i.e.\ very low.
Nevertheless, we find cells flowing in clusters as well as single cells.
For the present work we have analyzed only the latter.
To this end, previous experimental and theoretical results showed that the hydrodynamic interaction in a linear channel decays exponentially, and becomes negligible if the inter-particle distance is more than twice the channel width \cite{Cui2002,Diamant2009,McWhirter2011}.
Considering that our channel has the dimensions $\approx \SI{12}{\micro\meter} \times \SI{10}{\micro\meter}$, cells can be considered as being single for distances $\gtrsim \SI{25}{\micro\meter}$.
We only used cells that were at least $\SI{40}{\micro\meter}$ apart from other entities.

We perform measurements at two locations along the channel, namely at the entrance ($x = \SI{0}{mm}$) and at $x = \SI{10}{mm}$ downstream.
Vessel lengths in-between bifurcations in the microvascular system are less than $\SI{1}{mm}$, i.e.\ much shorter \cite{Koller1987}.
Nevertheless, this is not necessarily true for in-vitro experiments or lab-on-a-chip devices, and the long-time behavior also holds information about the general intrinsic properties.
The flowing RBCs are recorded by an inverted bright-field microscope (Nikon TE \mbox{2000-S}) with an oil-immersion objective (Nikon CFI Plan Fluor $60\times$, $\mathrm{NA} = 1.25$) and a high-resolution
camera (Fastec HiSpec~2G) at a frame rate of 400 frames per second.
The camera is aligned along the $z$-direction so that the photographs show the cells in the $x$-$y$-plane (compare figure~\ref{fig:Setup}).
Hence, determination of the $z$-position is not possible, but also not absolutely necessary as our simulations always show a $z$-position of nearly~$0$ (see section~\ref{sec:comparison}).
We analyze the recorded image sequence with a custom MATLAB script that detects each projected cell shape and the corresponding 2D center of mass position.  
It additionally tracks the cell position over the image sequence to obtain the individual cell velocity.
Considering the optical setup, we assume an uncertainty in the position measurements of $\pm s_\mathrm{P}$ with $s_\mathrm{P} = \SI{0.1}{\micro\meter}$.

\subsection{Simulation setup}\label{sec:method:sims}

\begin{figure}[t]
	\centering
	\includegraphics[width=\columnwidth]{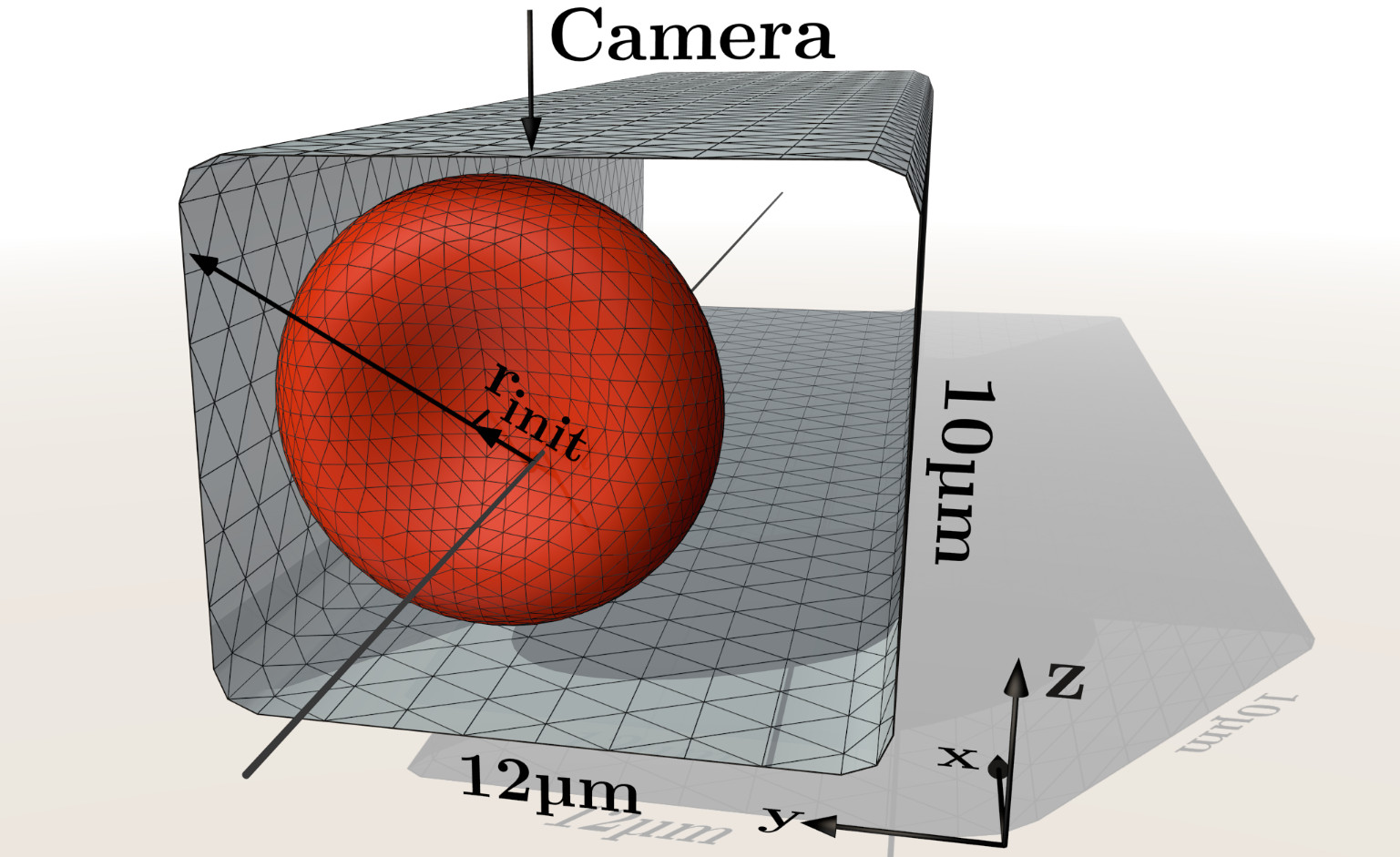}
	\caption{%
	Simulation setup: A single red blood cell is placed in a rectangular channel of width $L_y = \SI{12}{\micro\meter}$ and height $L_z = \SI{10}{\micro\meter}$.
	Periodic boundary conditions are employed.
	Initially, the centroid of the cell is offset from the center axis along the left black arrow by a distance $r_\mathrm{init}$.
	The depicted RBC illustrates the discocyte starting shape, although other shapes have been used, too.
	Furthermore, the black lines on the surfaces illustrate the employed meshes.
	The arrow at the top shows the view from the camera in the experiments (i.e.\ onto the $x$-$y$-plane) and the flow is in the positive $x$-direction.
	}
	\label{fig:Setup}
\end{figure}
The numerical simulations mimic our experimental setup as far as possible.
Hence, we place a single red blood cell in a rectangular channel as shown in figure~\ref{fig:Setup}.
The channel has a cross-section of width $L_y = \SI{12}{\micro\meter}$ and height $L_z = \SI{10}{\micro\meter}$.
Periodic boundary conditions are assumed in the $x$-direction with a periodicity of $L_x = \SI{42.7}{\micro\meter}$, in agreement with above estimates for the decay of hydrodynamic interactions.

We vary the initial $y$-$z$-position (relative to the channel center) of the RBC's centroid along the line $z_\mathrm{init} = 5 y_\mathrm{init} / 9$, which almost corresponds to the channel diagonal.
The corresponding initial radial position is thus simply given by $r_\mathrm{init} = \sqrt{y_\mathrm{init}^2 + z_\mathrm{init}^2}$.
When starting with the typical discocyte equilibrium shape \cite{Evans1972,Le2010a}, as depicted in figure~\ref{fig:TypicalShapes}(a), the RBC axis is aligned with the channel axis (as shown in figure~\ref{fig:Setup}).
Cell velocities are extracted by considering the difference of the centroids between successive time steps.
During the simulation, we monitor several quantities such as the radial, $y$- and $z$-positions, the RBC asphericity or the cell velocity as well as the full 3D shape to determine when a steady state has been reached.

Regarding the actual modeling of the constituents, the RBC is filled with a Newtonian fluid with a dynamic viscosity $\mu_\mathrm{RBC}$, whereas the ambient flow is a Newtonian fluid with the dynamic viscosity $\mu = \SI{1.2e-3}{kg/(s.m)}$ of blood plasma \cite{Chien1966, Skalak1989, Secomb2017}.
We set the viscosity ratio $\lambda = \mu_\mathrm{RBC} / \mu$ to a value of $5$ in all simulations.
The surface area of the RBC is set to $\SI{140}{\micro\meter^2}$ and the volume is set to $\SI{100}{\micro\meter^3}$ (see e.g.\ references~\citenum{Skalak1989} and \citenum{Kim2014a}), leading to a large radius of $R_\mathrm{RBC} = \SI{4}{\micro\meter}$ when the cell is in the typical discocyte equilibrium shape (figure~\ref{fig:TypicalShapes}(a)).
The mechanics of the infinitely thin membrane are governed by Skalak's law \cite{Skalak1973,Kruger2011} for the in-plane elasticity with a shear modulus of $\kappa_\mathrm{S} = \SI{5e-6}{N/m}$ \cite{Yoon2008, Freund2014} and an area dilatation modulus of $\kappa_\mathrm{A} = 100\,\kappa_\mathrm{S}$.
This value for $\kappa_\mathrm{A}$ ensures that the area changes remain below $\SI{2}{\percent}$ in all cases.
We take the reference state for the Skalak model to be the typical discocyte shape \cite{Evans1972,Le2010a}.
The membrane is additionally endowed with some bending resistance which is modeled according to the Canham-Helfrich law \cite{Canham1970, Helfrich1973, Guckenberger2017}, where the bending modulus is fixed to $\kappa_\mathrm{B} = \SI{3e-19}{N.m}$ \cite{Park2010, Freund2014}.
The spontaneous curvature is set to zero.

We use 2048 flat triangles to discretize the RBC in our numerical implementation.
The forces are computed as described by~\citet{Guckenberger2016}, with Method~C therein being used for the bending contribution.
An unavoidable artificial volume drift of the cell is countered by adjusting the velocity to obey the no-flux condition and by a subsequent rescaling of the object \cite{Farutin2014, BubblePreprint2}.
Moreover, the channel is represented by 2166 flat triangles.
The corners are rounded to prevent numerical problems (compare figure~\ref{fig:Setup}).
Rather than prescribing a zero velocity at the channel walls, we use a penalty method for efficiency reasons with a spring constant of $\kappa_\mathrm{W} = \SI{1.9e7}{N/m^3}$ \cite{BubblePreprint2, Tahiri2013}.
Increasing the triangle counts and the box length $L_x$ did not change the results significantly.

The Reynolds number in the considered system is defined as $\mathrm{Re} = 2 R_\mathrm{RBC} u_\mathrm{max} \rho / \mu$.
For a velocity of $u_\mathrm{max} \leqslant \SI{10}{mm/s}$ and the density $\rho \approx 10^3\,\si{kg/m^3}$ of the ambient and inner liquid we therefore have $\mathrm{Re} < 0.1$.
Hence, the flow can be appropriately described using the Stokes equation.
This allows us to employ the boundary integral method (BIM) \cite{Pozrikidis2001a} for 3D periodic systems~\cite{BubblePreprint2, Zhao2010}.
Note that this method requires to prescribe a certain average flow through the whole unit cell instead of a pressure drop within the channel.
The latter is unfortunately not easily accessible.
We therefore compare with experiments by means of cell velocities.
Continuing, the integrals are computed by a standard Gaussian quadrature with $7$ points per triangle in conjunction with linear interpolation of nodal quantities and appropriate singularity removal for the single- and double-layer potentials \cite{BubblePreprint2}.
Furthermore, we use the smooth particle mesh Ewald (SPME) method \cite{Saintillan2005} to accelerate the computation of the periodic Green's functions; cutoff errors are kept below $\num{5e-5}$.
The resulting linear system is solved via GMRES \cite{Saad1986} up to a residuum of $10^{-5}$, and the kinematic condition is integrated in time using the adaptive Bogacki-Shampine algorithm \cite{Bogacki1989} with the absolute tolerance set to $10^{-5} R_\mathrm{RBC}$.
When the run-times are normalized to a two-socket system with 28 cores, each simulation took 1 to 29 days, with an average of around 5 days.
The phase diagrams below are formed by 329 of such simulations in total.
Further details on the numerical method as well as verifications of the implementation can be found in our previous publications~\cite{BubblePreprint2, Guckenberger2016, Daddi-Moussa-Ider2016, Quint2017}.

\section{Experimental results}\label{sec:exp}

We classify cells in the experiments either as croissants, slippers or \enquote{other} not uniquely identifiable or completely different shapes.
Typical slipper and croissant shapes are shown in the photographs~(b) and~(e) of figure~\ref{fig:TypicalShapes}.
See the supplementary information (SI) for a collection of all images.

\begin{figure*}[htpb]
	\centering
	\hfill
	\includegraphics[width=0.45\linewidth]{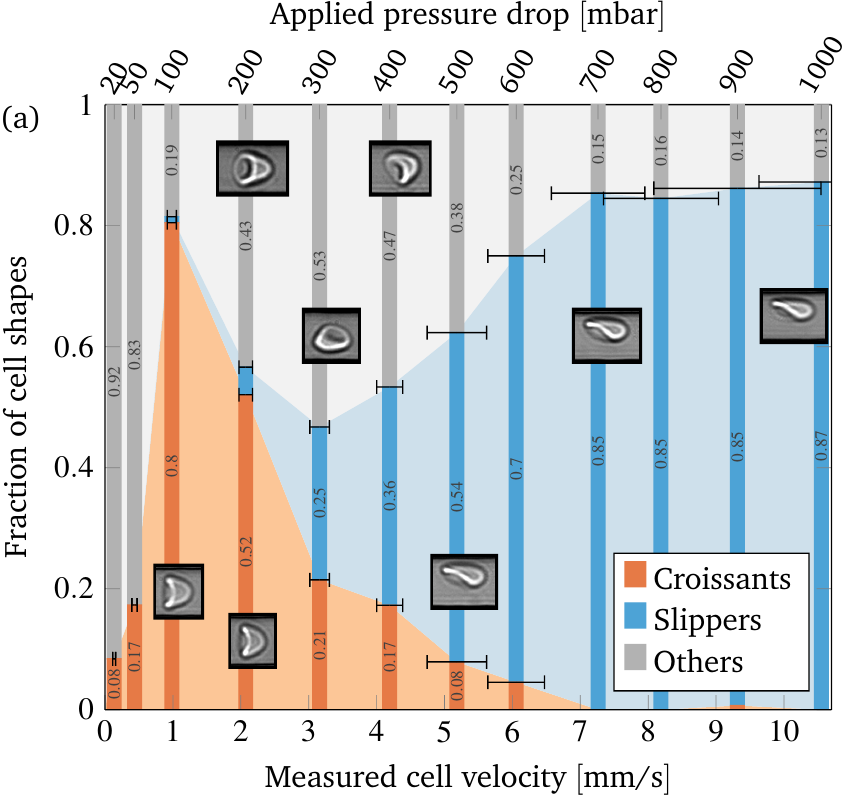}
	\hfill
	\includegraphics[width=0.9\columnwidth]{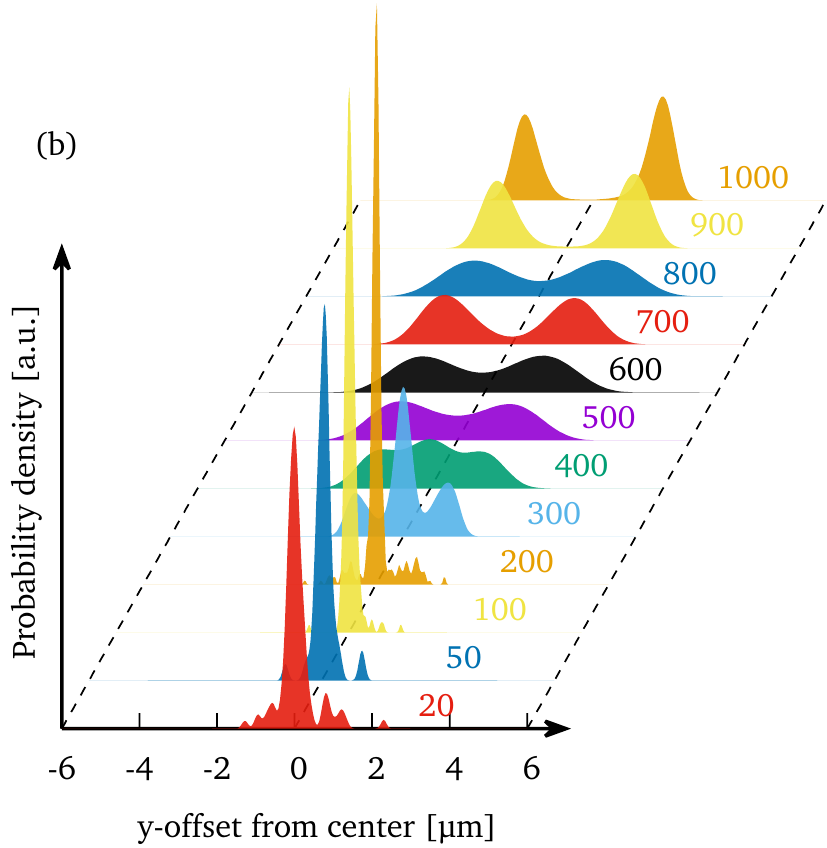}
	\hfill
	\caption{%
	Experimental results:
	(a)~Fraction of observed cell shapes as a function of the applied pressure drop (top axis) and mean cell velocity (bottom axis).
	The horizontal error bars depict the standard deviation of the measured cell velocities for each applied pressured drop.
	The shaded background is a guide to the eye.
	Furthermore, the insets show examples of experimental images (see the SI for a collection of all photographs).
	(b)~Estimated probability density function of the RBCs' center-of-mass $y$-position within the channel for various pressure drops (indicated as numbers on the right in millibar) for all shapes combined.
	We show the separated contributions of each shape to the distribution in the supplementary information.
	The area under the curves is normalized to one.
	The dashed lines illustrate the wall positions.
	Both figures are for the position \SI{10}{mm} downstream from the channel entrance.
	}	
	\label{exp10mm}
\end{figure*}
To systematically investigate the occurrence of the different shapes, we vary the imposed pressure drops from 20 to \SI{1000}{mbar}.
The corresponding cell velocities range from \SI{0.14}{mm/s} to \SI{10.6}{mm/s}, covering the whole physiological range in microchannels \cite{Pries1995, Popel2005, Baskurt2011}.
We consider the cells \SI{10}{mm} away from the channel entrance where most of the cells reached a steady state \cite{Claveria2016}.
Figure~\ref{exp10mm}(a) depicts the fraction of observed shapes as a function of the measured cell velocities, constituting our central result from the experiments.
This distribution was obtained by considering typically more than 100 cells per imposed pressure drop.
The average velocities were computed by averaging over all cells at a certain pressure drop, with the horizontal error bars showing the corresponding standard deviations $\sigma_u$ in cell velocity.
Not all velocities are the same because croissants and slippers have different velocities at otherwise identical flow conditions \cite{Quint2017}, and because of the natural variations of cell properties such as elasticity and size, as also noted by \citet{Tomaiuolo2009}.
See the supplementary information for more details.
Considering figure~\ref{exp10mm}(a), high velocities obviously favor slippers while croissants are the most prominent for medium velocities. A pronounced peak exists from around $1$ to $\SI{2}{mm/s}$.
Very small velocities produce mostly shapes that fall outside our simple two-state classification.

Figure~\ref{exp10mm}(b) illustrates the corresponding estimated probability density function of the center of mass $y$-position of the cells at the various pressure drops.
This estimate was obtained from the measured $y$-positions by using the kernel density estimator as implemented in MATLAB R2017a (\verb|ksdensity|) with a support of $[-6, 6] \, \si{\micro\meter}$ and otherwise default settings.
Thus, croissants and \enquote{others} occurring at lower velocities are centered in the channel, while slippers occurring at high velocities show a pronounced off-centered position.
The assumed shapes therefore imply a certain $y$-position within the channel with slippers being off-centered and croissants centered.
This is confirmed when analyzing the offset distribution separately for each shape class as shown in the supplementary information.

\begin{figure}[htpb]
	\centering
	\includegraphics[width=0.85\columnwidth]{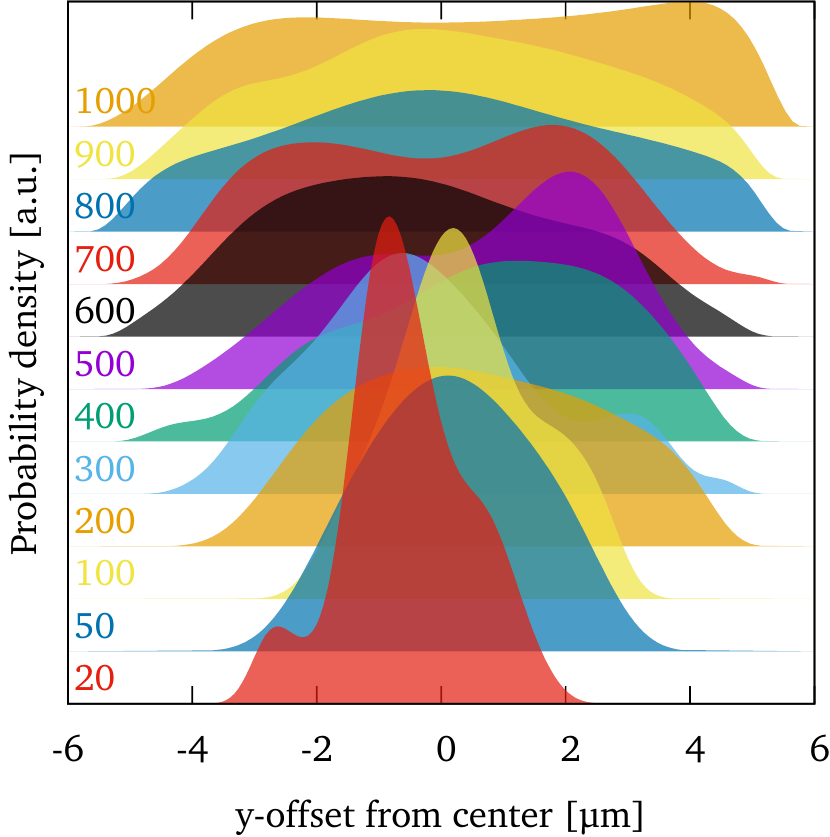}	
	\caption{%
	Experimental results:
	Estimated probability density function of the cells' center-of-mass $y$-position at the channel entrance (position $x = \SI{0}{mm}$).
	The pressure drops increase from the bottom (\SI{20}{mbar}) to the top (\SI{1000}{mbar}) with the numbers on the left side indicating the corresponding value in millibar.
	The area under the curves is normalized to one.
	The curves are offset in the vertical direction for illustration purpose.
	}
	\label{exp0mm}
\end{figure}

From figure~\ref{exp10mm}(a) it is tempting to conclude that the flow velocity is the major parameter that determines the RBC shape with low velocities favoring centered and high velocities favoring off-centered flow positions.
However, looking at the cell positions near the channel entrance (figure~\ref{exp0mm}) we find that already upon entering the channel RBCs are not homogeneously distributed. At low velocities we observe a clear bias towards a centered initial position, with the distribution becoming approximately homogeneous only at the highest measured velocities.
These experimental observations allow two distinct parameters as the reason for the dominance of the slipper shapes at high velocities: either the higher flow velocity itself or the more off-centered entry into the channel. 
To disentangle these two possibilities we now present numerical simulations whose geometry directly corresponds to the experimental setup.

\section{Numerical results}\label{sec:sims}

We numerically study the behavior of a single RBC in a rectangular microchannel by varying the imposed flow velocity, the initial shape and the initial offset $r_\mathrm{init}$ from the centerline of the tube (see section~\ref{sec:method:sims}).
After starting the flow, we wait until the RBC reaches the steady state where the shape as well as the radial position does no longer change, or alternatively until periodic motion is observed.

\begin{figure}[t]
	\centering
	\includegraphics[width=\columnwidth]{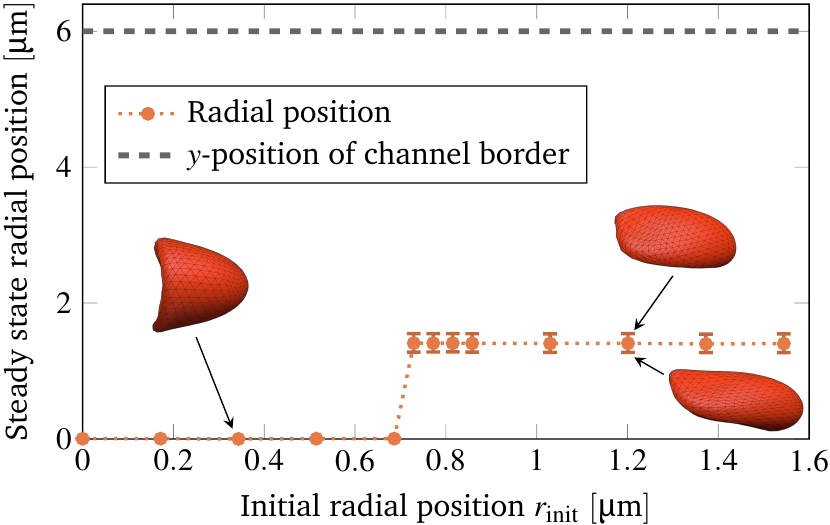}
	\caption{%
	Simulation results:
	Averaged radial position in the steady state as a function of the initial radial offset for a cell velocity of $\approx \SI{6.5}{mm/s}$. 
	The RBC starts in the typical discocyte shape with its rotation axis aligned with the tube's axis (figure~\ref{fig:Setup}).
	The dotted line is a guide to the eye.
	Half of the channel's extent along the $y$-direction (width) is shown as a dashed line at the top. 
	The extent in the $z$-direction (height) is of less significance here since the steady states are always almost centered in the $z$-direction.
	Furthermore, the radial position for the converged slippers oscillates around a mean value and their shapes show periodic \enquote{contractions} as indicated by the vertical error bars and the right two insets, respectively.
	}
	\label{fig:RadPosOffsetVaried}
\end{figure}

In the majority of cases, we observe two different states: A croissant shape (which moves as a rigid body, figure~\ref{fig:TypicalShapes}(c)) and a slipper shape (figure~\ref{fig:TypicalShapes}(f)).
The latter exhibits tank-treading (TT) and oscillatory contractions similar to the slippers seen by \citet{Fedosov2014} (see the SI for a movie and the insets in figure~\ref{fig:RadPosOffsetVaried}).
Tank-treading refers to the motion of the membrane around a (more or less) static shape.
Note that perfectly axisymmetric parachutes are suppressed by the rectangular channel flow, contrary to the situation for cylindrical tubes \cite{Fedosov2014} or unbounded Poiseuille flows \cite{Farutin2014a}.

To start the systematic study, we take a red blood cell that is initially in the typical discocyte shape with its rotation axis aligned along the tube's axis (cf.\ fig.~\ref{fig:Setup}).
We then vary the radial offset~$r_\mathrm{init}$ from the center line as described in section~\ref{sec:method:sims} and record the final radial position as well as the shape.
The mean of the radial position is extracted by a temporal average once the cell is in the steady state (see the supplementary information for more details).
Figure~\ref{fig:RadPosOffsetVaried} shows the result for a cell velocity of $\approx \SI{6.5}{mm/s}$.
A single sharp transition at $r_\mathrm{init} \approx \SI{0.7}{\micro\meter}$ from centered croissants to off-centered slippers is observed.
The final position of the slippers is mostly offset only along the wider width of the channel ($y$-direction), but not along the smaller height ($z$-direction).
Hence we find pronounced bistability: The result is significantly determined by the initial condition and two different shapes coexist.
This is consistent with the 2D simulations by Secomb \textit{et al.}~\cite{Secomb2007} and Tahiri \textit{et al.}~\cite{Tahiri2013}.
It also agrees qualitatively with observations by Farutin and Misbah for 3D simulations of vesicles in unbounded Poiseuille flow~\cite{Farutin2014a}.

\begin{figure*}[t]
	\centering
	\includegraphics[height=5.7cm]{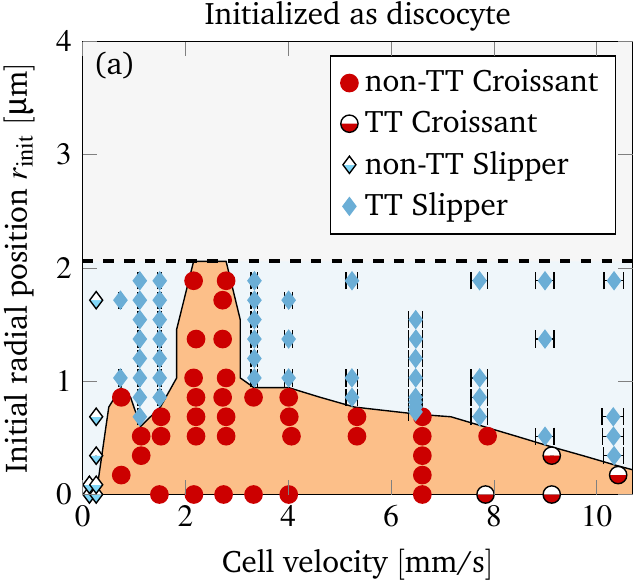}
	\hfill
	\includegraphics[height=5.7cm]{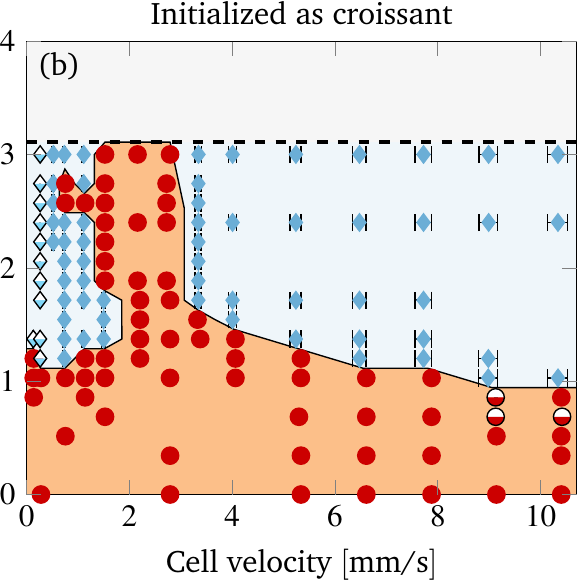}
	\hfill
	\includegraphics[height=5.7cm]{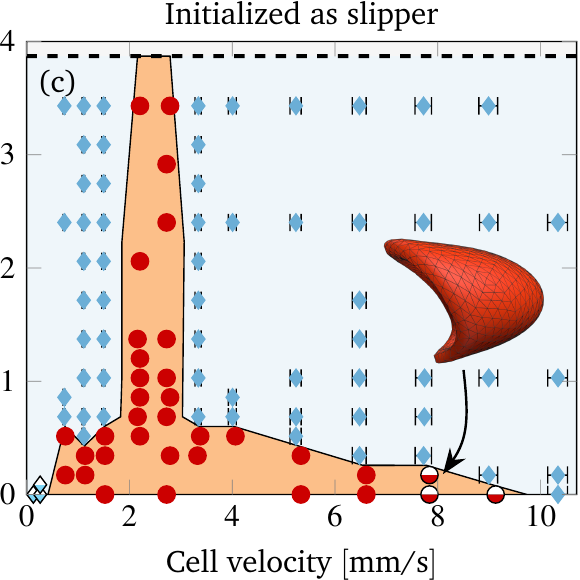}
	\caption{%
	Simulation results:
	Shapes obtained when varying the initial offset $r_\mathrm{init}$ and the velocity.
	Each symbol corresponds to one simulation.
	The horizontal axis shows the average cell velocity in the steady state, while horizontal error bars depict the minimal and maximal velocities in one period (variations for croissants nearly zero and thus not visible).
	The upper dashed line represents the maximal initial offset: Above this offset, the cell would overlap with the wall.
	The other lines and the colored areas are guides to the eye and illustrate the different regions in the phase diagram.
	Each figure corresponds to a different initial shape, namely 
	(a)~to the typical discocyte shape,
	(b)~to a croissant and 
	(c)~to a slipper.
	These shapes are shown in figure~\ref{fig:TypicalShapes}(a) and in the SI.
	The inset in the last figure depicts an example of a tank-treading croissant.
	Figure~\ref{fig:RadPosOffsetVaried} corresponds to the vertical column at $\approx \SI{6.5}{mm/s}$ in sub-figure~(a).
	}
	\label{fig:PhaseDiagNormal}
\end{figure*}

To study the bistability in more detail, we vary the imposed flow velocity as well as the initial offset $r_\mathrm{init}$ and characterize the behavior in the steady state.
This yields the shape phase diagram depicted in figure~\ref{fig:PhaseDiagNormal}(a).
The cell velocity is extracted in the steady state via a temporal average.
For slippers the velocity varies periodically (similar to the radial position): the minimum and maximum in one period is indicated by the horizontal error bars.
Overall, the mean cell velocity~$u$ ranges from $\SI{0.132}{mm/s}$ to $\SI{10.4}{mm/s}$, matching with the experimentally covered range.
The corresponding shear capillary number $\mathrm{Ca}_\mathrm{S} := \mu u / \kappa_\mathrm{S}$ varies therefore in the interval $\mathrm{Ca}_\mathrm{S} \in [0.0317, 2.50]$,
while the bending capillary number $\mathrm{Ca}_\mathrm{B} := \mu u R_\mathrm{RBC}^2 / \kappa_\mathrm{B}$ lies in the range $\mathrm{Ca}_\mathrm{B} \in [8.45, 666]$.
The reddish area illustrates the approximated region where croissants exist.
Furthermore, there is a maximal initial offset $r_\mathrm{init}$ above which overlapping with the vessel wall would occur.

The shape phase diagram in figure~\ref{fig:PhaseDiagNormal}(a) (together with~(b) and~(c) explained below) constitutes our main result from the simulations.
Starting near the channel center (in the reddish region) results in croissants, whereas higher initial offsets lead to slippers.
The transition is found to be sharp, and depends significantly on the velocity. Croissants are the only stable steady state in a small region ranging from around $2$ to $\SI{3}{mm/s}$, independently of the initial radial position.
Smaller and larger velocities tend to favor slippers.
Stable croissants do not appear below $\SI{0.25}{mm/s}$.
In the case of the slippers, the final periodic state is usually reached after roughly $2$ to $\SI{10}{s}$.
In contrast, the final croissant state is sometimes achieved only after more than $10$ to $\SI{30}{s}$, possibly after an intermediate slipper state that can last several seconds (see figure~S4 and the movie in the supplementary information).
Hence, shapes observed after less than one second often turn out to be transient, contrary to the interpretation of \citet{Ye2017} but in agreement with \citet{Prado2015}.

Considering our results in figure~\ref{fig:PhaseDiagNormal}(a) in more detail, we find that two different types of croissants and slippers are possible.
On the one hand, at very low velocities ($\lesssim \SI{0.7}{mm/s}$) the slippers no longer exhibit tank-treading motion of the membrane and instead show tumbling behavior: The cell rotates around the $z$-axis while approximately preserving its shape (similar to a rigid-body, see the SI for a movie).
The difference compared to the tumbling motion observed by \citet{Fedosov2014} is that the cell still exhibits a clear slipper-like instead of a proper discocyte shape.
Hence, we classify this mode still as slipper.
On the other hand, at very high velocities ($\gtrsim \SI{7}{mm/s}$) slightly asymmetric shapes strongly reminiscent of croissants with a distinct tank-treading motion can sometimes be observed (see the inset in figure~\ref{fig:PhaseDiagNormal}(c) for an example).
As the shape itself is very close to a croissant, we will nevertheless consider it to be a croissant below.

A natural question that occurs in light of the profound bistability is the influence of other initial shapes on the result.
To this end, we consider a typical croissant as well as a typical slipper as the starting shape.
Both were obtained from previous simulations that started with the discocyte form and are depicted in the supplementary information.
We once again construct the shape phase diagram as before and display the results in figures~\ref{fig:PhaseDiagNormal}(b) and~(c).
Note that the different starting shapes admit a larger initial radial position $r_\mathrm{init}$ of the centroid.
In short, starting with a croissant favors croissants in the steady state (the reddish area is larger than in figure~\ref{fig:PhaseDiagNormal}(a)).
For slippers it is the other way around: Starting with a slipper tends to produce more slippers (reddish area smaller than in figure~\ref{fig:PhaseDiagNormal}(a)).
Despite this, the croissant-only region from around $2$ to $\SI{3}{mm/s}$ still exists unscathed.
Overall, only two qualitative differences occur between the phase diagrams of different initial shapes, both at lower velocity when starting with the croissant shape (figure~\ref{fig:PhaseDiagNormal}(b)): First, stable croissants emerge at very low velocities ($\lesssim \SI{0.7}{mm/s}$) and second, the croissant-only peak exhibits a \enquote{protrusion} into the slipper space.
This observation suggests that slippers and croissants can be stable below \SI{2}{mm/s} for most $r_\mathrm{init}$ values, although in some cases a very precise croissant configuration is required in order to actually get a croissant in the steady state.

\begin{figure*}[htpb]
	\centering
	\includegraphics[height=6.5cm]{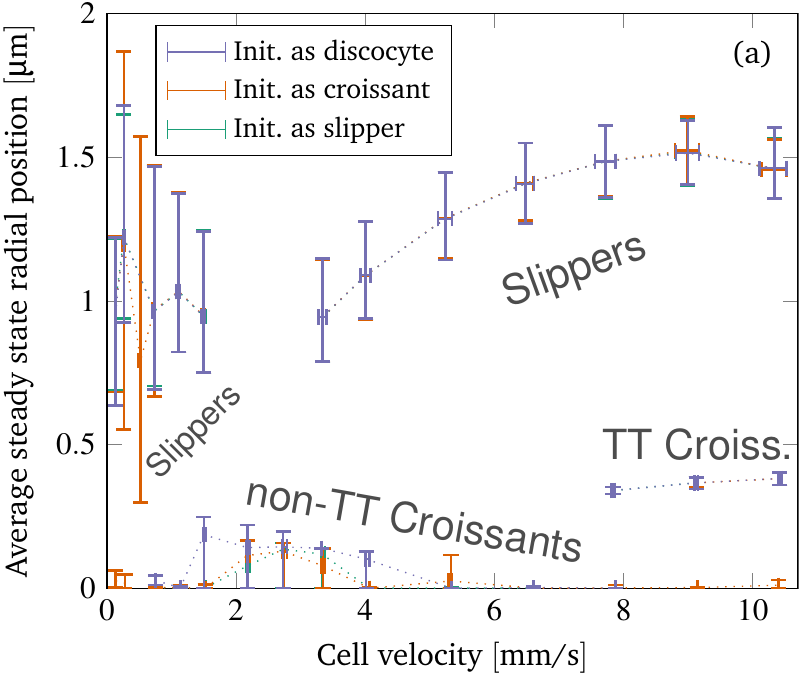}
	\hfill
	\includegraphics[height=6.5cm]{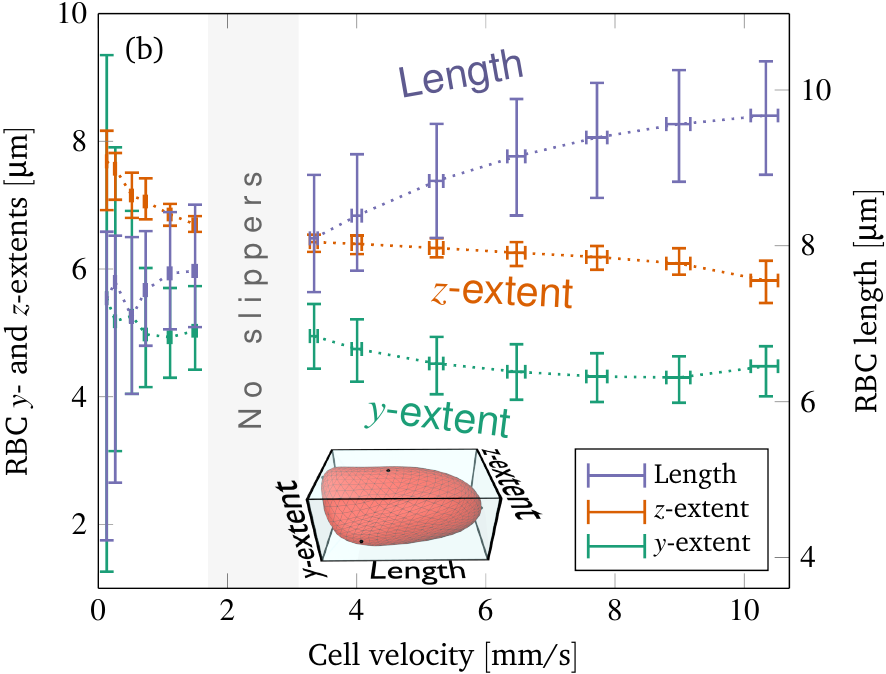}
	\caption{%
	Simulation results:
	(a)~Average radial positions of the steady states from figure~\ref{fig:PhaseDiagNormal} as a function of cell velocity for the three different starting shapes.
	The lower curves are for non-TT croissants and TT croissants, the upper curves are for (TT and non-TT) slippers. 
	We show on the vertical axis the weighted temporal mean of the radial centroid position of RBCs that assume the same shapes.
	The vertical error bars depict the total minimal and maximal position, while the horizontal error bars show the total minimal and maximal cell velocities (in each period of the steady states, respectively).
	(b)~Extents of the slipper shapes from figure~(a) in the flow ($x$-)direction (length) and along the other two axes, as illustrated by the inset showing the channel-aligned bounding box around a slipper.
	The vertical error bars depict the minimum and maximum extents during the periodic contractions, while the horizontal error bars are the same as in~(a).
	}
	\label{fig:AvgRadPos}
\end{figure*}

Another interesting aspect concerns the radial positions of the centroids in the final steady states.
The average values are obtained by computing the temporal average in the steady state first for each simulation, and then combining the results for identical shapes via a weighted arithmetic mean.
We use the observation time in the steady state as the weight.
This procedure leads to figure~\ref{fig:AvgRadPos}(a).
Obviously, the final radial positions are independent of the initial starting shape, i.e.\ a particular steady state shape at a certain velocity is always located at the same position.
Furthermore, non-tank-treading croissants are always almost centered, with only minor deviations away from zero.
These slight deviations in the range from 2 to \SI{4}{mm/s} are mainly due to some croissants exhibiting minuscule periodic shape deformations.
Moreover, the centroids of tank-treading croissants occurring at velocities $\gtrsim \SI{8}{mm/s}$ are located near but not directly in the center.
Their slight off-centered position is a result of their asymmetry.

In contrast to croissants, slippers are located $0.8$ to $\SI{1.5}{\micro\meter}$ away from the channel's axis.
The minimum position is attained for velocities near the border of the croissant-only region in the phase diagram (at around $2$ and $\SI{3}{mm/s}$, compare figure~\ref{fig:PhaseDiagNormal}).
Above, the off-center position increases and seems to converge to a value of around $\SI{1.5}{\micro\meter}$.
The reason for this increase is that slippers become more elongated and thinner at higher velocities (up to a certain degree), as shown in figure~\ref{fig:AvgRadPos}(b) and also observed in previous experiments \cite{Tomaiuolo2009}.
Thus, they effectively become smaller in the radial direction and their centroids can move closer to the wall.
We note that the distance between the wall and the upper side of the slipper approximately remains the same for all velocities.
This also hints at that the \enquote{optimal} off-center position for the slippers is more than $\SI{1.5}{\micro\meter}$ away from the center, and that this particular value is due to the smallness of the channel.

\section{Comparison between experiments and simulations}\label{sec:comparison}

\subsection{Comparison of shapes}
Considering figure~\ref{fig:TypicalShapes}, the croissants obtained from simulations and experiments look very similar, although the experimental shapes appear to be somewhat larger.
The reason is diffraction: The \enquote{true} cell border lies in the bright and not within the dark rim. 
However, the slippers appear to look qualitatively different.
This is due to the high magnification and numerical aperture of the objective which results in a small depth of field of around \SI{1}{\micro\meter}.
Cell borders above and below the middle plane are therefore blurred out and become invisible while the mid-plane cut becomes dominant.
Thus, for comparison we should use the middle cross-section of the numerically obtained shapes.
Here we find good agreement (compare figure~\ref{fig:TypicalShapes}(g) with~(e)).

\subsection{Comparison of the phase diagrams}
A qualitative comparison between the phase diagrams of steady states from the experiments (fig.~\ref{exp10mm}(a)) and the simulations (fig.~\ref{fig:PhaseDiagNormal}) shows a striking resemblance: Both exhibit a distinct peak in the number of croissants at lower velocities (1 to \SI{3}{mm/s}) at the expense of the number of slippers. The latter dominate the picture at high velocities ($> \SI{7}{mm/s}$).
At intermediate velocities both shapes coexist and can therefore be observed simultaneously in measurements.
Moreover, the simulations at very low velocities showed croissants only if the initial RBC was already prepared in that state, meaning that in the experiments this shape is highly unexpected. 
Indeed, we were not able to clearly classify most of the observed shapes in that regime as either croissants or slippers.

Obtaining a direct quantitative comparison requires a translation of the numerical threshold in figure~\ref{fig:PhaseDiagNormal} (which is in terms of the initial offset) into a \emph{prediction} regarding the fraction of shapes, because the experimental phase diagram is in terms of the observed fraction of shapes.
This is done by counting the fraction of croissants entering the channel with an offset below the numerical threshold. This fraction corresponds directly to the predicted fraction of croissant shapes.
More precisely, we first define $r_\mathrm{trans}$ as the initial radial offset which separates croissants from slippers in the simulations by using the black line in figure~\ref{fig:PhaseDiagNormal}.
An exception is the small croissant-only region (i.e.\ the interval of the topmost horizontal line in figure~\ref{fig:PhaseDiagNormal}) where we take $r_\mathrm{trans} \to \infty$.
This is consistent with our interpretation that only croissants exist in this particular interval.
One $r_\mathrm{trans}$ is computed for each experimental cell velocity from figure~\ref{exp10mm}~(a).
Second, each radial position $r_\mathrm{trans}$ is projected onto the $y$-axis to give $y_\mathrm{trans}$ (see sec.~\ref{sec:method:sims}) because only the $y$-offset is known from experiments.
Third, from the experimental offset distribution at the channel entrance (figure~\ref{exp0mm}) we can then estimate the fraction of cells~$\phi$ that enter the channel with an offset below $y_\mathrm{trans}$.
Accordingly, the simulations predict a fraction~$\phi$ of croissants in the steady state. The value of~$\phi$ can thus be directly compared with the experimental phase diagram from figure~\ref{exp10mm}~(a).
This is done once for every starting configuration employed in the simulations.

\begin{figure*}[t]
	\centering
	\includegraphics[scale=0.92]{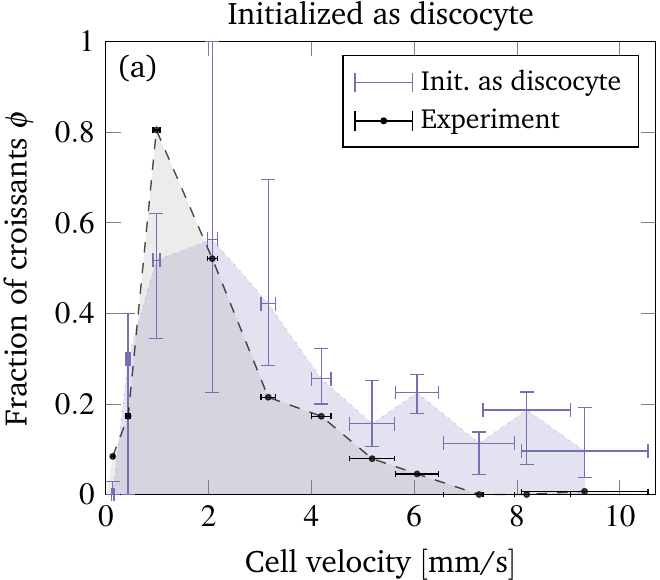}
	\hfill
	\includegraphics[scale=0.92]{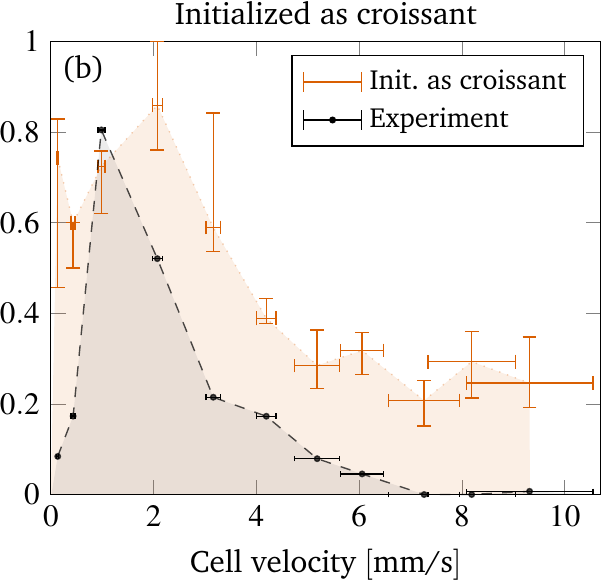}
	\hfill
	\includegraphics[scale=0.92]{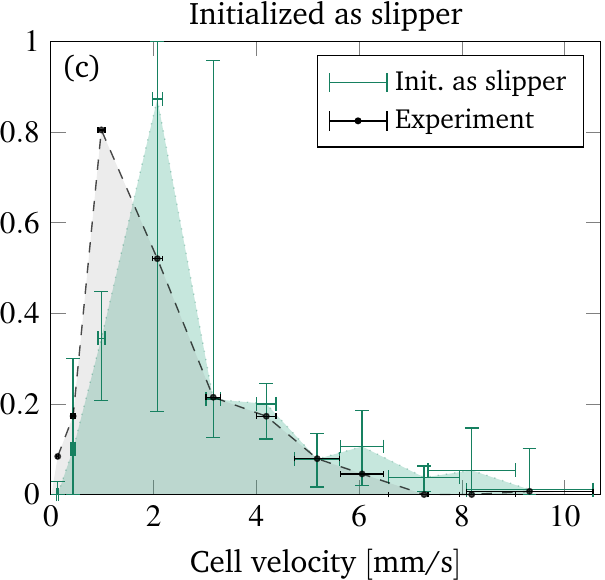}
	\caption{%
	Fraction of croissants $\phi$ predicted by the simulations, once for each starting configuration employed in the simulations: 
	(a)~Simulations started with the typical discocyte, (b)~with the croissant and (c)~with the slipper shape.
	To allow for a direct comparison, we included the experimental results from figure~\ref{exp10mm}(a) in each diagram (black dashed line).
	The horizontal error bars depict the standard deviation~$\sigma_u$ of the measured cell velocities (as in figure~\ref{exp10mm}(a)), while the vertical error bars show the uncertainty in the prediction as explained in the supplementary information.
	The lines and shaded areas serve as guides to the eye.
	See the main text for further details. 
	}
	\label{fig:cmpSimExp}
\end{figure*}

Figure~\ref{fig:cmpSimExp} shows this key result of our contribution, i.e.\ the predicted fraction of croissants $\phi$ as a function of the cell velocity for each starting shape.
The vertical error bars depict the uncertainty in the prediction, whose computation is explained in the supplementary information.
They are comparably large in the croissant-only region because the experimental velocities lie very near its sharp boundary.
The horizontal error bars illustrate the standard deviation~$\sigma_u$ of the experimentally measured cell velocities.
Clearly, we find very good agreement between the prediction from the simulation and the experimental observation when considering the slipper starting shape (figure~\ref{fig:cmpSimExp}(c)).
Starting with a discocyte or croissant leads to slightly more pronounced deviations (figures~\ref{fig:cmpSimExp}(a) and~(b)), but still a satisfactory semi-quantitative agreement is maintained.
This suggests the intuitive conclusion that the starting shapes in the experiment are closer to the rather asymmetric slippers than to the highly symmetric discocytes or croissants.
Indeed, as explicitly shown in the SI, we only observe non-classifiable and rather asymmetric \enquote{other} shapes at the channel entrance.

As mentioned in the introduction, experimental investigations with more detailed shape studies are rather scarce.
A comparison of the phase diagrams with the experimental literature is therefore limited to rough qualitative statements.
\citet{Tomaiuolo2009} found croissants and \enquote{others} for a cell velocity of \SI{1.1}{mm/s} using $\lambda \approx 5$ in a cylindrical tube with diameter \SI{10}{\micro\meter}.
This is in agreement with our results.
At \SI{36}{mm/s}, slippers but also croissants have been observed. 
Since we cannot reach velocities that high, we can neither confirm nor refute the occurrence of the latter.
Extrapolation of figure~\ref{fig:cmpSimExp} is dangerous since the Reynolds number at \SI{36}{mm/s} is around $\mathrm{Re} \approx 0.24$ and thus inertia effects might have noticeable contributions \cite{Kaoui2016a, Schaaf2017}.
Continuing, \citet{Cluitmans2014} found croissants and tumbling \enquote{others} at \SI{1.1}{mm/s} and slippers at \SI{13.6}{mm/s} in rectangular channels of $\SI{10}{\micro\meter}$ and $\SI{7}{\micro\meter}$ widths and a height of $\SI{10}{\micro\meter}$, which is consistent with our results. 
The experimental phase diagram presented in references~\citenum{Abkarian2008} and~\citenum{Faivre2006} also agrees with our results insofar that slippers occur at higher and croissants at lower velocities.
Yet, the considered velocities were higher than $\SI{10}{mm/s}$ and the viscosity ratio was $\lambda \lesssim 0.27$, i.e.\ much lower.
Furthermore, figure~3 in reference~\citenum{Suzuki1996} (cylindrical tube, $\lambda \approx 4$) also showed coexistence of croissants and slippers for velocities $\lesssim \SI{1}{mm/s}$ and only croissants roughly in the range $1$\,--\,$\SI{2}{mm/s}$, matching approximately with our results.

Regarding previous numerical studies, \citet{Fedosov2014} performed detailed 3D numerical simulations in cylindrical channels for $\lambda = 1$.
Taking a diameter of \SI{10}{\micro\meter} (translating into a confinement value of $\chi = 0.65$ in their work), they varied the average velocity from around \SI{0.05}{mm/s} to \SI{0.7}{mm/s}.
They observed a transition from snaking, to tumbling, to tank-treading slippers and finally to parachutes (which are very similar to croissants).
In our simulations we found tumbling and tank-treading slippers at velocities of the order of \SI{0.1}{mm/s}, and an increasing frequency of croissants above.
This matches at least qualitatively with \citeauthor{Fedosov2014}'s results. 
However, they did not vary the initial condition.

\subsection{Comparison of cell positions}

\begin{figure}[t]
	\centering
	\includegraphics{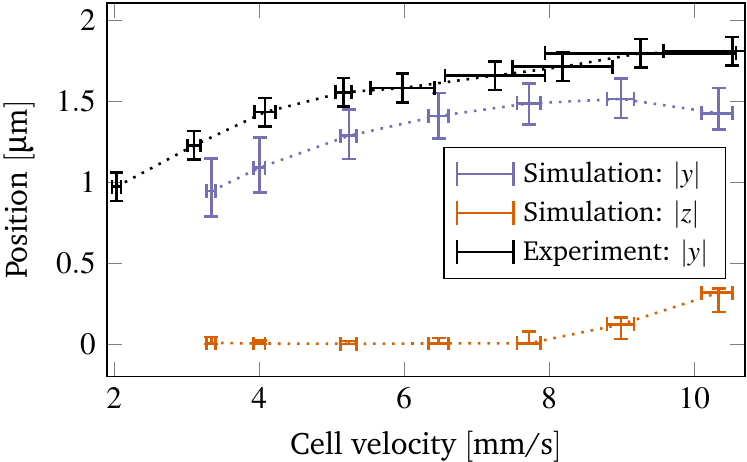}
	\caption{%
	Comparison between the centroid positions from the simulations (absolute values of the $y$- and $z$-coordinates) and experiments (absolute value of the $y$-coordinate) for cells that have a TT-slipper shape in the steady state.
	Error bars for the simulations as in figure~\ref{fig:AvgRadPos}(a).
	The horizontal error bars for the experimental data depict the standard deviation $\sigma_u$ of the cell velocities, while the vertical error bars represent the estimated uncertainty in the position determination.
	}
	\label{fig:cmpSimExpRad}
\end{figure}
Next, we compare the preferred position of the cells in the steady state.
The simulations predict a centered positioning of croissants (figure~\ref{fig:AvgRadPos}(a)), i.e.\ both the $y$- and the $z$-offsets are nearly zero.
This matches with figure~\ref{exp10mm}(b) where a very sharp peak at the channel center is found for the pressure drops within the croissant-peak region.

For slippers, the simulations showed an increase of the radial position of up to around $\SI{1.5}{\micro\meter}$ (figure~\ref{fig:AvgRadPos}(a)).
Considering the $y$- and $z$-coordinates separately in figure~\ref{fig:cmpSimExpRad}, we see that $z \approx 0$ and the major offset happens in the $y$-direction.
This is rather fortunate as the $y$-offset is also easily accessible in the experiments, contrary to the $z$-offset.
As can be seen in the measured $y$-distribution (figure~\ref{exp10mm}(b)), we have two off-centered peaks for slippers.
Taking the distribution function for only the slippers, we extract the positions $y_l$ and $y_r$ of the two peaks.
Exploiting the $\pm y$-symmetry of the channel, the off-centered position is then computed as $(y_r - y_l)/2$, i.e.\ in essence  as the average of the two peak distances to the central minimum.
Figure~\ref{fig:cmpSimExpRad} compares these values with the numerical results: The behavior is the same (an increase with velocity) and the predicted values show only a small systematic deviation of around $\approx \SI{0.3}{\micro\meter}$, i.e.\ of less than \SI{4}{\percent} of the RBC diameter $2 R_\mathrm{RBC}$.
A possible reason is that the optically recorded boundaries of the RBC and the channel walls are somewhat blurry (compare the experimental images in figure~\ref{fig:TypicalShapes}).

\subsection{Implications of the comparison}

There has been quite some debate in the literature if the croissant (or parachute) shapes observed via light microscopy are indeed what they appear to be.
\citet{Gaehtgens1980} (fig.~4 therein), for example, solidified the flowing RBCs with glutaraldehyde and found that the croissant-like shapes were actually slipper-like.
\citet{Skalak1969} pointed out that such shapes can also be \enquote{edge-on} discocytes with a flattened back.
Ultimately, to uniquely identify the forms one needs some method to record the full 3D geometry of the flowing cells (e.g.\ as in references~\citenum{Gorthi2012, Pegard2013, Pegard2014, Jagannadh2016, Kim2016, Merola2017, Quint2017}).
This is unfortunately very hard to implement in the present experimental setup.
However, this missing information is complemented here by the numerical simulations which are in good agreement with the experiments and thus our interpretation of the shapes as croissants should be correct.

The good agreement furthermore implies that our red blood cell model and simulation method is fully appropriate for describing the flow of RBCs in a straight microchannel.
More sophisticated methods including e.g.\ thermal fluctuations or surface viscosity \cite{Noguchi2005, McWhirter2011, Tomaiuolo2011a, Yazdani2013, Fedosov2014, Prado2015} are, at least for the present geometry, not required.
For croissants this is intuitive since membrane movement such as tank-treading is absent, for the tank-treading slippers it is somewhat less obvious.

\section{Summary \& conclusion}\label{sec:conclusion}

To summarize, we have performed in-vitro experiments and 3D simulations of healthy red blood cells flowing in a microchannel. 
The viscosity ratio was approximately $5$ and the flow velocities ranged from around $\SI{0.1}{mm/s}$ to $\SI{10}{mm/s}$ in both methodologies, corresponding to the typical conditions prevailing in the microvascular system.
We found that both the flow velocity as well as the initial starting configuration (offset from channel center, shape) have a major impact on the final steady state of the cells.
Using three different starting shapes (discocyte, croissant, slipper), we constructed the corresponding phase diagrams via simulations.
In most cases the cells assumed one out of two different forms: either a centered croissant or an off-centered slipper.
Interestingly, for most velocities bistability, i.e.\ a dependence of the final shape on the initial position, was observed.
Only in a small range of velocities (at around $\approx \SI{1}{mm/s}$) was the final shape found to be always a croissant.
The experimental diagram showed very good agreement with the numerical result, especially when considering the simulations that used the rather asymmetric slipper as starting shape.

We thus conclude that the employed numerical RBC model can sensibly describe the cell behavior in the presented setup.
Moreover, since we used physiological viscosity ratios and flow velocities, we speculate that croissants and slippers can occur in the microvasculature at the same set of system parameters not just as transients but rather that both are states which are intrinsically assumed by the cells.
Our results are important for applications where the cells should be in a specific state (e.g.\ in lab-on-a-chip devices) and allow for a comprehensive validation of numerical models.

\section*{Conflicts of interest}
There are no conflicts to declare.

	\begin{acknowledgments}
A.~Guckenberger and A.~Kihm contributed equally to this work.
S.~Gekle and C.~Wagner contributed equally to this work.

Funding from the Volkswagen Foundation and computing time granted by the Leibniz-Rechenzentrum on \mbox{SuperMUC} are gratefully acknowledged by A.~Guckenberger and S.~Gekle.
A.~Kihm, T.~John and C.~Wagner kindly acknowledge the support and funding of the \enquote{Deutsch-Franz\"osische-Hochschule} (DFH) DFDK \enquote{Living Fluids}.

	\end{acknowledgments}

	\bibliographystyle{apsrev4-1_mod}
	\bibliography{literature}

\begin{thebibliography}{96}%
\makeatletter
\providecommand \@ifxundefined [1]{%
 \@ifx{#1\undefined}
}%
\providecommand \@ifnum [1]{%
 \ifnum #1\expandafter \@firstoftwo
 \else \expandafter \@secondoftwo
 \fi
}%
\providecommand \@ifx [1]{%
 \ifx #1\expandafter \@firstoftwo
 \else \expandafter \@secondoftwo
 \fi
}%
\providecommand \natexlab [1]{#1}%
\providecommand \enquote  [1]{``#1''}%
\providecommand \bibnamefont  [1]{#1}%
\providecommand \bibfnamefont [1]{#1}%
\providecommand \citenamefont [1]{#1}%
\providecommand \href@noop [0]{\@secondoftwo}%
\providecommand \href [0]{\begingroup \@sanitize@url \@href}%
\providecommand \@href[1]{\@@startlink{#1}\@@href}%
\providecommand \@@href[1]{\endgroup#1\@@endlink}%
\providecommand \@sanitize@url [0]{\catcode `\\12\catcode `\$12\catcode
  `\&12\catcode `\#12\catcode `\^12\catcode `\_12\catcode `\%12\relax}%
\providecommand \@@startlink[1]{}%
\providecommand \@@endlink[0]{}%
\providecommand \url  [0]{\begingroup\@sanitize@url \@url }%
\providecommand \@url [1]{\endgroup\@href {#1}{\urlprefix }}%
\providecommand \urlprefix  [0]{URL }%
\providecommand \Eprint [0]{\href }%
\providecommand \doibase [0]{http://dx.doi.org/}%
\providecommand \selectlanguage [0]{\@gobble}%
\providecommand \bibinfo  [0]{\@secondoftwo}%
\providecommand \bibfield  [0]{\@secondoftwo}%
\providecommand \translation [1]{[#1]}%
\providecommand \BibitemOpen [0]{}%
\providecommand \bibitemStop [0]{}%
\providecommand \bibitemNoStop [0]{.\EOS\space}%
\providecommand \EOS [0]{\spacefactor3000\relax}%
\providecommand \BibitemShut  [1]{\csname bibitem#1\endcsname}%
\let\auto@bib@innerbib\@empty
%</preamble>
\bibitem [{\citenamefont {Freund}(2013)}]{Freund2013}%
  \BibitemOpen
  \bibfield  {author} {\bibinfo {author} {\bibfnamefont {J.~B.}\ \bibnamefont
  {Freund}},\ }\bibfield  {title} {\emph {\enquote {\bibinfo {title} {The flow
  of red blood cells through a narrow spleen-like slit},}\ }}\href {\doibase
  10.1063/1.4819341} {\bibfield  {journal} {\bibinfo  {journal} {Phys. Fluids}\
  }\textbf {\bibinfo {volume} {25}},\ \bibinfo {eid} {110807} (\bibinfo {year}
  {2013})}\BibitemShut {NoStop}%
\bibitem [{\citenamefont {Picot}\ \emph {et~al.}(2015)\citenamefont {Picot},
  \citenamefont {Ndour}, \citenamefont {Lefevre}, \citenamefont {El~Nemer},
  \citenamefont {Tawfik}, \citenamefont {Galimand}, \citenamefont {Da~Costa},
  \citenamefont {Ribeil}, \citenamefont {{de Montalembert}}, \citenamefont
  {Brousse}, \citenamefont {Le~Pioufle}, \citenamefont {Buffet}, \citenamefont
  {Le~Van~Kim},\ and\ \citenamefont {Fran{\c c}ais}}]{Picot2015}%
  \BibitemOpen
  \bibfield  {author} {\bibinfo {author} {\bibfnamefont {J.}~\bibnamefont
  {Picot}}, \bibinfo {author} {\bibfnamefont {P.~A.}\ \bibnamefont {Ndour}},
  \bibinfo {author} {\bibfnamefont {S.~D.}\ \bibnamefont {Lefevre}}, \bibinfo
  {author} {\bibfnamefont {W.}~\bibnamefont {El~Nemer}}, \bibinfo {author}
  {\bibfnamefont {H.}~\bibnamefont {Tawfik}}, \bibinfo {author} {\bibfnamefont
  {J.}~\bibnamefont {Galimand}}, \bibinfo {author} {\bibfnamefont
  {L.}~\bibnamefont {Da~Costa}}, \bibinfo {author} {\bibfnamefont {J.-A.}\
  \bibnamefont {Ribeil}}, \bibinfo {author} {\bibfnamefont {M.}~\bibnamefont
  {{de Montalembert}}}, \bibinfo {author} {\bibfnamefont {V.}~\bibnamefont
  {Brousse}}, \bibinfo {author} {\bibfnamefont {B.}~\bibnamefont {Le~Pioufle}},
  \bibinfo {author} {\bibfnamefont {P.}~\bibnamefont {Buffet}}, \bibinfo
  {author} {\bibfnamefont {C.}~\bibnamefont {Le~Van~Kim}}, \ and\ \bibinfo
  {author} {\bibfnamefont {O.}~\bibnamefont {Fran{\c c}ais}},\ }\bibfield
  {title} {\emph {\enquote {\bibinfo {title} {A biomimetic microfluidic chip to
  study the circulation and mechanical retention of red blood cells in the
  spleen},}\ }}\href {\doibase 10.1002/ajh.23941} {\bibfield  {journal}
  {\bibinfo  {journal} {Am. J. Hematol.}\ }\textbf {\bibinfo {volume} {90}},\
  \bibinfo {pages} {339} (\bibinfo {year} {2015})}\BibitemShut {NoStop}%
\bibitem [{\citenamefont {Salehyar}\ and\ \citenamefont
  {Zhu}(2016)}]{Salehyar2016}%
  \BibitemOpen
  \bibfield  {author} {\bibinfo {author} {\bibfnamefont {S.}~\bibnamefont
  {Salehyar}}\ and\ \bibinfo {author} {\bibfnamefont {Q.}~\bibnamefont {Zhu}},\
  }\bibfield  {title} {\emph {\enquote {\bibinfo {title} {Deformation and
  internal stress in a red blood cell as it is driven through a slit by an
  incoming flow},}\ }}\href {\doibase 10.1039/C5SM02933C} {\bibfield  {journal}
  {\bibinfo  {journal} {Soft Matter}\ }\textbf {\bibinfo {volume} {12}},\
  \bibinfo {pages} {3156} (\bibinfo {year} {2016})}\BibitemShut {NoStop}%
\bibitem [{\citenamefont {Fedosov}\ \emph {et~al.}(2014)\citenamefont
  {Fedosov}, \citenamefont {Peltom{\"a}ki},\ and\ \citenamefont
  {Gompper}}]{Fedosov2014}%
  \BibitemOpen
  \bibfield  {author} {\bibinfo {author} {\bibfnamefont {D.~A.}\ \bibnamefont
  {Fedosov}}, \bibinfo {author} {\bibfnamefont {M.}~\bibnamefont
  {Peltom{\"a}ki}}, \ and\ \bibinfo {author} {\bibfnamefont {G.}~\bibnamefont
  {Gompper}},\ }\bibfield  {title} {\emph {\enquote {\bibinfo {title}
  {Deformation and dynamics of red blood cells in flow through cylindrical
  microchannels},}\ }}\href {\doibase 10.1039/C4SM00248B} {\bibfield  {journal}
  {\bibinfo  {journal} {Soft Matter}\ }\textbf {\bibinfo {volume} {10}},\
  \bibinfo {pages} {4258} (\bibinfo {year} {2014})}\BibitemShut {NoStop}%
\bibitem [{\citenamefont {Aouane}\ \emph {et~al.}(2014)\citenamefont {Aouane},
  \citenamefont {Thi{\'e}baud}, \citenamefont {Benyoussef}, \citenamefont
  {Wagner},\ and\ \citenamefont {Misbah}}]{Aouane2014}%
  \BibitemOpen
  \bibfield  {author} {\bibinfo {author} {\bibfnamefont {O.}~\bibnamefont
  {Aouane}}, \bibinfo {author} {\bibfnamefont {M.}~\bibnamefont
  {Thi{\'e}baud}}, \bibinfo {author} {\bibfnamefont {A.}~\bibnamefont
  {Benyoussef}}, \bibinfo {author} {\bibfnamefont {C.}~\bibnamefont {Wagner}},
  \ and\ \bibinfo {author} {\bibfnamefont {C.}~\bibnamefont {Misbah}},\
  }\bibfield  {title} {\emph {\enquote {\bibinfo {title} {Vesicle dynamics in a
  confined {{Poiseuille}} flow: {{From}} steady state to chaos},}\ }}\href
  {\doibase 10.1103/PhysRevE.90.033011} {\bibfield  {journal} {\bibinfo
  {journal} {Phys. Rev. E}\ }\textbf {\bibinfo {volume} {90}},\ \bibinfo
  {pages} {033011} (\bibinfo {year} {2014})}\BibitemShut {NoStop}%
\bibitem [{\citenamefont {Tahiri}\ \emph {et~al.}(2013)\citenamefont {Tahiri},
  \citenamefont {Biben}, \citenamefont {Ez-Zahraouy}, \citenamefont
  {Benyoussef},\ and\ \citenamefont {Misbah}}]{Tahiri2013}%
  \BibitemOpen
  \bibfield  {author} {\bibinfo {author} {\bibfnamefont {N.}~\bibnamefont
  {Tahiri}}, \bibinfo {author} {\bibfnamefont {T.}~\bibnamefont {Biben}},
  \bibinfo {author} {\bibfnamefont {H.}~\bibnamefont {Ez-Zahraouy}}, \bibinfo
  {author} {\bibfnamefont {A.}~\bibnamefont {Benyoussef}}, \ and\ \bibinfo
  {author} {\bibfnamefont {C.}~\bibnamefont {Misbah}},\ }\bibfield  {title}
  {\emph {\enquote {\bibinfo {title} {On the problem of slipper shapes of red
  blood cells in the microvasculature},}\ }}\href {\doibase
  10.1016/j.mvr.2012.10.001} {\bibfield  {journal} {\bibinfo  {journal}
  {Microvasc. Res.}\ }\textbf {\bibinfo {volume} {85}},\ \bibinfo {pages} {40}
  (\bibinfo {year} {2013})}\BibitemShut {NoStop}%
\bibitem [{\citenamefont {Vitkova}\ \emph {et~al.}(2008)\citenamefont
  {Vitkova}, \citenamefont {Mader}, \citenamefont {Polack}, \citenamefont
  {Misbah},\ and\ \citenamefont {Podgorski}}]{Vitkova2008}%
  \BibitemOpen
  \bibfield  {author} {\bibinfo {author} {\bibfnamefont {V.}~\bibnamefont
  {Vitkova}}, \bibinfo {author} {\bibfnamefont {M.-A.}\ \bibnamefont {Mader}},
  \bibinfo {author} {\bibfnamefont {B.}~\bibnamefont {Polack}}, \bibinfo
  {author} {\bibfnamefont {C.}~\bibnamefont {Misbah}}, \ and\ \bibinfo {author}
  {\bibfnamefont {T.}~\bibnamefont {Podgorski}},\ }\bibfield  {title} {\emph
  {\enquote {\bibinfo {title} {Micro-{{Macro Link}} in {{Rheology}} of
  {{Erythrocyte}} and {{Vesicle Suspensions}}},}\ }}\href {\doibase
  10.1529/biophysj.108.138826} {\bibfield  {journal} {\bibinfo  {journal}
  {Biophys. J.}\ }\textbf {\bibinfo {volume} {95}},\ \bibinfo {pages} {L33}
  (\bibinfo {year} {2008})}\BibitemShut {NoStop}%
\bibitem [{\citenamefont {Fedosov}\ \emph {et~al.}(2011)\citenamefont
  {Fedosov}, \citenamefont {Pan}, \citenamefont {Caswell}, \citenamefont
  {Gompper},\ and\ \citenamefont {Karniadakis}}]{Fedosov2011}%
  \BibitemOpen
  \bibfield  {author} {\bibinfo {author} {\bibfnamefont {D.~A.}\ \bibnamefont
  {Fedosov}}, \bibinfo {author} {\bibfnamefont {W.}~\bibnamefont {Pan}},
  \bibinfo {author} {\bibfnamefont {B.}~\bibnamefont {Caswell}}, \bibinfo
  {author} {\bibfnamefont {G.}~\bibnamefont {Gompper}}, \ and\ \bibinfo
  {author} {\bibfnamefont {G.~E.}\ \bibnamefont {Karniadakis}},\ }\bibfield
  {title} {\emph {\enquote {\bibinfo {title} {Predicting human blood viscosity
  in silico},}\ }}\href {\doibase 10.1073/pnas.1101210108} {\bibfield
  {journal} {\bibinfo  {journal} {Proc. Natl. Acad. Sci.}\ }\textbf {\bibinfo
  {volume} {108}},\ \bibinfo {pages} {11772} (\bibinfo {year}
  {2011})}\BibitemShut {NoStop}%
\bibitem [{\citenamefont {Kr{\"u}ger}\ \emph {et~al.}(2013)\citenamefont
  {Kr{\"u}ger}, \citenamefont {Gross}, \citenamefont {Raabe},\ and\
  \citenamefont {Varnik}}]{Kruger2013}%
  \BibitemOpen
  \bibfield  {author} {\bibinfo {author} {\bibfnamefont {T.}~\bibnamefont
  {Kr{\"u}ger}}, \bibinfo {author} {\bibfnamefont {M.}~\bibnamefont {Gross}},
  \bibinfo {author} {\bibfnamefont {D.}~\bibnamefont {Raabe}}, \ and\ \bibinfo
  {author} {\bibfnamefont {F.}~\bibnamefont {Varnik}},\ }\bibfield  {title}
  {\emph {\enquote {\bibinfo {title} {Crossover from tumbling to
  tank-treading-like motion in dense simulated suspensions of red blood
  cells},}\ }}\href {\doibase 10.1039/C3SM51645H} {\bibfield  {journal}
  {\bibinfo  {journal} {Soft Matter}\ }\textbf {\bibinfo {volume} {9}},\
  \bibinfo {pages} {9008} (\bibinfo {year} {2013})}\BibitemShut {NoStop}%
\bibitem [{\citenamefont {Thi{\'e}baud}\ \emph {et~al.}(2014)\citenamefont
  {Thi{\'e}baud}, \citenamefont {Shen}, \citenamefont {Harting},\ and\
  \citenamefont {Misbah}}]{Thiebaud2014}%
  \BibitemOpen
  \bibfield  {author} {\bibinfo {author} {\bibfnamefont {M.}~\bibnamefont
  {Thi{\'e}baud}}, \bibinfo {author} {\bibfnamefont {Z.}~\bibnamefont {Shen}},
  \bibinfo {author} {\bibfnamefont {J.}~\bibnamefont {Harting}}, \ and\
  \bibinfo {author} {\bibfnamefont {C.}~\bibnamefont {Misbah}},\ }\bibfield
  {title} {\emph {\enquote {\bibinfo {title} {Prediction of {{Anomalous Blood
  Viscosity}} in {{Confined Shear Flow}}},}\ }}\href {\doibase
  10.1103/PhysRevLett.112.238304} {\bibfield  {journal} {\bibinfo  {journal}
  {Phys. Rev. Lett.}\ }\textbf {\bibinfo {volume} {112}},\ \bibinfo {pages}
  {238304} (\bibinfo {year} {2014})}\BibitemShut {NoStop}%
\bibitem [{\citenamefont {Katanov}\ \emph {et~al.}(2015)\citenamefont
  {Katanov}, \citenamefont {Gompper},\ and\ \citenamefont
  {Fedosov}}]{Katanov2015}%
  \BibitemOpen
  \bibfield  {author} {\bibinfo {author} {\bibfnamefont {D.}~\bibnamefont
  {Katanov}}, \bibinfo {author} {\bibfnamefont {G.}~\bibnamefont {Gompper}}, \
  and\ \bibinfo {author} {\bibfnamefont {D.~A.}\ \bibnamefont {Fedosov}},\
  }\bibfield  {title} {\emph {\enquote {\bibinfo {title} {Microvascular blood
  flow resistance: {{Role}} of red blood cell migration and dispersion},}\
  }}\href {\doibase 10.1016/j.mvr.2015.02.006} {\bibfield  {journal} {\bibinfo
  {journal} {Microvasc. Res.}\ }\textbf {\bibinfo {volume} {99}},\ \bibinfo
  {pages} {57} (\bibinfo {year} {2015})}\BibitemShut {NoStop}%
\bibitem [{\citenamefont {Lanotte}\ \emph {et~al.}(2016)\citenamefont
  {Lanotte}, \citenamefont {Mauer}, \citenamefont {Mendez}, \citenamefont
  {Fedosov}, \citenamefont {Fromental}, \citenamefont {Claveria}, \citenamefont
  {Nicoud}, \citenamefont {Gompper},\ and\ \citenamefont
  {Abkarian}}]{Lanotte2016}%
  \BibitemOpen
  \bibfield  {author} {\bibinfo {author} {\bibfnamefont {L.}~\bibnamefont
  {Lanotte}}, \bibinfo {author} {\bibfnamefont {J.}~\bibnamefont {Mauer}},
  \bibinfo {author} {\bibfnamefont {S.}~\bibnamefont {Mendez}}, \bibinfo
  {author} {\bibfnamefont {D.~A.}\ \bibnamefont {Fedosov}}, \bibinfo {author}
  {\bibfnamefont {J.-M.}\ \bibnamefont {Fromental}}, \bibinfo {author}
  {\bibfnamefont {V.}~\bibnamefont {Claveria}}, \bibinfo {author}
  {\bibfnamefont {F.}~\bibnamefont {Nicoud}}, \bibinfo {author} {\bibfnamefont
  {G.}~\bibnamefont {Gompper}}, \ and\ \bibinfo {author} {\bibfnamefont
  {M.}~\bibnamefont {Abkarian}},\ }\bibfield  {title} {\emph {\enquote
  {\bibinfo {title} {Red cells' dynamic morphologies govern blood shear
  thinning under microcirculatory flow conditions},}\ }}\href {\doibase
  10.1073/pnas.1608074113} {\bibfield  {journal} {\bibinfo  {journal} {Proc.
  Natl. Acad. Sci.}\ }\textbf {\bibinfo {volume} {113}},\ \bibinfo {pages}
  {13289} (\bibinfo {year} {2016})}\BibitemShut {NoStop}%
\bibitem [{\citenamefont {Henry}\ \emph {et~al.}(2016)\citenamefont {Henry},
  \citenamefont {Holm}, \citenamefont {Zhang}, \citenamefont {Beech},
  \citenamefont {Tegenfeldt}, \citenamefont {Fedosov},\ and\ \citenamefont
  {Gompper}}]{Henry2016}%
  \BibitemOpen
  \bibfield  {author} {\bibinfo {author} {\bibfnamefont {E.}~\bibnamefont
  {Henry}}, \bibinfo {author} {\bibfnamefont {S.~H.}\ \bibnamefont {Holm}},
  \bibinfo {author} {\bibfnamefont {Z.}~\bibnamefont {Zhang}}, \bibinfo
  {author} {\bibfnamefont {J.~P.}\ \bibnamefont {Beech}}, \bibinfo {author}
  {\bibfnamefont {J.~O.}\ \bibnamefont {Tegenfeldt}}, \bibinfo {author}
  {\bibfnamefont {D.~A.}\ \bibnamefont {Fedosov}}, \ and\ \bibinfo {author}
  {\bibfnamefont {G.}~\bibnamefont {Gompper}},\ }\bibfield  {title} {\emph
  {\enquote {\bibinfo {title} {Sorting cells by their dynamical properties},}\
  }}\href {\doibase 10.1038/srep34375} {\bibfield  {journal} {\bibinfo
  {journal} {Sci. Rep.}\ }\textbf {\bibinfo {volume} {6}},\ \bibinfo {pages}
  {34375} (\bibinfo {year} {2016})}\BibitemShut {NoStop}%
\bibitem [{\citenamefont {Otto}\ \emph {et~al.}(2015)\citenamefont {Otto},
  \citenamefont {Rosendahl}, \citenamefont {Mietke}, \citenamefont {Golfier},
  \citenamefont {Herold}, \citenamefont {Klaue}, \citenamefont {Girardo},
  \citenamefont {Pagliara}, \citenamefont {Ekpenyong}, \citenamefont {Jacobi},
  \citenamefont {Wobus}, \citenamefont {T{\"o}pfner}, \citenamefont {Keyser},
  \citenamefont {Mansfeld}, \citenamefont {Fischer-Friedrich},\ and\
  \citenamefont {Guck}}]{Otto2015}%
  \BibitemOpen
  \bibfield  {author} {\bibinfo {author} {\bibfnamefont {O.}~\bibnamefont
  {Otto}}, \bibinfo {author} {\bibfnamefont {P.}~\bibnamefont {Rosendahl}},
  \bibinfo {author} {\bibfnamefont {A.}~\bibnamefont {Mietke}}, \bibinfo
  {author} {\bibfnamefont {S.}~\bibnamefont {Golfier}}, \bibinfo {author}
  {\bibfnamefont {C.}~\bibnamefont {Herold}}, \bibinfo {author} {\bibfnamefont
  {D.}~\bibnamefont {Klaue}}, \bibinfo {author} {\bibfnamefont
  {S.}~\bibnamefont {Girardo}}, \bibinfo {author} {\bibfnamefont
  {S.}~\bibnamefont {Pagliara}}, \bibinfo {author} {\bibfnamefont
  {A.}~\bibnamefont {Ekpenyong}}, \bibinfo {author} {\bibfnamefont
  {A.}~\bibnamefont {Jacobi}}, \bibinfo {author} {\bibfnamefont
  {M.}~\bibnamefont {Wobus}}, \bibinfo {author} {\bibfnamefont
  {N.}~\bibnamefont {T{\"o}pfner}}, \bibinfo {author} {\bibfnamefont {U.~F.}\
  \bibnamefont {Keyser}}, \bibinfo {author} {\bibfnamefont {J.}~\bibnamefont
  {Mansfeld}}, \bibinfo {author} {\bibfnamefont {E.}~\bibnamefont
  {Fischer-Friedrich}}, \ and\ \bibinfo {author} {\bibfnamefont
  {J.}~\bibnamefont {Guck}},\ }\bibfield  {title} {\emph {\enquote {\bibinfo
  {title} {Real-time deformability cytometry: On-the-fly cell mechanical
  phenotyping},}\ }}\href {\doibase 10.1038/nmeth.3281} {\bibfield  {journal}
  {\bibinfo  {journal} {Nat. Methods}\ }\textbf {\bibinfo {volume} {12}},\
  \bibinfo {pages} {199} (\bibinfo {year} {2015})}\BibitemShut {NoStop}%
\bibitem [{\citenamefont {Merola}\ \emph {et~al.}(2017)\citenamefont {Merola},
  \citenamefont {Memmolo}, \citenamefont {Miccio}, \citenamefont {Savoia},
  \citenamefont {Mugnano}, \citenamefont {Fontana}, \citenamefont {D'Ippolito},
  \citenamefont {Sardo}, \citenamefont {Iolascon}, \citenamefont {Gambale},\
  and\ \citenamefont {Ferraro}}]{Merola2017}%
  \BibitemOpen
  \bibfield  {author} {\bibinfo {author} {\bibfnamefont {F.}~\bibnamefont
  {Merola}}, \bibinfo {author} {\bibfnamefont {P.}~\bibnamefont {Memmolo}},
  \bibinfo {author} {\bibfnamefont {L.}~\bibnamefont {Miccio}}, \bibinfo
  {author} {\bibfnamefont {R.}~\bibnamefont {Savoia}}, \bibinfo {author}
  {\bibfnamefont {M.}~\bibnamefont {Mugnano}}, \bibinfo {author} {\bibfnamefont
  {A.}~\bibnamefont {Fontana}}, \bibinfo {author} {\bibfnamefont
  {G.}~\bibnamefont {D'Ippolito}}, \bibinfo {author} {\bibfnamefont
  {A.}~\bibnamefont {Sardo}}, \bibinfo {author} {\bibfnamefont
  {A.}~\bibnamefont {Iolascon}}, \bibinfo {author} {\bibfnamefont
  {A.}~\bibnamefont {Gambale}}, \ and\ \bibinfo {author} {\bibfnamefont
  {P.}~\bibnamefont {Ferraro}},\ }\bibfield  {title} {\emph {\enquote {\bibinfo
  {title} {Tomographic {{Flow Cytometry}} by {{Digital Holography}}},}\ }}\href
  {\doibase 10.1038/lsa.2016.241} {\bibfield  {journal} {\bibinfo  {journal}
  {Light Sci. Appl.}\ }\textbf {\bibinfo {volume} {6}},\ \bibinfo {pages}
  {e16241} (\bibinfo {year} {2017})}\BibitemShut {NoStop}%
\bibitem [{\citenamefont {Farutin}\ and\ \citenamefont
  {Misbah}(2014)}]{Farutin2014a}%
  \BibitemOpen
  \bibfield  {author} {\bibinfo {author} {\bibfnamefont {A.}~\bibnamefont
  {Farutin}}\ and\ \bibinfo {author} {\bibfnamefont {C.}~\bibnamefont
  {Misbah}},\ }\bibfield  {title} {\emph {\enquote {\bibinfo {title} {Symmetry
  breaking and cross-streamline migration of three-dimensional vesicles in an
  axial {{Poiseuille}} flow},}\ }}\href {\doibase 10.1103/PhysRevE.89.042709}
  {\bibfield  {journal} {\bibinfo  {journal} {Phys. Rev. E}\ }\textbf {\bibinfo
  {volume} {89}},\ \bibinfo {pages} {042709} (\bibinfo {year}
  {2014})}\BibitemShut {NoStop}%
\bibitem [{\citenamefont {Gaehtgens}\ \emph {et~al.}(1980)\citenamefont
  {Gaehtgens}, \citenamefont {D{\"u}hrssen},\ and\ \citenamefont
  {Albrecht}}]{Gaehtgens1980}%
  \BibitemOpen
  \bibfield  {author} {\bibinfo {author} {\bibfnamefont {P.}~\bibnamefont
  {Gaehtgens}}, \bibinfo {author} {\bibfnamefont {C.}~\bibnamefont
  {D{\"u}hrssen}}, \ and\ \bibinfo {author} {\bibfnamefont {K.~H.}\
  \bibnamefont {Albrecht}},\ }\bibfield  {title} {\emph {\enquote {\bibinfo
  {title} {Motion, deformation, and interaction of blood cells and plasma
  during flow through narrow capillary tubes},}\ }}\href@noop {} {\bibfield
  {journal} {\bibinfo  {journal} {Blood Cells}\ }\textbf {\bibinfo {volume}
  {6}},\ \bibinfo {pages} {799} (\bibinfo {year} {1980})}\BibitemShut {NoStop}%
\bibitem [{\citenamefont {Suzuki}\ \emph {et~al.}(1996)\citenamefont {Suzuki},
  \citenamefont {Tateishi}, \citenamefont {Soutani},\ and\ \citenamefont
  {Maeda}}]{Suzuki1996}%
  \BibitemOpen
  \bibfield  {author} {\bibinfo {author} {\bibfnamefont {Y.}~\bibnamefont
  {Suzuki}}, \bibinfo {author} {\bibfnamefont {N.}~\bibnamefont {Tateishi}},
  \bibinfo {author} {\bibfnamefont {M.}~\bibnamefont {Soutani}}, \ and\
  \bibinfo {author} {\bibfnamefont {N.}~\bibnamefont {Maeda}},\ }\bibfield
  {title} {\emph {\enquote {\bibinfo {title} {Deformation of {{Erythrocytes}}
  in {{Microvessels}} and {{Glass Capillaries}}: {{Effects}} of {{Erythrocyte
  Deformability}}},}\ }}\href {\doibase 10.3109/10739689609146782} {\bibfield
  {journal} {\bibinfo  {journal} {Microcirculation}\ }\textbf {\bibinfo
  {volume} {3}},\ \bibinfo {pages} {49} (\bibinfo {year} {1996})}\BibitemShut
  {NoStop}%
\bibitem [{\citenamefont {Secomb}\ \emph {et~al.}(2007)\citenamefont {Secomb},
  \citenamefont {Styp-Rekowska},\ and\ \citenamefont {Pries}}]{Secomb2007}%
  \BibitemOpen
  \bibfield  {author} {\bibinfo {author} {\bibfnamefont {T.~W.}\ \bibnamefont
  {Secomb}}, \bibinfo {author} {\bibfnamefont {B.}~\bibnamefont
  {Styp-Rekowska}}, \ and\ \bibinfo {author} {\bibfnamefont {A.~R.}\
  \bibnamefont {Pries}},\ }\bibfield  {title} {\emph {\enquote {\bibinfo
  {title} {Two-{{Dimensional Simulation}} of {{Red Blood Cell Deformation}} and
  {{Lateral Migration}} in {{Microvessels}}},}\ }}\href {\doibase
  10.1007/s10439-007-9275-0} {\bibfield  {journal} {\bibinfo  {journal} {Ann.
  Biomed. Eng.}\ }\textbf {\bibinfo {volume} {35}},\ \bibinfo {pages} {755}
  (\bibinfo {year} {2007})}\BibitemShut {NoStop}%
\bibitem [{\citenamefont {Faivre}(2006)}]{Faivre2006}%
  \BibitemOpen
  \bibfield  {author} {\bibinfo {author} {\bibfnamefont {M.}~\bibnamefont
  {Faivre}},\ }\emph {\bibinfo {title} {Drops, Vesicles and Red Blood Cells:
  {{Deformability}} and Behavior under Flow}},\ \href@noop {} {Ph.D. thesis},\
  \bibinfo  {school} {Universit{\'e} Joseph-Fourier - Grenoble I}, \bibinfo
  {address} {Grenoble} (\bibinfo {year} {2006})\BibitemShut {NoStop}%
\bibitem [{\citenamefont {Abkarian}\ \emph {et~al.}(2008)\citenamefont
  {Abkarian}, \citenamefont {Faivre}, \citenamefont {Horton}, \citenamefont
  {Smistrup}, \citenamefont {Best-Popescu},\ and\ \citenamefont
  {Stone}}]{Abkarian2008}%
  \BibitemOpen
  \bibfield  {author} {\bibinfo {author} {\bibfnamefont {M.}~\bibnamefont
  {Abkarian}}, \bibinfo {author} {\bibfnamefont {M.}~\bibnamefont {Faivre}},
  \bibinfo {author} {\bibfnamefont {R.}~\bibnamefont {Horton}}, \bibinfo
  {author} {\bibfnamefont {K.}~\bibnamefont {Smistrup}}, \bibinfo {author}
  {\bibfnamefont {C.~A.}\ \bibnamefont {Best-Popescu}}, \ and\ \bibinfo
  {author} {\bibfnamefont {H.~A.}\ \bibnamefont {Stone}},\ }\bibfield  {title}
  {\emph {\enquote {\bibinfo {title} {Cellular-scale hydrodynamics},}\ }}\href
  {\doibase 10.1088/1748-6041/3/3/034011} {\bibfield  {journal} {\bibinfo
  {journal} {Biomed. Mater.}\ }\textbf {\bibinfo {volume} {3}},\ \bibinfo
  {pages} {034011} (\bibinfo {year} {2008})}\BibitemShut {NoStop}%
\bibitem [{\citenamefont {Tomaiuolo}\ \emph {et~al.}(2009)\citenamefont
  {Tomaiuolo}, \citenamefont {Simeone}, \citenamefont {Martinelli},
  \citenamefont {Rotoli},\ and\ \citenamefont {Guido}}]{Tomaiuolo2009}%
  \BibitemOpen
  \bibfield  {author} {\bibinfo {author} {\bibfnamefont {G.}~\bibnamefont
  {Tomaiuolo}}, \bibinfo {author} {\bibfnamefont {M.}~\bibnamefont {Simeone}},
  \bibinfo {author} {\bibfnamefont {V.}~\bibnamefont {Martinelli}}, \bibinfo
  {author} {\bibfnamefont {B.}~\bibnamefont {Rotoli}}, \ and\ \bibinfo {author}
  {\bibfnamefont {S.}~\bibnamefont {Guido}},\ }\bibfield  {title} {\emph
  {\enquote {\bibinfo {title} {Red blood cell deformation in microconfined
  flow},}\ }}\href {\doibase 10.1039/B904584H} {\bibfield  {journal} {\bibinfo
  {journal} {Soft Matter}\ }\textbf {\bibinfo {volume} {5}},\ \bibinfo {pages}
  {3736} (\bibinfo {year} {2009})}\BibitemShut {NoStop}%
\bibitem [{\citenamefont {Tomaiuolo}\ and\ \citenamefont
  {Guido}(2011)}]{Tomaiuolo2011}%
  \BibitemOpen
  \bibfield  {author} {\bibinfo {author} {\bibfnamefont {G.}~\bibnamefont
  {Tomaiuolo}}\ and\ \bibinfo {author} {\bibfnamefont {S.}~\bibnamefont
  {Guido}},\ }\bibfield  {title} {\emph {\enquote {\bibinfo {title} {Start-up
  shape dynamics of red blood cells in microcapillary flow},}\ }}\href
  {\doibase 10.1016/j.mvr.2011.03.004} {\bibfield  {journal} {\bibinfo
  {journal} {Microvasc. Res.}\ }\textbf {\bibinfo {volume} {82}},\ \bibinfo
  {pages} {35} (\bibinfo {year} {2011})}\BibitemShut {NoStop}%
\bibitem [{\citenamefont {Prado}\ \emph {et~al.}(2015)\citenamefont {Prado},
  \citenamefont {Farutin}, \citenamefont {Misbah},\ and\ \citenamefont
  {Bureau}}]{Prado2015}%
  \BibitemOpen
  \bibfield  {author} {\bibinfo {author} {\bibfnamefont {G.}~\bibnamefont
  {Prado}}, \bibinfo {author} {\bibfnamefont {A.}~\bibnamefont {Farutin}},
  \bibinfo {author} {\bibfnamefont {C.}~\bibnamefont {Misbah}}, \ and\ \bibinfo
  {author} {\bibfnamefont {L.}~\bibnamefont {Bureau}},\ }\bibfield  {title}
  {\emph {\enquote {\bibinfo {title} {Viscoelastic {{Transient}} of {{Confined
  Red Blood Cells}}},}\ }}\href {\doibase 10.1016/j.bpj.2015.03.046} {\bibfield
   {journal} {\bibinfo  {journal} {Biophys. J.}\ }\textbf {\bibinfo {volume}
  {108}},\ \bibinfo {pages} {2126} (\bibinfo {year} {2015})}\BibitemShut
  {NoStop}%
\bibitem [{\citenamefont {Cluitmans}\ \emph {et~al.}(2014)\citenamefont
  {Cluitmans}, \citenamefont {Chokkalingam}, \citenamefont {Janssen},
  \citenamefont {Brock}, \citenamefont {Huck},\ and\ \citenamefont
  {Bosman}}]{Cluitmans2014}%
  \BibitemOpen
  \bibfield  {author} {\bibinfo {author} {\bibfnamefont {J.~C.~A.}\
  \bibnamefont {Cluitmans}}, \bibinfo {author} {\bibfnamefont {V.}~\bibnamefont
  {Chokkalingam}}, \bibinfo {author} {\bibfnamefont {A.~M.}\ \bibnamefont
  {Janssen}}, \bibinfo {author} {\bibfnamefont {R.}~\bibnamefont {Brock}},
  \bibinfo {author} {\bibfnamefont {W.~T.~S.}\ \bibnamefont {Huck}}, \ and\
  \bibinfo {author} {\bibfnamefont {G.~J. C. G.~M.}\ \bibnamefont {Bosman}},\
  }\bibfield  {title} {\emph {\enquote {\bibinfo {title} {Alterations in {{Red
  Blood Cell Deformability}} during {{Storage}}: {{A Microfluidic
  Approach}}},}\ }}\href {\doibase 10.1155/2014/764268} {\bibfield  {journal}
  {\bibinfo  {journal} {BioMed Res. Int.}\ }\textbf {\bibinfo {volume}
  {2014}},\ \bibinfo {pages} {e764268} (\bibinfo {year} {2014})}\BibitemShut
  {NoStop}%
\bibitem [{\citenamefont {Quint}\ \emph {et~al.}(2017)\citenamefont {Quint},
  \citenamefont {Christ}, \citenamefont {Guckenberger}, \citenamefont
  {Himbert}, \citenamefont {Kaestner}, \citenamefont {Gekle},\ and\
  \citenamefont {Wagner}}]{Quint2017}%
  \BibitemOpen
  \bibfield  {author} {\bibinfo {author} {\bibfnamefont {S.}~\bibnamefont
  {Quint}}, \bibinfo {author} {\bibfnamefont {A.~F.}\ \bibnamefont {Christ}},
  \bibinfo {author} {\bibfnamefont {A.}~\bibnamefont {Guckenberger}}, \bibinfo
  {author} {\bibfnamefont {S.}~\bibnamefont {Himbert}}, \bibinfo {author}
  {\bibfnamefont {L.}~\bibnamefont {Kaestner}}, \bibinfo {author}
  {\bibfnamefont {S.}~\bibnamefont {Gekle}}, \ and\ \bibinfo {author}
  {\bibfnamefont {C.}~\bibnamefont {Wagner}},\ }\bibfield  {title} {\emph
  {\enquote {\bibinfo {title} {{{3D}} tomography of cells in micro-channels},}\
  }}\href {\doibase 10.1063/1.4986392} {\bibfield  {journal} {\bibinfo
  {journal} {Appl. Phys. Lett.}\ }\textbf {\bibinfo {volume} {111}},\ \bibinfo
  {pages} {103701} (\bibinfo {year} {2017})}\BibitemShut {NoStop}%
\bibitem [{\citenamefont {Hochmuth}\ \emph {et~al.}(1970)\citenamefont
  {Hochmuth}, \citenamefont {Marple},\ and\ \citenamefont
  {Sutera}}]{Hochmuth1970}%
  \BibitemOpen
  \bibfield  {author} {\bibinfo {author} {\bibfnamefont {R.~M.}\ \bibnamefont
  {Hochmuth}}, \bibinfo {author} {\bibfnamefont {R.~N.}\ \bibnamefont
  {Marple}}, \ and\ \bibinfo {author} {\bibfnamefont {S.~P.}\ \bibnamefont
  {Sutera}},\ }\bibfield  {title} {\emph {\enquote {\bibinfo {title} {Capillary
  blood flow: {{I}}. {{Erythrocyte}} deformation in glass capillaries},}\
  }}\href {\doibase 10.1016/0026-2862(70)90034-8} {\bibfield  {journal}
  {\bibinfo  {journal} {Microvascular Research}\ }\textbf {\bibinfo {volume}
  {2}},\ \bibinfo {pages} {409} (\bibinfo {year} {1970})}\BibitemShut {NoStop}%
\bibitem [{\citenamefont {Seshadri}\ \emph {et~al.}(1970)\citenamefont
  {Seshadri}, \citenamefont {Hochmuth}, \citenamefont {Croce},\ and\
  \citenamefont {Sutera}}]{Seshadri1970}%
  \BibitemOpen
  \bibfield  {author} {\bibinfo {author} {\bibfnamefont {V.}~\bibnamefont
  {Seshadri}}, \bibinfo {author} {\bibfnamefont {R.~M.}\ \bibnamefont
  {Hochmuth}}, \bibinfo {author} {\bibfnamefont {P.~A.}\ \bibnamefont {Croce}},
  \ and\ \bibinfo {author} {\bibfnamefont {S.~P.}\ \bibnamefont {Sutera}},\
  }\bibfield  {title} {\emph {\enquote {\bibinfo {title} {Capillary blood flow:
  {{III}}. {{Deformable}} model cells compared to erythrocytes in vitro},}\
  }}\href {\doibase 10.1016/0026-2862(70)90036-1} {\bibfield  {journal}
  {\bibinfo  {journal} {Microvascular Research}\ }\textbf {\bibinfo {volume}
  {2}},\ \bibinfo {pages} {434} (\bibinfo {year} {1970})}\BibitemShut {NoStop}%
\bibitem [{\citenamefont {Zharov}\ \emph {et~al.}(2006)\citenamefont {Zharov},
  \citenamefont {Galanzha}, \citenamefont {Menyaev},\ and\ \citenamefont
  {Tuchin}}]{Zharov2006}%
  \BibitemOpen
  \bibfield  {author} {\bibinfo {author} {\bibfnamefont {V.~P.}\ \bibnamefont
  {Zharov}}, \bibinfo {author} {\bibfnamefont {E.~I.}\ \bibnamefont
  {Galanzha}}, \bibinfo {author} {\bibfnamefont {Y.}~\bibnamefont {Menyaev}}, \
  and\ \bibinfo {author} {\bibfnamefont {V.~V.}\ \bibnamefont {Tuchin}},\
  }\bibfield  {title} {\emph {\enquote {\bibinfo {title} {In vivo high-speed
  imaging of individual cells in fast blood flow},}\ }}\href {\doibase
  10.1117/1.2355666} {\bibfield  {journal} {\bibinfo  {journal} {J. Biomed.
  Opt}\ }\textbf {\bibinfo {volume} {11}},\ \bibinfo {pages} {054034} (\bibinfo
  {year} {2006})}\BibitemShut {NoStop}%
\bibitem [{\citenamefont {Tomaiuolo}\ \emph {et~al.}(2007)\citenamefont
  {Tomaiuolo}, \citenamefont {Preziosi}, \citenamefont {Simeone}, \citenamefont
  {Guido}, \citenamefont {Ciancia}, \citenamefont {Martinelli}, \citenamefont
  {Rinaldi},\ and\ \citenamefont {Rotoli}}]{Tomaiuolo2007}%
  \BibitemOpen
  \bibfield  {author} {\bibinfo {author} {\bibfnamefont {G.}~\bibnamefont
  {Tomaiuolo}}, \bibinfo {author} {\bibfnamefont {V.}~\bibnamefont {Preziosi}},
  \bibinfo {author} {\bibfnamefont {M.}~\bibnamefont {Simeone}}, \bibinfo
  {author} {\bibfnamefont {S.}~\bibnamefont {Guido}}, \bibinfo {author}
  {\bibfnamefont {R.}~\bibnamefont {Ciancia}}, \bibinfo {author} {\bibfnamefont
  {V.}~\bibnamefont {Martinelli}}, \bibinfo {author} {\bibfnamefont
  {C.}~\bibnamefont {Rinaldi}}, \ and\ \bibinfo {author} {\bibfnamefont
  {B.}~\bibnamefont {Rotoli}},\ }\bibfield  {title} {\emph {\enquote {\bibinfo
  {title} {A methodology to study the deformability of red blood cells flowing
  in microcapillaries in vitro},}\ }}\href@noop {} {\bibfield  {journal}
  {\bibinfo  {journal} {Ann. Ist. Super. Sanit{\`a}}\ }\textbf {\bibinfo
  {volume} {43}},\ \bibinfo {pages} {186} (\bibinfo {year} {2007})}\BibitemShut
  {NoStop}%
\bibitem [{\citenamefont {Guido}\ and\ \citenamefont
  {Tomaiuolo}(2009)}]{Guido2009}%
  \BibitemOpen
  \bibfield  {author} {\bibinfo {author} {\bibfnamefont {S.}~\bibnamefont
  {Guido}}\ and\ \bibinfo {author} {\bibfnamefont {G.}~\bibnamefont
  {Tomaiuolo}},\ }\bibfield  {title} {\emph {\enquote {\bibinfo {title}
  {Microconfined flow behavior of red blood cells in vitro},}\ }}\href
  {\doibase 10.1016/j.crhy.2009.10.002} {\bibfield  {journal} {\bibinfo
  {journal} {Comptes Rendus Phys.}\ }\textbf {\bibinfo {volume} {10}},\
  \bibinfo {pages} {751} (\bibinfo {year} {2009})}\BibitemShut {NoStop}%
\bibitem [{\citenamefont {Gorthi}\ and\ \citenamefont
  {Schonbrun}(2012)}]{Gorthi2012}%
  \BibitemOpen
  \bibfield  {author} {\bibinfo {author} {\bibfnamefont {S.~S.}\ \bibnamefont
  {Gorthi}}\ and\ \bibinfo {author} {\bibfnamefont {E.}~\bibnamefont
  {Schonbrun}},\ }\bibfield  {title} {\emph {\enquote {\bibinfo {title} {Phase
  imaging flow cytometry using a focus-stack collecting microscope},}\ }}\href
  {\doibase 10.1364/OL.37.000707} {\bibfield  {journal} {\bibinfo  {journal}
  {Opt. Lett.}\ }\textbf {\bibinfo {volume} {37}},\ \bibinfo {pages} {707}
  (\bibinfo {year} {2012})}\BibitemShut {NoStop}%
\bibitem [{\citenamefont {Lanotte}\ \emph {et~al.}(2014)\citenamefont
  {Lanotte}, \citenamefont {Tomaiuolo}, \citenamefont {Misbah}, \citenamefont
  {Bureau},\ and\ \citenamefont {Guido}}]{Lanotte2014}%
  \BibitemOpen
  \bibfield  {author} {\bibinfo {author} {\bibfnamefont {L.}~\bibnamefont
  {Lanotte}}, \bibinfo {author} {\bibfnamefont {G.}~\bibnamefont {Tomaiuolo}},
  \bibinfo {author} {\bibfnamefont {C.}~\bibnamefont {Misbah}}, \bibinfo
  {author} {\bibfnamefont {L.}~\bibnamefont {Bureau}}, \ and\ \bibinfo {author}
  {\bibfnamefont {S.}~\bibnamefont {Guido}},\ }\bibfield  {title} {\emph
  {\enquote {\bibinfo {title} {Red blood cell dynamics in polymer brush-coated
  microcapillaries: {{A}} model of endothelial glycocalyx in vitro},}\ }}\href
  {\doibase 10.1063/1.4863723} {\bibfield  {journal} {\bibinfo  {journal}
  {Biomicrofluidics}\ }\textbf {\bibinfo {volume} {8}},\ \bibinfo {pages}
  {014104} (\bibinfo {year} {2014})}\BibitemShut {NoStop}%
\bibitem [{\citenamefont {Tomaiuolo}\ \emph {et~al.}(2016)\citenamefont
  {Tomaiuolo}, \citenamefont {Lanotte}, \citenamefont {D'Apolito},
  \citenamefont {Cassinese},\ and\ \citenamefont {Guido}}]{Tomaiuolo2016}%
  \BibitemOpen
  \bibfield  {author} {\bibinfo {author} {\bibfnamefont {G.}~\bibnamefont
  {Tomaiuolo}}, \bibinfo {author} {\bibfnamefont {L.}~\bibnamefont {Lanotte}},
  \bibinfo {author} {\bibfnamefont {R.}~\bibnamefont {D'Apolito}}, \bibinfo
  {author} {\bibfnamefont {A.}~\bibnamefont {Cassinese}}, \ and\ \bibinfo
  {author} {\bibfnamefont {S.}~\bibnamefont {Guido}},\ }\bibfield  {title}
  {\emph {\enquote {\bibinfo {title} {Microconfined flow behavior of red blood
  cells},}\ }}\href {\doibase 10.1016/j.medengphy.2015.05.007} {\bibfield
  {journal} {\bibinfo  {journal} {Med. Eng. Phys.}\ }\textbf {\bibinfo {volume}
  {38}},\ \bibinfo {pages} {11} (\bibinfo {year} {2016})}\BibitemShut {NoStop}%
\bibitem [{\citenamefont {Claver{\'\i}a}\ \emph {et~al.}(2016)\citenamefont
  {Claver{\'\i}a}, \citenamefont {Aouane}, \citenamefont {Thi{\'e}baud},
  \citenamefont {Abkarian}, \citenamefont {Coupier}, \citenamefont {Misbah},
  \citenamefont {John},\ and\ \citenamefont {Wagner}}]{Claveria2016}%
  \BibitemOpen
  \bibfield  {author} {\bibinfo {author} {\bibfnamefont {V.}~\bibnamefont
  {Claver{\'\i}a}}, \bibinfo {author} {\bibfnamefont {O.}~\bibnamefont
  {Aouane}}, \bibinfo {author} {\bibfnamefont {M.}~\bibnamefont
  {Thi{\'e}baud}}, \bibinfo {author} {\bibfnamefont {M.}~\bibnamefont
  {Abkarian}}, \bibinfo {author} {\bibfnamefont {G.}~\bibnamefont {Coupier}},
  \bibinfo {author} {\bibfnamefont {C.}~\bibnamefont {Misbah}}, \bibinfo
  {author} {\bibfnamefont {T.}~\bibnamefont {John}}, \ and\ \bibinfo {author}
  {\bibfnamefont {C.}~\bibnamefont {Wagner}},\ }\bibfield  {title} {\emph
  {\enquote {\bibinfo {title} {Clusters of red blood cells in microcapillary
  flow: Hydrodynamic versus macromolecule induced interaction},}\ }}\href
  {\doibase 10.1039/C6SM01165A} {\bibfield  {journal} {\bibinfo  {journal}
  {Soft Matter}\ }\textbf {\bibinfo {volume} {12}},\ \bibinfo {pages} {8235}
  (\bibinfo {year} {2016})}\BibitemShut {NoStop}%
\bibitem [{\citenamefont {Guest}\ \emph {et~al.}(1963)\citenamefont {Guest},
  \citenamefont {Bond}, \citenamefont {Cooper},\ and\ \citenamefont
  {Derrick}}]{Guest1963}%
  \BibitemOpen
  \bibfield  {author} {\bibinfo {author} {\bibfnamefont {M.~M.}\ \bibnamefont
  {Guest}}, \bibinfo {author} {\bibfnamefont {T.~P.}\ \bibnamefont {Bond}},
  \bibinfo {author} {\bibfnamefont {R.~G.}\ \bibnamefont {Cooper}}, \ and\
  \bibinfo {author} {\bibfnamefont {J.~R.}\ \bibnamefont {Derrick}},\
  }\bibfield  {title} {\emph {\enquote {\bibinfo {title} {Red {{Blood Cells}}:
  {{Change}} in {{Shape}} in {{Capillaries}}},}\ }}\href {\doibase
  10.1126/science.142.3597.1319} {\bibfield  {journal} {\bibinfo  {journal}
  {Science}\ }\textbf {\bibinfo {volume} {142}},\ \bibinfo {pages} {1319}
  (\bibinfo {year} {1963})}\BibitemShut {NoStop}%
\bibitem [{\citenamefont {Skalak}\ and\ \citenamefont
  {Branemark}(1969)}]{Skalak1969}%
  \BibitemOpen
  \bibfield  {author} {\bibinfo {author} {\bibfnamefont {R.}~\bibnamefont
  {Skalak}}\ and\ \bibinfo {author} {\bibfnamefont {P.~I.}\ \bibnamefont
  {Branemark}},\ }\bibfield  {title} {\emph {\enquote {\bibinfo {title}
  {Deformation of {{Red Blood Cells}} in {{Capillaries}}},}\ }}\href {\doibase
  10.1126/science.164.3880.717} {\bibfield  {journal} {\bibinfo  {journal}
  {Science}\ }\textbf {\bibinfo {volume} {164}},\ \bibinfo {pages} {717}
  (\bibinfo {year} {1969})}\BibitemShut {NoStop}%
\bibitem [{\citenamefont {Kubota}\ \emph {et~al.}(1996)\citenamefont {Kubota},
  \citenamefont {Tamura}, \citenamefont {Shirakura}, \citenamefont {K~Imura},
  \citenamefont {Y~Amanaka}, \citenamefont {Isozaki},\ and\ \citenamefont
  {Nishio}}]{Kubota1996}%
  \BibitemOpen
  \bibfield  {author} {\bibinfo {author} {\bibfnamefont {K.}~\bibnamefont
  {Kubota}}, \bibinfo {author} {\bibfnamefont {J.}~\bibnamefont {Tamura}},
  \bibinfo {author} {\bibfnamefont {T.}~\bibnamefont {Shirakura}}, \bibinfo
  {author} {\bibfnamefont {M.}~\bibnamefont {K~Imura}}, \bibinfo {author}
  {\bibfnamefont {K.}~\bibnamefont {Y~Amanaka}}, \bibinfo {author}
  {\bibfnamefont {T.}~\bibnamefont {Isozaki}}, \ and\ \bibinfo {author}
  {\bibfnamefont {I.}~\bibnamefont {Nishio}},\ }\bibfield  {title} {\emph
  {\enquote {\bibinfo {title} {The behaviour of red cells in narrow tubes in
  vitro as a model of the microcirculation},}\ }}\href {\doibase
  10.1046/j.1365-2141.1996.d01-1794.x} {\bibfield  {journal} {\bibinfo
  {journal} {Br. J. Haematol.}\ }\textbf {\bibinfo {volume} {94}},\ \bibinfo
  {pages} {266} (\bibinfo {year} {1996})}\BibitemShut {NoStop}%
\bibitem [{\citenamefont {Tomaiuolo}\ \emph {et~al.}(2012)\citenamefont
  {Tomaiuolo}, \citenamefont {Lanotte}, \citenamefont {Ghigliotti},
  \citenamefont {Misbah},\ and\ \citenamefont {Guido}}]{Tomaiuolo2012}%
  \BibitemOpen
  \bibfield  {author} {\bibinfo {author} {\bibfnamefont {G.}~\bibnamefont
  {Tomaiuolo}}, \bibinfo {author} {\bibfnamefont {L.}~\bibnamefont {Lanotte}},
  \bibinfo {author} {\bibfnamefont {G.}~\bibnamefont {Ghigliotti}}, \bibinfo
  {author} {\bibfnamefont {C.}~\bibnamefont {Misbah}}, \ and\ \bibinfo {author}
  {\bibfnamefont {S.}~\bibnamefont {Guido}},\ }\bibfield  {title} {\emph
  {\enquote {\bibinfo {title} {Red blood cell clustering in {{Poiseuille}}
  microcapillary flow},}\ }}\href {\doibase 10.1063/1.4721811} {\bibfield
  {journal} {\bibinfo  {journal} {Phys. Fluids}\ }\textbf {\bibinfo {volume}
  {24}},\ \bibinfo {eid} {051903} (\bibinfo {year} {2012})}\BibitemShut
  {NoStop}%
\bibitem [{\citenamefont {Wagner}\ \emph {et~al.}(2013)\citenamefont {Wagner},
  \citenamefont {Steffen},\ and\ \citenamefont {Svetina}}]{Wagner2013}%
  \BibitemOpen
  \bibfield  {author} {\bibinfo {author} {\bibfnamefont {C.}~\bibnamefont
  {Wagner}}, \bibinfo {author} {\bibfnamefont {P.}~\bibnamefont {Steffen}}, \
  and\ \bibinfo {author} {\bibfnamefont {S.}~\bibnamefont {Svetina}},\
  }\bibfield  {title} {\emph {\enquote {\bibinfo {title} {Aggregation of red
  blood cells: {{From}} rouleaux to clot formation},}\ }}\href {\doibase
  10.1016/j.crhy.2013.04.004} {\bibfield  {journal} {\bibinfo  {journal}
  {Comptes Rendus Physique}\ }\bibinfo {series} {Living fluids / Fluides
  vivants},\ \textbf {\bibinfo {volume} {14}},\ \bibinfo {pages} {459}
  (\bibinfo {year} {2013})}\BibitemShut {NoStop}%
\bibitem [{\citenamefont {Brust}\ \emph {et~al.}(2014)\citenamefont {Brust},
  \citenamefont {Aouane}, \citenamefont {Thi{\'e}baud}, \citenamefont
  {Flormann}, \citenamefont {Verdier}, \citenamefont {Kaestner}, \citenamefont
  {Laschke}, \citenamefont {Selmi}, \citenamefont {Benyoussef}, \citenamefont
  {Podgorski}, \citenamefont {Coupier}, \citenamefont {Misbah},\ and\
  \citenamefont {Wagner}}]{Brust2014}%
  \BibitemOpen
  \bibfield  {author} {\bibinfo {author} {\bibfnamefont {M.}~\bibnamefont
  {Brust}}, \bibinfo {author} {\bibfnamefont {O.}~\bibnamefont {Aouane}},
  \bibinfo {author} {\bibfnamefont {M.}~\bibnamefont {Thi{\'e}baud}}, \bibinfo
  {author} {\bibfnamefont {D.}~\bibnamefont {Flormann}}, \bibinfo {author}
  {\bibfnamefont {C.}~\bibnamefont {Verdier}}, \bibinfo {author} {\bibfnamefont
  {L.}~\bibnamefont {Kaestner}}, \bibinfo {author} {\bibfnamefont {M.~W.}\
  \bibnamefont {Laschke}}, \bibinfo {author} {\bibfnamefont {H.}~\bibnamefont
  {Selmi}}, \bibinfo {author} {\bibfnamefont {A.}~\bibnamefont {Benyoussef}},
  \bibinfo {author} {\bibfnamefont {T.}~\bibnamefont {Podgorski}}, \bibinfo
  {author} {\bibfnamefont {G.}~\bibnamefont {Coupier}}, \bibinfo {author}
  {\bibfnamefont {C.}~\bibnamefont {Misbah}}, \ and\ \bibinfo {author}
  {\bibfnamefont {C.}~\bibnamefont {Wagner}},\ }\bibfield  {title} {\emph
  {\enquote {\bibinfo {title} {The plasma protein fibrinogen stabilizes
  clusters of red blood cells in microcapillary flows},}\ }}\href {\doibase
  10.1038/srep04348} {\bibfield  {journal} {\bibinfo  {journal} {Sci. Rep.}\
  }\textbf {\bibinfo {volume} {4}},\ \bibinfo {pages} {4348} (\bibinfo {year}
  {2014})}\BibitemShut {NoStop}%
\bibitem [{\citenamefont {Goldsmith}\ and\ \citenamefont
  {Marlow}(1972)}]{Goldsmith1972}%
  \BibitemOpen
  \bibfield  {author} {\bibinfo {author} {\bibfnamefont {H.~L.}\ \bibnamefont
  {Goldsmith}}\ and\ \bibinfo {author} {\bibfnamefont {J.}~\bibnamefont
  {Marlow}},\ }\bibfield  {title} {\emph {\enquote {\bibinfo {title} {Flow
  {{Behaviour}} of {{Erythrocytes}}. {{I}}. {{Rotation}} and {{Deformation}} in
  {{Dilute Suspensions}}},}\ }}\href {\doibase 10.1098/rspb.1972.0084}
  {\bibfield  {journal} {\bibinfo  {journal} {Proc. R. Soc. Lond. B Biol.
  Sci.}\ }\textbf {\bibinfo {volume} {182}},\ \bibinfo {pages} {351} (\bibinfo
  {year} {1972})}\BibitemShut {NoStop}%
\bibitem [{\citenamefont {Secomb}\ \emph {et~al.}(1986)\citenamefont {Secomb},
  \citenamefont {Skalak}, \citenamefont {{\"O}zkaya},\ and\ \citenamefont
  {Gross}}]{Secomb1986}%
  \BibitemOpen
  \bibfield  {author} {\bibinfo {author} {\bibfnamefont {T.~W.}\ \bibnamefont
  {Secomb}}, \bibinfo {author} {\bibfnamefont {R.}~\bibnamefont {Skalak}},
  \bibinfo {author} {\bibfnamefont {N.}~\bibnamefont {{\"O}zkaya}}, \ and\
  \bibinfo {author} {\bibfnamefont {J.~F.}\ \bibnamefont {Gross}},\ }\bibfield
  {title} {\emph {\enquote {\bibinfo {title} {Flow of axisymmetric red blood
  cells in narrow capillaries},}\ }}\href {\doibase 10.1017/S0022112086002355}
  {\bibfield  {journal} {\bibinfo  {journal} {J. Fluid Mech.}\ }\textbf
  {\bibinfo {volume} {163}},\ \bibinfo {pages} {405} (\bibinfo {year}
  {1986})}\BibitemShut {NoStop}%
\bibitem [{\citenamefont {Secomb}(1987)}]{Secomb1987}%
  \BibitemOpen
  \bibfield  {author} {\bibinfo {author} {\bibfnamefont {T.~W.}\ \bibnamefont
  {Secomb}},\ }\bibfield  {title} {\emph {\enquote {\bibinfo {title}
  {Flow-dependent rheological properties of blood in capillaries},}\ }}\href
  {\doibase 10.1016/0026-2862(87)90078-1} {\bibfield  {journal} {\bibinfo
  {journal} {Microvascular Research}\ }\textbf {\bibinfo {volume} {34}},\
  \bibinfo {pages} {46} (\bibinfo {year} {1987})}\BibitemShut {NoStop}%
\bibitem [{\citenamefont {Secomb}\ \emph {et~al.}(2001)\citenamefont {Secomb},
  \citenamefont {Hsu},\ and\ \citenamefont {Pries}}]{Secomb2001}%
  \BibitemOpen
  \bibfield  {author} {\bibinfo {author} {\bibfnamefont {T.~W.}\ \bibnamefont
  {Secomb}}, \bibinfo {author} {\bibfnamefont {R.}~\bibnamefont {Hsu}}, \ and\
  \bibinfo {author} {\bibfnamefont {A.~R.}\ \bibnamefont {Pries}},\ }\bibfield
  {title} {\emph {\enquote {\bibinfo {title} {Motion of red blood cells in a
  capillary with an endothelial surface layer: Effect of flow velocity},}\
  }}\href@noop {} {\bibfield  {journal} {\bibinfo  {journal} {Am. J. Physiol. -
  Heart Circ. Physiol.}\ }\textbf {\bibinfo {volume} {281}},\ \bibinfo {pages}
  {H629} (\bibinfo {year} {2001})}\BibitemShut {NoStop}%
\bibitem [{\citenamefont {Secomb}\ and\ \citenamefont
  {Skalak}(1982)}]{Secomb1982}%
  \BibitemOpen
  \bibfield  {author} {\bibinfo {author} {\bibfnamefont {T.~W.}\ \bibnamefont
  {Secomb}}\ and\ \bibinfo {author} {\bibfnamefont {R.}~\bibnamefont
  {Skalak}},\ }\bibfield  {title} {\emph {\enquote {\bibinfo {title} {A
  two-dimensional model for capillary flow of an asymmetric cell},}\ }}\href
  {\doibase 10.1016/0026-2862(82)90056-5} {\bibfield  {journal} {\bibinfo
  {journal} {Microvascular Research}\ }\textbf {\bibinfo {volume} {24}},\
  \bibinfo {pages} {194} (\bibinfo {year} {1982})}\BibitemShut {NoStop}%
\bibitem [{\citenamefont {Kaoui}\ \emph
  {et~al.}(2009{\natexlab{a}})\citenamefont {Kaoui}, \citenamefont {Biros},\
  and\ \citenamefont {Misbah}}]{Kaoui2009}%
  \BibitemOpen
  \bibfield  {author} {\bibinfo {author} {\bibfnamefont {B.}~\bibnamefont
  {Kaoui}}, \bibinfo {author} {\bibfnamefont {G.}~\bibnamefont {Biros}}, \ and\
  \bibinfo {author} {\bibfnamefont {C.}~\bibnamefont {Misbah}},\ }\bibfield
  {title} {\emph {\enquote {\bibinfo {title} {Why {{Do Red Blood Cells Have
  Asymmetric Shapes Even}} in a {{Symmetric Flow}}?}}\ }}\href {\doibase
  10.1103/PhysRevLett.103.188101} {\bibfield  {journal} {\bibinfo  {journal}
  {Phys. Rev. Lett.}\ }\textbf {\bibinfo {volume} {103}},\ \bibinfo {pages}
  {188101} (\bibinfo {year} {2009}{\natexlab{a}})}\BibitemShut {NoStop}%
\bibitem [{\citenamefont {Kaoui}\ \emph {et~al.}(2011)\citenamefont {Kaoui},
  \citenamefont {Tahiri}, \citenamefont {Biben}, \citenamefont {Ez-Zahraouy},
  \citenamefont {Benyoussef}, \citenamefont {Biros},\ and\ \citenamefont
  {Misbah}}]{Kaoui2011}%
  \BibitemOpen
  \bibfield  {author} {\bibinfo {author} {\bibfnamefont {B.}~\bibnamefont
  {Kaoui}}, \bibinfo {author} {\bibfnamefont {N.}~\bibnamefont {Tahiri}},
  \bibinfo {author} {\bibfnamefont {T.}~\bibnamefont {Biben}}, \bibinfo
  {author} {\bibfnamefont {H.}~\bibnamefont {Ez-Zahraouy}}, \bibinfo {author}
  {\bibfnamefont {A.}~\bibnamefont {Benyoussef}}, \bibinfo {author}
  {\bibfnamefont {G.}~\bibnamefont {Biros}}, \ and\ \bibinfo {author}
  {\bibfnamefont {C.}~\bibnamefont {Misbah}},\ }\bibfield  {title} {\emph
  {\enquote {\bibinfo {title} {Complexity of vesicle microcirculation},}\
  }}\href {\doibase 10.1103/PhysRevE.84.041906} {\bibfield  {journal} {\bibinfo
   {journal} {Phys. Rev. E}\ }\textbf {\bibinfo {volume} {84}},\ \bibinfo
  {pages} {041906} (\bibinfo {year} {2011})}\BibitemShut {NoStop}%
\bibitem [{\citenamefont {Kaoui}\ \emph {et~al.}(2012)\citenamefont {Kaoui},
  \citenamefont {Kr{\"u}ger},\ and\ \citenamefont {Harting}}]{Kaoui2012}%
  \BibitemOpen
  \bibfield  {author} {\bibinfo {author} {\bibfnamefont {B.}~\bibnamefont
  {Kaoui}}, \bibinfo {author} {\bibfnamefont {T.}~\bibnamefont {Kr{\"u}ger}}, \
  and\ \bibinfo {author} {\bibfnamefont {J.}~\bibnamefont {Harting}},\
  }\bibfield  {title} {\emph {\enquote {\bibinfo {title} {How does confinement
  affect the dynamics of viscous vesicles and red blood cells?}}\ }}\href
  {\doibase 10.1039/C2SM26289D} {\bibfield  {journal} {\bibinfo  {journal}
  {Soft Matter}\ }\textbf {\bibinfo {volume} {8}},\ \bibinfo {pages} {9246}
  (\bibinfo {year} {2012})}\BibitemShut {NoStop}%
\bibitem [{\citenamefont {Shi}\ \emph {et~al.}(2012)\citenamefont {Shi},
  \citenamefont {Pan},\ and\ \citenamefont {Glowinski}}]{Shi2012a}%
  \BibitemOpen
  \bibfield  {author} {\bibinfo {author} {\bibfnamefont {L.}~\bibnamefont
  {Shi}}, \bibinfo {author} {\bibfnamefont {T.-W.}\ \bibnamefont {Pan}}, \ and\
  \bibinfo {author} {\bibfnamefont {R.}~\bibnamefont {Glowinski}},\ }\bibfield
  {title} {\emph {\enquote {\bibinfo {title} {Deformation of a single red blood
  cell in bounded {{Poiseuille}} flows},}\ }}\href {\doibase
  10.1103/PhysRevE.85.016307} {\bibfield  {journal} {\bibinfo  {journal} {Phys.
  Rev. E}\ }\textbf {\bibinfo {volume} {85}},\ \bibinfo {eid} {016307}
  (\bibinfo {year} {2012}),\ 10.1103/PhysRevE.85.016307}\BibitemShut {NoStop}%
\bibitem [{\citenamefont {L{\'a}zaro}\ \emph {et~al.}(2014)\citenamefont
  {L{\'a}zaro}, \citenamefont {Hern{\'a}ndez-Machado},\ and\ \citenamefont
  {Pagonabarraga}}]{Lazaro2014}%
  \BibitemOpen
  \bibfield  {author} {\bibinfo {author} {\bibfnamefont {G.~R.}\ \bibnamefont
  {L{\'a}zaro}}, \bibinfo {author} {\bibfnamefont {A.}~\bibnamefont
  {Hern{\'a}ndez-Machado}}, \ and\ \bibinfo {author} {\bibfnamefont
  {I.}~\bibnamefont {Pagonabarraga}},\ }\bibfield  {title} {\emph {\enquote
  {\bibinfo {title} {Rheology of red blood cells under flow in highly confined
  microchannels: {{I}}. effect of elasticity},}\ }}\href {\doibase
  10.1039/C4SM00894D} {\bibfield  {journal} {\bibinfo  {journal} {Soft Matter}\
  }\textbf {\bibinfo {volume} {10}},\ \bibinfo {pages} {7195} (\bibinfo {year}
  {2014})}\BibitemShut {NoStop}%
\bibitem [{\citenamefont {Noguchi}\ and\ \citenamefont
  {Gompper}(2005)}]{Noguchi2005}%
  \BibitemOpen
  \bibfield  {author} {\bibinfo {author} {\bibfnamefont {H.}~\bibnamefont
  {Noguchi}}\ and\ \bibinfo {author} {\bibfnamefont {G.}~\bibnamefont
  {Gompper}},\ }\bibfield  {title} {\emph {\enquote {\bibinfo {title} {Shape
  transitions of fluid vesicles and red blood cells in capillary flows},}\
  }}\href {\doibase 10.1073/pnas.0504243102} {\bibfield  {journal} {\bibinfo
  {journal} {Proc. Natl. Acad. Sci.}\ }\textbf {\bibinfo {volume} {102}},\
  \bibinfo {pages} {14159} (\bibinfo {year} {2005})}\BibitemShut {NoStop}%
\bibitem [{\citenamefont {McWhirter}\ \emph {et~al.}(2011)\citenamefont
  {McWhirter}, \citenamefont {Noguchi},\ and\ \citenamefont
  {Gompper}}]{McWhirter2011}%
  \BibitemOpen
  \bibfield  {author} {\bibinfo {author} {\bibfnamefont {J.~L.}\ \bibnamefont
  {McWhirter}}, \bibinfo {author} {\bibfnamefont {H.}~\bibnamefont {Noguchi}},
  \ and\ \bibinfo {author} {\bibfnamefont {G.}~\bibnamefont {Gompper}},\
  }\bibfield  {title} {\emph {\enquote {\bibinfo {title} {Deformation and
  clustering of red blood cells in microcapillary flows},}\ }}\href {\doibase
  10.1039/C1SM05794D} {\bibfield  {journal} {\bibinfo  {journal} {Soft Matter}\
  }\textbf {\bibinfo {volume} {7}},\ \bibinfo {pages} {10967} (\bibinfo {year}
  {2011})}\BibitemShut {NoStop}%
\bibitem [{\citenamefont {Ye}\ \emph {et~al.}(2017)\citenamefont {Ye},
  \citenamefont {Shi}, \citenamefont {Peng},\ and\ \citenamefont
  {Li}}]{Ye2017}%
  \BibitemOpen
  \bibfield  {author} {\bibinfo {author} {\bibfnamefont {T.}~\bibnamefont
  {Ye}}, \bibinfo {author} {\bibfnamefont {H.}~\bibnamefont {Shi}}, \bibinfo
  {author} {\bibfnamefont {L.}~\bibnamefont {Peng}}, \ and\ \bibinfo {author}
  {\bibfnamefont {Y.}~\bibnamefont {Li}},\ }\bibfield  {title} {\emph {\enquote
  {\bibinfo {title} {Numerical studies of a red blood cell in rectangular
  microchannels},}\ }}\href {\doibase 10.1063/1.5000357} {\bibfield  {journal}
  {\bibinfo  {journal} {J. Appl. Phys.}\ }\textbf {\bibinfo {volume} {122}},\
  \bibinfo {pages} {084701} (\bibinfo {year} {2017})}\BibitemShut {NoStop}%
\bibitem [{\citenamefont {Kaoui}\ \emph
  {et~al.}(2009{\natexlab{b}})\citenamefont {Kaoui}, \citenamefont {Coupier},
  \citenamefont {Misbah},\ and\ \citenamefont {Podgorski}}]{Kaoui2009a}%
  \BibitemOpen
  \bibfield  {author} {\bibinfo {author} {\bibfnamefont {B.}~\bibnamefont
  {Kaoui}}, \bibinfo {author} {\bibfnamefont {G.}~\bibnamefont {Coupier}},
  \bibinfo {author} {\bibfnamefont {C.}~\bibnamefont {Misbah}}, \ and\ \bibinfo
  {author} {\bibfnamefont {T.}~\bibnamefont {Podgorski}},\ }\bibfield  {title}
  {\emph {\enquote {\bibinfo {title} {Lateral migration of vesicles in
  microchannels: Effects of walls and shear gradient},}\ }}\href {\doibase
  10.1051/lhb/2009063} {\bibfield  {journal} {\bibinfo  {journal} {Houille
  Blanche}\ ,\ \bibinfo {pages} {112}} (\bibinfo {year}
  {2009}{\natexlab{b}})}\BibitemShut {NoStop}%
\bibitem [{\citenamefont {Cordasco}\ \emph {et~al.}(2014)\citenamefont
  {Cordasco}, \citenamefont {Yazdani},\ and\ \citenamefont
  {Bagchi}}]{Cordasco2014}%
  \BibitemOpen
  \bibfield  {author} {\bibinfo {author} {\bibfnamefont {D.}~\bibnamefont
  {Cordasco}}, \bibinfo {author} {\bibfnamefont {A.}~\bibnamefont {Yazdani}}, \
  and\ \bibinfo {author} {\bibfnamefont {P.}~\bibnamefont {Bagchi}},\
  }\bibfield  {title} {\emph {\enquote {\bibinfo {title} {Comparison of
  erythrocyte dynamics in shear flow under different stress-free
  configurations},}\ }}\href {\doibase 10.1063/1.4871300} {\bibfield  {journal}
  {\bibinfo  {journal} {Phys. Fluids}\ }\textbf {\bibinfo {volume} {26}},\
  \bibinfo {pages} {041902} (\bibinfo {year} {2014})}\BibitemShut {NoStop}%
\bibitem [{\citenamefont {Peng}\ \emph {et~al.}(2014)\citenamefont {Peng},
  \citenamefont {Mashayekh},\ and\ \citenamefont {Zhu}}]{Peng2014}%
  \BibitemOpen
  \bibfield  {author} {\bibinfo {author} {\bibfnamefont {Z.}~\bibnamefont
  {Peng}}, \bibinfo {author} {\bibfnamefont {A.}~\bibnamefont {Mashayekh}}, \
  and\ \bibinfo {author} {\bibfnamefont {Q.}~\bibnamefont {Zhu}},\ }\bibfield
  {title} {\emph {\enquote {\bibinfo {title} {Erythrocyte responses in
  low-shear-rate flows: Effects of non-biconcave stress-free state in the
  cytoskeleton},}\ }}\href {\doibase 10.1017/jfm.2014.14} {\bibfield  {journal}
  {\bibinfo  {journal} {J. Fluid Mech.}\ }\textbf {\bibinfo {volume} {742}},\
  \bibinfo {pages} {96} (\bibinfo {year} {2014})}\BibitemShut {NoStop}%
\bibitem [{\citenamefont {Sinha}\ and\ \citenamefont
  {Graham}(2015)}]{Sinha2015}%
  \BibitemOpen
  \bibfield  {author} {\bibinfo {author} {\bibfnamefont {K.}~\bibnamefont
  {Sinha}}\ and\ \bibinfo {author} {\bibfnamefont {M.~D.}\ \bibnamefont
  {Graham}},\ }\bibfield  {title} {\emph {\enquote {\bibinfo {title} {Dynamics
  of a single red blood cell in simple shear flow},}\ }}\href {\doibase
  10.1103/PhysRevE.92.042710} {\bibfield  {journal} {\bibinfo  {journal} {Phys.
  Rev. E}\ }\textbf {\bibinfo {volume} {92}},\ \bibinfo {pages} {042710}
  (\bibinfo {year} {2015})}\BibitemShut {NoStop}%
\bibitem [{\citenamefont {Cokelet}\ and\ \citenamefont
  {Meiselman}(1968)}]{Cokelet1968}%
  \BibitemOpen
  \bibfield  {author} {\bibinfo {author} {\bibfnamefont {G.~R.}\ \bibnamefont
  {Cokelet}}\ and\ \bibinfo {author} {\bibfnamefont {H.~J.}\ \bibnamefont
  {Meiselman}},\ }\bibfield  {title} {\emph {\enquote {\bibinfo {title}
  {Rheological {{Comparison}} of {{Hemoglobin Solutions}} and {{Erythrocyte
  Suspensions}}},}\ }}\href {\doibase 10.1126/science.162.3850.275} {\bibfield
  {journal} {\bibinfo  {journal} {Science}\ }\textbf {\bibinfo {volume}
  {162}},\ \bibinfo {pages} {275} (\bibinfo {year} {1968})}\BibitemShut
  {NoStop}%
\bibitem [{\citenamefont {Popel}\ and\ \citenamefont
  {Johnson}(2005)}]{Popel2005}%
  \BibitemOpen
  \bibfield  {author} {\bibinfo {author} {\bibfnamefont {A.~S.}\ \bibnamefont
  {Popel}}\ and\ \bibinfo {author} {\bibfnamefont {P.~C.}\ \bibnamefont
  {Johnson}},\ }\bibfield  {title} {\emph {\enquote {\bibinfo {title}
  {Microcirculation and {{Hemorheology}}},}\ }}\href {\doibase
  10.1146/annurev.fluid.37.042604.133933} {\bibfield  {journal} {\bibinfo
  {journal} {Annu. Rev. Fluid Mech.}\ }\textbf {\bibinfo {volume} {37}},\
  \bibinfo {pages} {43} (\bibinfo {year} {2005})}\BibitemShut {NoStop}%
\bibitem [{\citenamefont {Cui}\ \emph {et~al.}(2002)\citenamefont {Cui},
  \citenamefont {Diamant},\ and\ \citenamefont {Lin}}]{Cui2002}%
  \BibitemOpen
  \bibfield  {author} {\bibinfo {author} {\bibfnamefont {B.}~\bibnamefont
  {Cui}}, \bibinfo {author} {\bibfnamefont {H.}~\bibnamefont {Diamant}}, \ and\
  \bibinfo {author} {\bibfnamefont {B.}~\bibnamefont {Lin}},\ }\bibfield
  {title} {\emph {\enquote {\bibinfo {title} {Screened {{Hydrodynamic
  Interaction}} in a {{Narrow Channel}}},}\ }}\href {\doibase
  10.1103/PhysRevLett.89.188302} {\bibfield  {journal} {\bibinfo  {journal}
  {Phys. Rev. Lett.}\ }\textbf {\bibinfo {volume} {89}},\ \bibinfo {pages}
  {188302} (\bibinfo {year} {2002})}\BibitemShut {NoStop}%
\bibitem [{\citenamefont {Diamant}(2009)}]{Diamant2009}%
  \BibitemOpen
  \bibfield  {author} {\bibinfo {author} {\bibfnamefont {H.}~\bibnamefont
  {Diamant}},\ }\bibfield  {title} {\emph {\enquote {\bibinfo {title}
  {Hydrodynamic {{Interaction}} in {{Confined Geometries}}},}\ }}\href
  {\doibase 10.1143/JPSJ.78.041002} {\bibfield  {journal} {\bibinfo  {journal}
  {J. Phys. Soc. Jpn.}\ }\textbf {\bibinfo {volume} {78}},\ \bibinfo {pages}
  {041002} (\bibinfo {year} {2009})}\BibitemShut {NoStop}%
\bibitem [{\citenamefont {Koller}\ \emph {et~al.}(1987)\citenamefont {Koller},
  \citenamefont {Dawant}, \citenamefont {Liu}, \citenamefont {Popel},\ and\
  \citenamefont {Johnson}}]{Koller1987}%
  \BibitemOpen
  \bibfield  {author} {\bibinfo {author} {\bibfnamefont {A.}~\bibnamefont
  {Koller}}, \bibinfo {author} {\bibfnamefont {B.}~\bibnamefont {Dawant}},
  \bibinfo {author} {\bibfnamefont {A.}~\bibnamefont {Liu}}, \bibinfo {author}
  {\bibfnamefont {A.~S.}\ \bibnamefont {Popel}}, \ and\ \bibinfo {author}
  {\bibfnamefont {P.~C.}\ \bibnamefont {Johnson}},\ }\bibfield  {title} {\emph
  {\enquote {\bibinfo {title} {Quantitative analysis of arteriolar network
  architecture in cat sartorius muscle},}\ }}\href@noop {} {\bibfield
  {journal} {\bibinfo  {journal} {Am. J. Physiol. - Heart Circ. Physiol.}\
  }\textbf {\bibinfo {volume} {253}},\ \bibinfo {pages} {H154} (\bibinfo {year}
  {1987})}\BibitemShut {NoStop}%
\bibitem [{\citenamefont {Evans}\ and\ \citenamefont {Fung}(1972)}]{Evans1972}%
  \BibitemOpen
  \bibfield  {author} {\bibinfo {author} {\bibfnamefont {E.}~\bibnamefont
  {Evans}}\ and\ \bibinfo {author} {\bibfnamefont {Y.-C.}\ \bibnamefont
  {Fung}},\ }\bibfield  {title} {\emph {\enquote {\bibinfo {title} {Improved
  measurements of the erythrocyte geometry},}\ }}\href {\doibase
  10.1016/0026-2862(72)90069-6} {\bibfield  {journal} {\bibinfo  {journal}
  {Microvasc. Res.}\ }\textbf {\bibinfo {volume} {4}},\ \bibinfo {pages} {335}
  (\bibinfo {year} {1972})}\BibitemShut {NoStop}%
\bibitem [{\citenamefont {Le}(2010)}]{Le2010a}%
  \BibitemOpen
  \bibfield  {author} {\bibinfo {author} {\bibfnamefont {D.-V.}\ \bibnamefont
  {Le}},\ }\bibfield  {title} {\emph {\enquote {\bibinfo {title} {Subdivision
  elements for large deformation of liquid capsules enclosed by thin shells},}\
  }}\href {\doibase 10.1016/j.cma.2010.04.014} {\bibfield  {journal} {\bibinfo
  {journal} {Comput. Methods Appl. Mech. Eng.}\ }\textbf {\bibinfo {volume}
  {199}},\ \bibinfo {pages} {2622} (\bibinfo {year} {2010})}\BibitemShut
  {NoStop}%
\bibitem [{\citenamefont {Chien}\ \emph {et~al.}(1966)\citenamefont {Chien},
  \citenamefont {Usami}, \citenamefont {Taylor}, \citenamefont {Lundberg},\
  and\ \citenamefont {Gregersen}}]{Chien1966}%
  \BibitemOpen
  \bibfield  {author} {\bibinfo {author} {\bibfnamefont {S.}~\bibnamefont
  {Chien}}, \bibinfo {author} {\bibfnamefont {S.}~\bibnamefont {Usami}},
  \bibinfo {author} {\bibfnamefont {H.~M.}\ \bibnamefont {Taylor}}, \bibinfo
  {author} {\bibfnamefont {J.~L.}\ \bibnamefont {Lundberg}}, \ and\ \bibinfo
  {author} {\bibfnamefont {M.~I.}\ \bibnamefont {Gregersen}},\ }\bibfield
  {title} {\emph {\enquote {\bibinfo {title} {Effects of hematocrit and plasma
  proteins on human blood rheology at low shear rates.}}\ }}\href@noop {}
  {\bibfield  {journal} {\bibinfo  {journal} {J. Appl. Physiol.}\ }\textbf
  {\bibinfo {volume} {21}},\ \bibinfo {pages} {81} (\bibinfo {year}
  {1966})}\BibitemShut {NoStop}%
\bibitem [{\citenamefont {Skalak}\ \emph {et~al.}(1989)\citenamefont {Skalak},
  \citenamefont {Ozkaya},\ and\ \citenamefont {Skalak}}]{Skalak1989}%
  \BibitemOpen
  \bibfield  {author} {\bibinfo {author} {\bibfnamefont {R.}~\bibnamefont
  {Skalak}}, \bibinfo {author} {\bibfnamefont {N.}~\bibnamefont {Ozkaya}}, \
  and\ \bibinfo {author} {\bibfnamefont {T.~C.}\ \bibnamefont {Skalak}},\
  }\bibfield  {title} {\emph {\enquote {\bibinfo {title} {Biofluid
  {{Mechanics}}},}\ }}\href {\doibase 10.1146/annurev.fl.21.010189.001123}
  {\bibfield  {journal} {\bibinfo  {journal} {Annu. Rev. Fluid Mech.}\ }\textbf
  {\bibinfo {volume} {21}},\ \bibinfo {pages} {167} (\bibinfo {year}
  {1989})}\BibitemShut {NoStop}%
\bibitem [{\citenamefont {Secomb}(2017)}]{Secomb2017}%
  \BibitemOpen
  \bibfield  {author} {\bibinfo {author} {\bibfnamefont {T.~W.}\ \bibnamefont
  {Secomb}},\ }\bibfield  {title} {\emph {\enquote {\bibinfo {title} {Blood
  {{Flow}} in the {{Microcirculation}}},}\ }}\href {\doibase
  10.1146/annurev-fluid-010816-060302} {\bibfield  {journal} {\bibinfo
  {journal} {Annu. Rev. Fluid Mech.}\ }\textbf {\bibinfo {volume} {49}},\
  \bibinfo {pages} {443} (\bibinfo {year} {2017})}\BibitemShut {NoStop}%
\bibitem [{\citenamefont {Kim}\ \emph {et~al.}(2014)\citenamefont {Kim},
  \citenamefont {Shim}, \citenamefont {Kim}, \citenamefont {Park},
  \citenamefont {Jang},\ and\ \citenamefont {Park}}]{Kim2014a}%
  \BibitemOpen
  \bibfield  {author} {\bibinfo {author} {\bibfnamefont {Y.}~\bibnamefont
  {Kim}}, \bibinfo {author} {\bibfnamefont {H.}~\bibnamefont {Shim}}, \bibinfo
  {author} {\bibfnamefont {K.}~\bibnamefont {Kim}}, \bibinfo {author}
  {\bibfnamefont {H.}~\bibnamefont {Park}}, \bibinfo {author} {\bibfnamefont
  {S.}~\bibnamefont {Jang}}, \ and\ \bibinfo {author} {\bibfnamefont
  {Y.}~\bibnamefont {Park}},\ }\bibfield  {title} {\emph {\enquote {\bibinfo
  {title} {Profiling individual human red blood cells using common-path
  diffraction optical tomography},}\ }}\href {\doibase 10.1038/srep06659}
  {\bibfield  {journal} {\bibinfo  {journal} {Sci. Rep.}\ }\textbf {\bibinfo
  {volume} {4}},\ \bibinfo {pages} {6659} (\bibinfo {year} {2014})}\BibitemShut
  {NoStop}%
\bibitem [{\citenamefont {Skalak}\ \emph {et~al.}(1973)\citenamefont {Skalak},
  \citenamefont {Tozeren}, \citenamefont {Zarda},\ and\ \citenamefont
  {Chien}}]{Skalak1973}%
  \BibitemOpen
  \bibfield  {author} {\bibinfo {author} {\bibfnamefont {R.}~\bibnamefont
  {Skalak}}, \bibinfo {author} {\bibfnamefont {A.}~\bibnamefont {Tozeren}},
  \bibinfo {author} {\bibfnamefont {R.~P.}\ \bibnamefont {Zarda}}, \ and\
  \bibinfo {author} {\bibfnamefont {S.}~\bibnamefont {Chien}},\ }\bibfield
  {title} {\emph {\enquote {\bibinfo {title} {Strain {{Energy Function}} of
  {{Red Blood Cell Membranes}}},}\ }}\href {\doibase
  10.1016/S0006-3495(73)85983-1} {\bibfield  {journal} {\bibinfo  {journal}
  {Biophys. J.}\ }\textbf {\bibinfo {volume} {13}},\ \bibinfo {pages} {245}
  (\bibinfo {year} {1973})}\BibitemShut {NoStop}%
\bibitem [{\citenamefont {Kr{\"u}ger}\ \emph {et~al.}(2011)\citenamefont
  {Kr{\"u}ger}, \citenamefont {Varnik},\ and\ \citenamefont
  {Raabe}}]{Kruger2011}%
  \BibitemOpen
  \bibfield  {author} {\bibinfo {author} {\bibfnamefont {T.}~\bibnamefont
  {Kr{\"u}ger}}, \bibinfo {author} {\bibfnamefont {F.}~\bibnamefont {Varnik}},
  \ and\ \bibinfo {author} {\bibfnamefont {D.}~\bibnamefont {Raabe}},\
  }\bibfield  {title} {\emph {\enquote {\bibinfo {title} {Efficient and
  accurate simulations of deformable particles immersed in a fluid using a
  combined immersed boundary lattice {{Boltzmann}} finite element method},}\
  }}\href {\doibase 10.1016/j.camwa.2010.03.057} {\bibfield  {journal}
  {\bibinfo  {journal} {Comput. Math. Appl.}\ }\textbf {\bibinfo {volume}
  {61}},\ \bibinfo {pages} {3485} (\bibinfo {year} {2011})}\BibitemShut
  {NoStop}%
\bibitem [{\citenamefont {Yoon}\ \emph {et~al.}(2008)\citenamefont {Yoon},
  \citenamefont {Kotar}, \citenamefont {Yoon},\ and\ \citenamefont
  {Cicuta}}]{Yoon2008}%
  \BibitemOpen
  \bibfield  {author} {\bibinfo {author} {\bibfnamefont {Y.-Z.}\ \bibnamefont
  {Yoon}}, \bibinfo {author} {\bibfnamefont {J.}~\bibnamefont {Kotar}},
  \bibinfo {author} {\bibfnamefont {G.}~\bibnamefont {Yoon}}, \ and\ \bibinfo
  {author} {\bibfnamefont {P.}~\bibnamefont {Cicuta}},\ }\bibfield  {title}
  {\emph {\enquote {\bibinfo {title} {The nonlinear mechanical response of the
  red blood cell},}\ }}\href {\doibase 10.1088/1478-3975/5/3/036007} {\bibfield
   {journal} {\bibinfo  {journal} {Phys. Biol.}\ }\textbf {\bibinfo {volume}
  {5}},\ \bibinfo {pages} {036007} (\bibinfo {year} {2008})}\BibitemShut
  {NoStop}%
\bibitem [{\citenamefont {Freund}(2014)}]{Freund2014}%
  \BibitemOpen
  \bibfield  {author} {\bibinfo {author} {\bibfnamefont {J.~B.}\ \bibnamefont
  {Freund}},\ }\bibfield  {title} {\emph {\enquote {\bibinfo {title} {Numerical
  {{Simulation}} of {{Flowing Blood Cells}}},}\ }}\href {\doibase
  10.1146/annurev-fluid-010313-141349} {\bibfield  {journal} {\bibinfo
  {journal} {Annu. Rev. Fluid Mech.}\ }\textbf {\bibinfo {volume} {46}},\
  \bibinfo {pages} {67} (\bibinfo {year} {2014})}\BibitemShut {NoStop}%
\bibitem [{\citenamefont {Canham}(1970)}]{Canham1970}%
  \BibitemOpen
  \bibfield  {author} {\bibinfo {author} {\bibfnamefont {P.~B.}\ \bibnamefont
  {Canham}},\ }\bibfield  {title} {\emph {\enquote {\bibinfo {title} {The
  minimum energy of bending as a possible explanation of the biconcave shape of
  the human red blood cell},}\ }}\href {\doibase 10.1016/S0022-5193(70)80032-7}
  {\bibfield  {journal} {\bibinfo  {journal} {J. Theor. Biol.}\ }\textbf
  {\bibinfo {volume} {26}},\ \bibinfo {pages} {61} (\bibinfo {year}
  {1970})}\BibitemShut {NoStop}%
\bibitem [{\citenamefont {Helfrich}(1973)}]{Helfrich1973}%
  \BibitemOpen
  \bibfield  {author} {\bibinfo {author} {\bibfnamefont {W.}~\bibnamefont
  {Helfrich}},\ }\bibfield  {title} {\emph {\enquote {\bibinfo {title} {Elastic
  {{Properties}} of {{Lipid Bilayers}}: {{Theory}} and {{Possible
  Experiments}}},}\ }}\href {\doibase 10.1515/znc-1973-11-1209} {\bibfield
  {journal} {\bibinfo  {journal} {Z. Naturforsch. C}\ }\textbf {\bibinfo
  {volume} {28}},\ \bibinfo {pages} {693} (\bibinfo {year} {1973})}\BibitemShut
  {NoStop}%
\bibitem [{\citenamefont {Guckenberger}\ and\ \citenamefont
  {Gekle}(2017{\natexlab{a}})}]{Guckenberger2017}%
  \BibitemOpen
  \bibfield  {author} {\bibinfo {author} {\bibfnamefont {A.}~\bibnamefont
  {Guckenberger}}\ and\ \bibinfo {author} {\bibfnamefont {S.}~\bibnamefont
  {Gekle}},\ }\bibfield  {title} {\emph {\enquote {\bibinfo {title} {Theory and
  algorithms to compute {{Helfrich}} bending forces: A review},}\ }}\href
  {\doibase 10.1088/1361-648X/aa6313} {\bibfield  {journal} {\bibinfo
  {journal} {J. Phys. Condens. Matter}\ }\textbf {\bibinfo {volume} {29}},\
  \bibinfo {pages} {203001} (\bibinfo {year} {2017}{\natexlab{a}})}\BibitemShut
  {NoStop}%
\bibitem [{\citenamefont {Park}\ \emph {et~al.}(2010)\citenamefont {Park},
  \citenamefont {Best}, \citenamefont {Badizadegan}, \citenamefont {Dasari},
  \citenamefont {Feld}, \citenamefont {Kuriabova}, \citenamefont {Henle},
  \citenamefont {Levine},\ and\ \citenamefont {Popescu}}]{Park2010}%
  \BibitemOpen
  \bibfield  {author} {\bibinfo {author} {\bibfnamefont {Y.}~\bibnamefont
  {Park}}, \bibinfo {author} {\bibfnamefont {C.~A.}\ \bibnamefont {Best}},
  \bibinfo {author} {\bibfnamefont {K.}~\bibnamefont {Badizadegan}}, \bibinfo
  {author} {\bibfnamefont {R.~R.}\ \bibnamefont {Dasari}}, \bibinfo {author}
  {\bibfnamefont {M.~S.}\ \bibnamefont {Feld}}, \bibinfo {author}
  {\bibfnamefont {T.}~\bibnamefont {Kuriabova}}, \bibinfo {author}
  {\bibfnamefont {M.~L.}\ \bibnamefont {Henle}}, \bibinfo {author}
  {\bibfnamefont {A.~J.}\ \bibnamefont {Levine}}, \ and\ \bibinfo {author}
  {\bibfnamefont {G.}~\bibnamefont {Popescu}},\ }\bibfield  {title} {\emph
  {\enquote {\bibinfo {title} {Measurement of red blood cell mechanics during
  morphological changes},}\ }}\href {\doibase 10.1073/pnas.0909533107}
  {\bibfield  {journal} {\bibinfo  {journal} {Proc. Natl. Acad. Sci.}\ }\textbf
  {\bibinfo {volume} {107}},\ \bibinfo {pages} {6731} (\bibinfo {year}
  {2010})}\BibitemShut {NoStop}%
\bibitem [{\citenamefont {Guckenberger}\ \emph {et~al.}(2016)\citenamefont
  {Guckenberger}, \citenamefont {Schraml}, \citenamefont {Chen}, \citenamefont
  {Leonetti},\ and\ \citenamefont {Gekle}}]{Guckenberger2016}%
  \BibitemOpen
  \bibfield  {author} {\bibinfo {author} {\bibfnamefont {A.}~\bibnamefont
  {Guckenberger}}, \bibinfo {author} {\bibfnamefont {M.~P.}\ \bibnamefont
  {Schraml}}, \bibinfo {author} {\bibfnamefont {P.~G.}\ \bibnamefont {Chen}},
  \bibinfo {author} {\bibfnamefont {M.}~\bibnamefont {Leonetti}}, \ and\
  \bibinfo {author} {\bibfnamefont {S.}~\bibnamefont {Gekle}},\ }\bibfield
  {title} {\emph {\enquote {\bibinfo {title} {On the bending algorithms for
  soft objects in flows},}\ }}\href {\doibase 10.1016/j.cpc.2016.04.018}
  {\bibfield  {journal} {\bibinfo  {journal} {Comput. Phys. Comm.}\ }\textbf
  {\bibinfo {volume} {207}},\ \bibinfo {pages} {1} (\bibinfo {year}
  {2016})}\BibitemShut {NoStop}%
\bibitem [{\citenamefont {Farutin}\ \emph {et~al.}(2014)\citenamefont
  {Farutin}, \citenamefont {Biben},\ and\ \citenamefont
  {Misbah}}]{Farutin2014}%
  \BibitemOpen
  \bibfield  {author} {\bibinfo {author} {\bibfnamefont {A.}~\bibnamefont
  {Farutin}}, \bibinfo {author} {\bibfnamefont {T.}~\bibnamefont {Biben}}, \
  and\ \bibinfo {author} {\bibfnamefont {C.}~\bibnamefont {Misbah}},\
  }\bibfield  {title} {\emph {\enquote {\bibinfo {title} {{{3D}} numerical
  simulations of vesicle and inextensible capsule dynamics},}\ }}\href
  {\doibase 10.1016/j.jcp.2014.07.008} {\bibfield  {journal} {\bibinfo
  {journal} {J. Comput. Phys.}\ }\textbf {\bibinfo {volume} {275}},\ \bibinfo
  {pages} {539} (\bibinfo {year} {2014})}\BibitemShut {NoStop}%
\bibitem [{\citenamefont {Guckenberger}\ and\ \citenamefont
  {Gekle}(2017{\natexlab{b}})}]{BubblePreprint2}%
  \BibitemOpen
  \bibfield  {author} {\bibinfo {author} {\bibfnamefont {A.}~\bibnamefont
  {Guckenberger}}\ and\ \bibinfo {author} {\bibfnamefont {S.}~\bibnamefont
  {Gekle}},\ }\bibfield  {title} {\emph {\enquote {\bibinfo {title} {A boundary
  integral method with volume-changing objects for ultrasound-triggered
  margination of microbubbles},}\ }}\href@noop {} {\bibfield  {journal}
  {\bibinfo  {journal} {arXiv:1608.05196 [physics]}\ } (\bibinfo {year}
  {2017}{\natexlab{b}})},\ \Eprint {http://arxiv.org/abs/1608.05196}
  {arXiv:1608.05196 [physics]} \BibitemShut {NoStop}%
\bibitem [{\citenamefont {Pozrikidis}(2001)}]{Pozrikidis2001a}%
  \BibitemOpen
  \bibfield  {author} {\bibinfo {author} {\bibfnamefont {C.}~\bibnamefont
  {Pozrikidis}},\ }\bibfield  {title} {\emph {\enquote {\bibinfo {title}
  {Interfacial {{Dynamics}} for {{Stokes Flow}}},}\ }}\href {\doibase
  10.1006/jcph.2000.6582} {\bibfield  {journal} {\bibinfo  {journal} {J.
  Comput. Phys.}\ }\textbf {\bibinfo {volume} {169}},\ \bibinfo {pages} {250}
  (\bibinfo {year} {2001})}\BibitemShut {NoStop}%
\bibitem [{\citenamefont {Zhao}\ \emph {et~al.}(2010)\citenamefont {Zhao},
  \citenamefont {Isfahani}, \citenamefont {Olson},\ and\ \citenamefont
  {Freund}}]{Zhao2010}%
  \BibitemOpen
  \bibfield  {author} {\bibinfo {author} {\bibfnamefont {H.}~\bibnamefont
  {Zhao}}, \bibinfo {author} {\bibfnamefont {A.~H.}\ \bibnamefont {Isfahani}},
  \bibinfo {author} {\bibfnamefont {L.~N.}\ \bibnamefont {Olson}}, \ and\
  \bibinfo {author} {\bibfnamefont {J.~B.}\ \bibnamefont {Freund}},\ }\bibfield
   {title} {\emph {\enquote {\bibinfo {title} {A spectral boundary integral
  method for flowing blood cells},}\ }}\href {\doibase
  10.1016/j.jcp.2010.01.024} {\bibfield  {journal} {\bibinfo  {journal} {J.
  Comput. Phys.}\ }\textbf {\bibinfo {volume} {229}},\ \bibinfo {pages} {3726}
  (\bibinfo {year} {2010})}\BibitemShut {NoStop}%
\bibitem [{\citenamefont {Saintillan}\ \emph {et~al.}(2005)\citenamefont
  {Saintillan}, \citenamefont {Darve},\ and\ \citenamefont
  {Shaqfeh}}]{Saintillan2005}%
  \BibitemOpen
  \bibfield  {author} {\bibinfo {author} {\bibfnamefont {D.}~\bibnamefont
  {Saintillan}}, \bibinfo {author} {\bibfnamefont {E.}~\bibnamefont {Darve}}, \
  and\ \bibinfo {author} {\bibfnamefont {E.~S.~G.}\ \bibnamefont {Shaqfeh}},\
  }\bibfield  {title} {\emph {\enquote {\bibinfo {title} {A smooth
  particle-mesh {{Ewald}} algorithm for {{Stokes}} suspension simulations:
  {{The}} sedimentation of fibers},}\ }}\href {\doibase 10.1063/1.1862262}
  {\bibfield  {journal} {\bibinfo  {journal} {Phys. Fluids}\ }\textbf {\bibinfo
  {volume} {17}},\ \bibinfo {eid} {033301} (\bibinfo {year}
  {2005})}\BibitemShut {NoStop}%
\bibitem [{\citenamefont {Saad}\ and\ \citenamefont
  {Schultz}(1986)}]{Saad1986}%
  \BibitemOpen
  \bibfield  {author} {\bibinfo {author} {\bibfnamefont {Y.}~\bibnamefont
  {Saad}}\ and\ \bibinfo {author} {\bibfnamefont {M.}~\bibnamefont {Schultz}},\
  }\bibfield  {title} {\emph {\enquote {\bibinfo {title} {{{GMRES}}: {{A
  Generalized Minimal Residual Algorithm}} for {{Solving Nonsymmetric Linear
  Systems}}},}\ }}\href {\doibase 10.1137/0907058} {\bibfield  {journal}
  {\bibinfo  {journal} {SIAM J. Sci. Stat. Comput.}\ }\textbf {\bibinfo
  {volume} {7}},\ \bibinfo {pages} {856} (\bibinfo {year} {1986})}\BibitemShut
  {NoStop}%
\bibitem [{\citenamefont {Bogacki}\ and\ \citenamefont
  {Shampine}(1989)}]{Bogacki1989}%
  \BibitemOpen
  \bibfield  {author} {\bibinfo {author} {\bibfnamefont {P.}~\bibnamefont
  {Bogacki}}\ and\ \bibinfo {author} {\bibfnamefont {L.~F.}\ \bibnamefont
  {Shampine}},\ }\bibfield  {title} {\emph {\enquote {\bibinfo {title} {A 3(2)
  pair of {{Runge}} - {{Kutta}} formulas},}\ }}\href {\doibase
  10.1016/0893-9659(89)90079-7} {\bibfield  {journal} {\bibinfo  {journal}
  {Appl. Math. Lett.}\ }\textbf {\bibinfo {volume} {2}},\ \bibinfo {pages}
  {321} (\bibinfo {year} {1989})}\BibitemShut {NoStop}%
\bibitem [{\citenamefont {Daddi-Moussa-Ider}\ \emph {et~al.}(2016)\citenamefont
  {Daddi-Moussa-Ider}, \citenamefont {Guckenberger},\ and\ \citenamefont
  {Gekle}}]{Daddi-Moussa-Ider2016}%
  \BibitemOpen
  \bibfield  {author} {\bibinfo {author} {\bibfnamefont {A.}~\bibnamefont
  {Daddi-Moussa-Ider}}, \bibinfo {author} {\bibfnamefont {A.}~\bibnamefont
  {Guckenberger}}, \ and\ \bibinfo {author} {\bibfnamefont {S.}~\bibnamefont
  {Gekle}},\ }\bibfield  {title} {\emph {\enquote {\bibinfo {title} {Long-lived
  anomalous thermal diffusion induced by elastic cell membranes on nearby
  particles},}\ }}\href {\doibase 10.1103/PhysRevE.93.012612} {\bibfield
  {journal} {\bibinfo  {journal} {Phys. Rev. E}\ }\textbf {\bibinfo {volume}
  {93}},\ \bibinfo {eid} {012612} (\bibinfo {year} {2016})}\BibitemShut
  {NoStop}%
\bibitem [{\citenamefont {Pries}\ \emph {et~al.}(1995)\citenamefont {Pries},
  \citenamefont {Secomb},\ and\ \citenamefont {Gaehtgens}}]{Pries1995}%
  \BibitemOpen
  \bibfield  {author} {\bibinfo {author} {\bibfnamefont {A.~R.}\ \bibnamefont
  {Pries}}, \bibinfo {author} {\bibfnamefont {T.~W.}\ \bibnamefont {Secomb}}, \
  and\ \bibinfo {author} {\bibfnamefont {P.}~\bibnamefont {Gaehtgens}},\
  }\bibfield  {title} {\emph {\enquote {\bibinfo {title} {Structure and
  hemodynamics of microvascular networks: Heterogeneity and correlations},}\
  }}\href@noop {} {\bibfield  {journal} {\bibinfo  {journal} {Am. J. Physiol. -
  Heart Circ. Physiol.}\ }\textbf {\bibinfo {volume} {269}},\ \bibinfo {pages}
  {H1713} (\bibinfo {year} {1995})}\BibitemShut {NoStop}%
\bibitem [{\citenamefont {Baskurt}\ \emph {et~al.}(2011)\citenamefont
  {Baskurt}, \citenamefont {Neu},\ and\ \citenamefont
  {Meiselman}}]{Baskurt2011}%
  \BibitemOpen
  \bibfield  {author} {\bibinfo {author} {\bibfnamefont {O.}~\bibnamefont
  {Baskurt}}, \bibinfo {author} {\bibfnamefont {B.}~\bibnamefont {Neu}}, \ and\
  \bibinfo {author} {\bibfnamefont {H.}~\bibnamefont {Meiselman}},\ }\href
  {\doibase 10.1201/b11221} {\emph {\bibinfo {title} {Red {{Blood Cell
  Aggregation}}}}}\ (\bibinfo  {publisher} {{CRC Press}},\ \bibinfo {year}
  {2011})\BibitemShut {NoStop}%
\bibitem [{\citenamefont {Kaoui}\ and\ \citenamefont
  {Harting}(2016)}]{Kaoui2016a}%
  \BibitemOpen
  \bibfield  {author} {\bibinfo {author} {\bibfnamefont {B.}~\bibnamefont
  {Kaoui}}\ and\ \bibinfo {author} {\bibfnamefont {J.}~\bibnamefont
  {Harting}},\ }\bibfield  {title} {\emph {\enquote {\bibinfo {title}
  {Two-dimensional lattice {{Boltzmann}} simulations of vesicles with viscosity
  contrast},}\ }}\href {\doibase 10.1007/s00397-015-0867-6} {\bibfield
  {journal} {\bibinfo  {journal} {Rheol Acta}\ }\textbf {\bibinfo {volume}
  {55}},\ \bibinfo {pages} {465} (\bibinfo {year} {2016})}\BibitemShut
  {NoStop}%
\bibitem [{\citenamefont {Schaaf}\ and\ \citenamefont
  {Stark}(2017)}]{Schaaf2017}%
  \BibitemOpen
  \bibfield  {author} {\bibinfo {author} {\bibfnamefont {C.}~\bibnamefont
  {Schaaf}}\ and\ \bibinfo {author} {\bibfnamefont {H.}~\bibnamefont {Stark}},\
  }\bibfield  {title} {\emph {\enquote {\bibinfo {title} {Inertial migration
  and axial control of deformable capsules},}\ }}\href {\doibase
  10.1039/C7SM00339K} {\bibfield  {journal} {\bibinfo  {journal} {Soft Matter}\
  }\textbf {\bibinfo {volume} {13}},\ \bibinfo {pages} {3544} (\bibinfo {year}
  {2017})}\BibitemShut {NoStop}%
\bibitem [{\citenamefont {P{\'e}gard}\ and\ \citenamefont
  {Fleischer}(2013)}]{Pegard2013}%
  \BibitemOpen
  \bibfield  {author} {\bibinfo {author} {\bibfnamefont {N.~C.}\ \bibnamefont
  {P{\'e}gard}}\ and\ \bibinfo {author} {\bibfnamefont {J.~W.}\ \bibnamefont
  {Fleischer}},\ }\bibfield  {title} {\emph {\enquote {\bibinfo {title}
  {Three-dimensional deconvolution microfluidic microscopy using a tilted
  channel},}\ }}\href {\doibase 10.1117/1.JBO.18.4.040503} {\bibfield
  {journal} {\bibinfo  {journal} {J. Biomed. Opt.}\ }\textbf {\bibinfo {volume}
  {18}},\ \bibinfo {pages} {040503} (\bibinfo {year} {2013})}\BibitemShut
  {NoStop}%
\bibitem [{\citenamefont {P{\'e}gard}\ \emph {et~al.}(2014)\citenamefont
  {P{\'e}gard}, \citenamefont {Toth}, \citenamefont {Driscoll},\ and\
  \citenamefont {Fleischer}}]{Pegard2014}%
  \BibitemOpen
  \bibfield  {author} {\bibinfo {author} {\bibfnamefont {N.~C.}\ \bibnamefont
  {P{\'e}gard}}, \bibinfo {author} {\bibfnamefont {M.~L.}\ \bibnamefont
  {Toth}}, \bibinfo {author} {\bibfnamefont {M.}~\bibnamefont {Driscoll}}, \
  and\ \bibinfo {author} {\bibfnamefont {J.~W.}\ \bibnamefont {Fleischer}},\
  }\bibfield  {title} {\emph {\enquote {\bibinfo {title} {Flow-scanning optical
  tomography},}\ }}\href {\doibase 10.1039/C4LC00701H} {\bibfield  {journal}
  {\bibinfo  {journal} {Lab Chip}\ }\textbf {\bibinfo {volume} {14}},\ \bibinfo
  {pages} {4447} (\bibinfo {year} {2014})}\BibitemShut {NoStop}%
\bibitem [{\citenamefont {Jagannadh}\ \emph {et~al.}(2016)\citenamefont
  {Jagannadh}, \citenamefont {Mackenzie}, \citenamefont {Pal}, \citenamefont
  {Kar},\ and\ \citenamefont {Gorthi}}]{Jagannadh2016}%
  \BibitemOpen
  \bibfield  {author} {\bibinfo {author} {\bibfnamefont {V.~K.}\ \bibnamefont
  {Jagannadh}}, \bibinfo {author} {\bibfnamefont {M.~D.}\ \bibnamefont
  {Mackenzie}}, \bibinfo {author} {\bibfnamefont {P.}~\bibnamefont {Pal}},
  \bibinfo {author} {\bibfnamefont {A.~K.}\ \bibnamefont {Kar}}, \ and\
  \bibinfo {author} {\bibfnamefont {S.~S.}\ \bibnamefont {Gorthi}},\ }\bibfield
   {title} {\emph {\enquote {\bibinfo {title} {Slanted channel microfluidic
  chip for {{3D}} fluorescence imaging of cells in flow},}\ }}\href {\doibase
  10.1364/OE.24.022144} {\bibfield  {journal} {\bibinfo  {journal} {Opt.
  Express}\ }\textbf {\bibinfo {volume} {24}},\ \bibinfo {pages} {22144}
  (\bibinfo {year} {2016})}\BibitemShut {NoStop}%
\bibitem [{\citenamefont {Kim}\ \emph {et~al.}(2016)\citenamefont {Kim},
  \citenamefont {Choe}, \citenamefont {Park}, \citenamefont {Kim},\ and\
  \citenamefont {Park}}]{Kim2016}%
  \BibitemOpen
  \bibfield  {author} {\bibinfo {author} {\bibfnamefont {K.}~\bibnamefont
  {Kim}}, \bibinfo {author} {\bibfnamefont {K.}~\bibnamefont {Choe}}, \bibinfo
  {author} {\bibfnamefont {I.}~\bibnamefont {Park}}, \bibinfo {author}
  {\bibfnamefont {P.}~\bibnamefont {Kim}}, \ and\ \bibinfo {author}
  {\bibfnamefont {Y.}~\bibnamefont {Park}},\ }\bibfield  {title} {\emph
  {\enquote {\bibinfo {title} {Holographic intravital microscopy for 2-{{D}}
  and 3-{{D}} imaging intact circulating blood cells in microcapillaries of
  live mice},}\ }}\href {\doibase 10.1038/srep33084} {\bibfield  {journal}
  {\bibinfo  {journal} {Sci. Rep.}\ }\textbf {\bibinfo {volume} {6}},\ \bibinfo
  {pages} {33084} (\bibinfo {year} {2016})}\BibitemShut {NoStop}%
\bibitem [{\citenamefont {Tomaiuolo}\ \emph {et~al.}(2011)\citenamefont
  {Tomaiuolo}, \citenamefont {Barra}, \citenamefont {Preziosi}, \citenamefont
  {Cassinese}, \citenamefont {Rotoli},\ and\ \citenamefont
  {Guido}}]{Tomaiuolo2011a}%
  \BibitemOpen
  \bibfield  {author} {\bibinfo {author} {\bibfnamefont {G.}~\bibnamefont
  {Tomaiuolo}}, \bibinfo {author} {\bibfnamefont {M.}~\bibnamefont {Barra}},
  \bibinfo {author} {\bibfnamefont {V.}~\bibnamefont {Preziosi}}, \bibinfo
  {author} {\bibfnamefont {A.}~\bibnamefont {Cassinese}}, \bibinfo {author}
  {\bibfnamefont {B.}~\bibnamefont {Rotoli}}, \ and\ \bibinfo {author}
  {\bibfnamefont {S.}~\bibnamefont {Guido}},\ }\bibfield  {title} {\emph
  {\enquote {\bibinfo {title} {Microfluidics analysis of red blood cell
  membrane viscoelasticity},}\ }}\href {\doibase 10.1039/C0LC00348D} {\bibfield
   {journal} {\bibinfo  {journal} {Lab Chip}\ }\textbf {\bibinfo {volume}
  {11}},\ \bibinfo {pages} {449} (\bibinfo {year} {2011})}\BibitemShut
  {NoStop}%
\bibitem [{\citenamefont {Yazdani}\ and\ \citenamefont
  {Bagchi}(2013)}]{Yazdani2013}%
  \BibitemOpen
  \bibfield  {author} {\bibinfo {author} {\bibfnamefont {A.}~\bibnamefont
  {Yazdani}}\ and\ \bibinfo {author} {\bibfnamefont {P.}~\bibnamefont
  {Bagchi}},\ }\bibfield  {title} {\emph {\enquote {\bibinfo {title} {Influence
  of membrane viscosity on capsule dynamics in shear flow},}\ }}\href {\doibase
  10.1017/jfm.2012.637} {\bibfield  {journal} {\bibinfo  {journal} {J. Fluid
  Mech.}\ }\textbf {\bibinfo {volume} {718}},\ \bibinfo {pages} {569} (\bibinfo
  {year} {2013})}\BibitemShut {NoStop}%
\end{thebibliography}%

\clearpage
\newpage\hspace{0cm}\includepdf[pages=1,fitpaper]{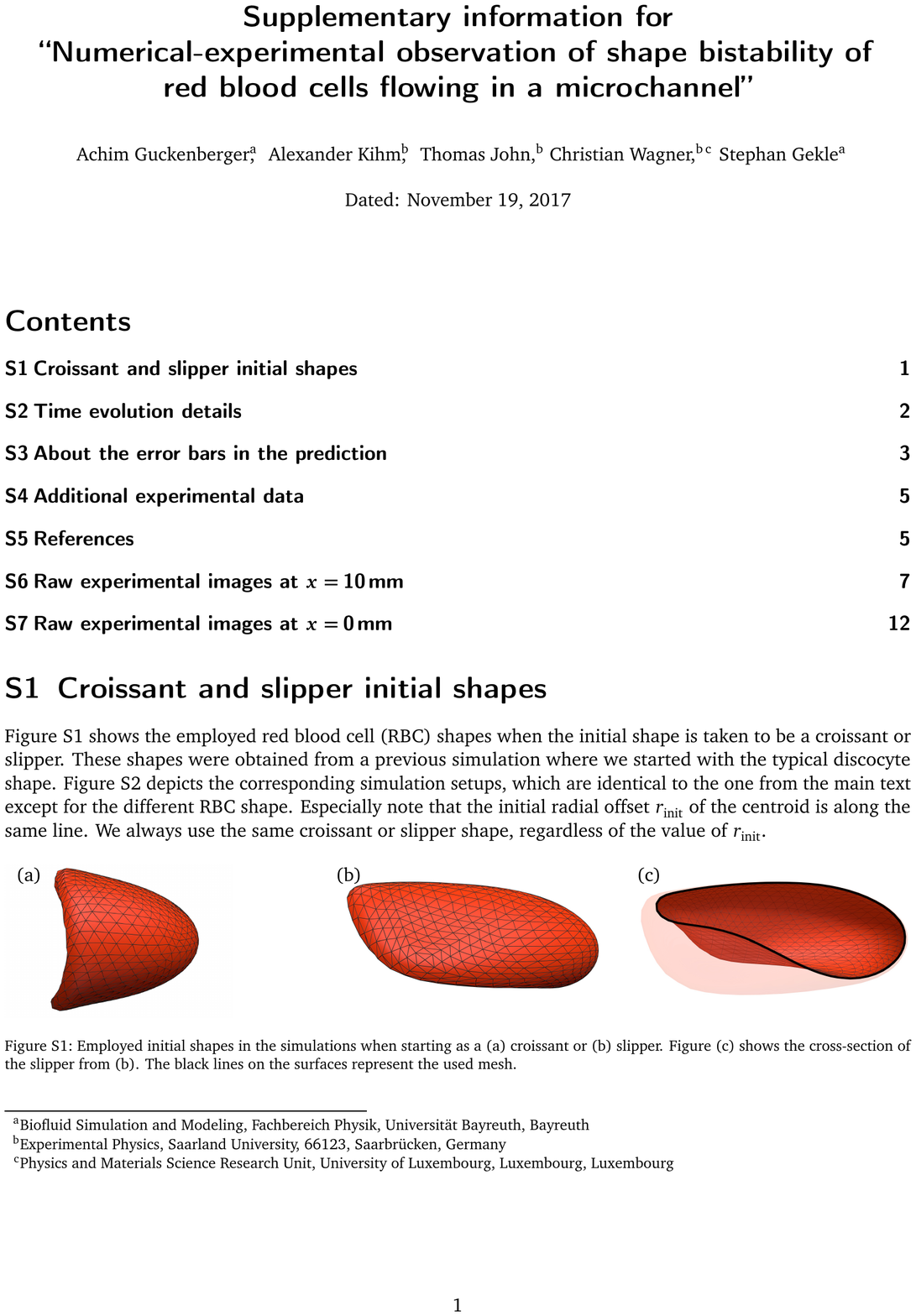}
\newpage\hspace{0cm}\includepdf[pages=2,fitpaper]{SI.pdf}
\newpage\hspace{0cm}\includepdf[pages=3,fitpaper]{SI.pdf}
\newpage\hspace{0cm}\includepdf[pages=4,fitpaper]{SI.pdf}
\newpage\hspace{0cm}\includepdf[pages=5,fitpaper]{SI.pdf}
\newpage\hspace{0cm}\includepdf[pages=6,fitpaper]{SI.pdf}
\begin{landscape}
\newpage\hspace{0cm}\includepdf[pages=7,fitpaper,angle=90, offset=0 -2cm]{SI.pdf}
\end{landscape}
\begin{landscape}
	\newpage\hspace{0cm}\includepdf[pages=8,fitpaper,angle=90, offset=0 -2cm]{SI.pdf}
\end{landscape}
\begin{landscape}
	\newpage\hspace{0cm}\includepdf[pages=9,fitpaper,angle=90, offset=0 -2cm]{SI.pdf}
\end{landscape}
\begin{landscape}
	\newpage\hspace{0cm}\includepdf[pages=10,fitpaper,angle=90, offset=0 -2cm]{SI.pdf}
\end{landscape}
\begin{landscape}
	\newpage\hspace{0cm}\includepdf[pages=11,fitpaper,angle=90, offset=0 -2cm]{SI.pdf}
\end{landscape}
\begin{landscape}
	\newpage\hspace{0cm}\includepdf[pages=12,fitpaper,angle=90, offset=0 -2cm]{SI.pdf}
\end{landscape}
\begin{landscape}
	\newpage\hspace{0cm}\includepdf[pages=13,fitpaper,angle=90, offset=0 -2cm]{SI.pdf}
\end{landscape}

\end{document}